\documentclass[10pt,twocolumn,letterpaper]{article}

\usepackage{cvpr}
\usepackage{times}
\usepackage{epsfig}
\usepackage{graphicx}
\usepackage{amsmath}
\usepackage{amssymb}
\usepackage{xcolor}
\usepackage{booktabs}
\usepackage{multirow}
\usepackage{makecell}

\usepackage{array}
\newcolumntype{C}[1]{>{\centering\let\newline\\\arraybackslash\hspace{0pt}}m{#1}}

\newcommand{\parsection}[1]{\noindent\textbf{#1} }

\usepackage[pagebackref=true,breaklinks=true,letterpaper=true,colorlinks,bookmarks=false]{hyperref}

\cvprfinalcopy 

\ifcvprfinal\fi
\begin{document}

\title{NTIRE 2020 Challenge on Real-World Image Super-Resolution:\\ Methods and Results}

\author{Andreas Lugmayr$^*$ \and Martin Danelljan$^*$ \and Radu Timofte$^*$ \and Namhyuk Ahn \and Dongwoon Bai \and Jie Cai \and Yun Cao \and Junyang Chen \and Kaihua Cheng \and SeYoung Chun \and Wei Deng \and Mostafa El-Khamy \and Chiu Man Ho \and Xiaozhong Ji \and Amin Kheradmand \and Gwantae Kim \and Hanseok Ko \and Kanghyu Lee \and Jungwon Lee \and Hao Li \and Ziluan Liu \and Zhi-Song Liu \and Shuai Liu \and Yunhua Lu \and Zibo Meng \and Pablo Navarrete Michelini \and Christian Micheloni \and Kalpesh Prajapati \and Haoyu Ren \and Yong Hyeok Seo \and Wan-Chi Siu \and Kyung-Ah Sohn \and Ying Tai \and Rao Muhammad Umer \and Shuangquan Wang \and Huibing Wang \and Timothy Haoning Wu \and Haoning Wu \and Biao Yang \and Fuzhi Yang \and Jaejun Yoo \and Tongtong Zhao \and Yuanbo Zhou \and Haijie Zhuo \and Ziyao Zong \and Xueyi Zou}

\maketitle
\thispagestyle{empty}

\begin{abstract}
This paper reviews the NTIRE 2020 challenge on real world super-resolution. It focuses on the participating methods and final results. The challenge addresses the real world setting, where paired true high and low-resolution images are unavailable. For training, only one set of source input images is therefore provided along with a set of unpaired high-quality target images. In \emph{Track~1: Image Processing artifacts}, the aim is to super-resolve images with synthetically generated image processing artifacts. This allows for quantitative benchmarking of the approaches \wrt a ground-truth image. In \emph{Track~2: Smartphone Images}, real low-quality smart phone images have to be super-resolved. In both tracks, the ultimate goal is to achieve the best perceptual quality, evaluated using a human study.
This is the second challenge on the subject, following AIM 2019, targeting to advance the state-of-the-art in super-resolution. To measure the performance we use the benchmark protocol from AIM 2019. In total 22 teams competed in the final testing phase, demonstrating new and innovative solutions to the problem.

\end{abstract}
{\let\thefootnote\relax\footnotetext{%
\hspace{-5mm}$^*$Andreas Lugmayr (\texttt{andreas.lugmayr@vision.ee.ethz.ch}), Martin Danelljan, and Radu Timofte at ETH Z\"urich are the NTIRE 2020 challenge organizers. The other authors participated in the challenge.\\
\ref{sec:affiliation} contains the authors' team names and affiliations.\\ \url{https://data.vision.ee.ethz.ch/cvl/ntire20/}
}}

\begin{figure}[t]
\centering\vspace{-3mm}%
	\newcommand{\wid}{\linewidth}
	\begin{minipage}{0.07\linewidth}
	\resizebox{0.6cm}{!}{
	\rotatebox{90}{
		\begin{tabular}{ C{0.0cm} C{2cm} C{2cm}}
       & Track 2 & Track 1~~~~ \\
		\end{tabular}%
		}}%
	\end{minipage}%
	\begin{minipage}{0.93\linewidth}
	\resizebox{\linewidth}{!}{%
		\begin{tabular}{C{2.5cm} C{2.5cm}}
			Source Domain &  Target Domain \\
		\end{tabular}%
	}\\
	\includegraphics[width=\wid]{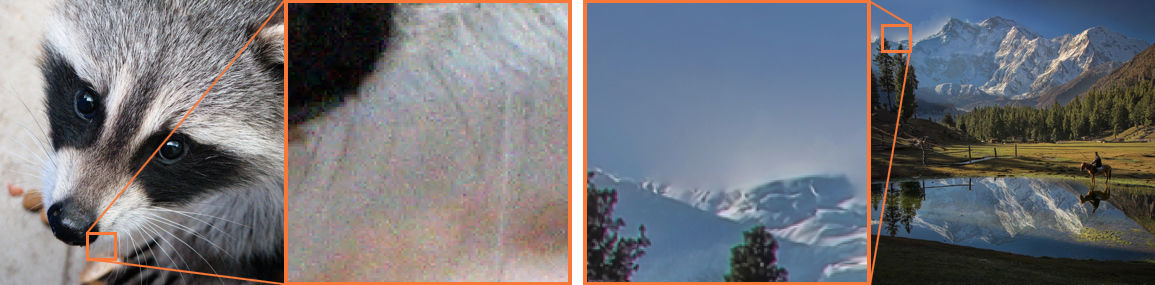}\vspace{-1mm}
	\vspace{0mm}
	\resizebox{\linewidth}{!}{%
		\begin{tabular}{C{4cm} C{4cm}}
			\small{Image Processing Artifacts} &  \small{Clean High Quality Image} \\
		\end{tabular}
	}\vspace{2mm}\\
	\includegraphics[width=\wid]{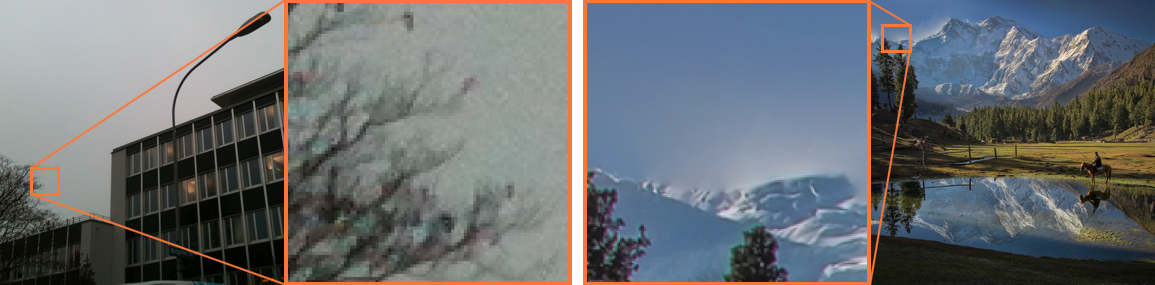}\vspace{-1mm}
	\resizebox{\linewidth}{!}{%
		\begin{tabular}{C{4cm} C{4cm}}
			\small{Smartphone Image} &  \small{Clean High Quality Image} \\
		\end{tabular}
	}\\
	\end{minipage}\vspace{-5mm}
	\caption{Visual example of the input LR images and ground truth HR images used in the challenge. For Track 1 the input is generated with a common image signal processing operation to simulate the real-world SR case where we can measure against a undisclosed ground truth. For Track 2 we the input are untouched iPhone3 images. Both tracks have the goal to super-resolve to a clean target domain.}%
	\vspace{-4mm}
	\label{fig:intro}
\end{figure}

\section{Introduction}

Single image Super-Resolution (SR) is the task of increasing the resolution of a given image by filling in additional high-frequency content. It has been a popular research topic for decades \cite{irani1991improving,freeman2002example,park2003super,Timofte13AnchNeighReg,Timofte2014a+,YangY13SimpleFuncSR,YangWHM08SRAsSparseRepresentationOfRawPatches,YangWHM10SRViaSparseRep,SunH12SRFromInternetScaleSceneMatching,DaiTG15JointlyOptimizedRegressorsForSR,huang2015single,Timofte16SevenWays,dong2014learning,dong2016image,kim2016accurate,lai2017deep,lim2017EDSR,fan2017balanced,ahn2018fast,ahn2018image,haris2018deep,huang2018densely,AIM2019ESR} due to its many applications. 
The current trend addresses the ill-posed SR problem using deep Convolutional Neural Networks (CNNs). While initial methods focused on achieving high fidelity in terms of PSNR~\cite{dong2014learning,dong2016image,kim2016accurate,lai2017deep,lim2017EDSR}. Recent work has put further emphasis on generating perceptually more appealing predictions using for instance adversarial losses~\cite{URDGN,ledig2017photo,wang2018esrgan}.

Deep learning based SR methods are known to consume large quantities of training data. Most current approaches rely on paired low and high-resolution images to train the network in a fully supervised manner. However, such image pairs are not available in real-world applications. To circumvent this fact, the conventional approach has been to downscale images, often with a bicubic kernel, to artificially generate corresponding LR images. This strategy significantly changes the low-level characteristics of the image, by \eg severely reducing the sensor noise. Super-resolution networks trained on downscaled images therefore often struggle to generalize to natural images. The research direction of blind super-resolution~\cite{michaeli2013nonparametric, gu2019blind,begin2004blind} does not fully address this setting since it often relies on paired data and constrained image formation models.
In this challenge, the aim is instead to learn super-resolution from unpaired data and without any restricting assumptions on the input image formation. This scenario has recently attracted significant interest due to its high relevance in applications~\cite{yuan2018unsupervised,kim2018task,bulat2018learn,lugmayrICCVW2019}.

The NTIRE 2020 Challenge on Real-World Image Super-Resolution aims to stimulate research in the direction of real-world super-resolution. No paired reference HR images are available for training. Instead, the participants are only provided the source input images, along with an unpaired set of high-quality images that act as the target quality domain. The challenge consists of two tracks. The source images for Track 1 are generated by performing a degradation operation that is unknown to the participants. This degradation arise from image signal processing methods similar to those found on low-end devices (see Figure~\ref{fig:intro} an example). A synthetic degradation allows us to compute reference-based metrics for evaluation. Track 2 employs images taken from a low-quality smartphone camera, with no available ground-truth. 
For both tracks, the goal is to achieve perceptually pleasing results. The final ranking is therefore performed using a human study.

This challenge is one of the NTIRE 2020 associated challenges on: deblurring~\cite{nah2020ntire}, nonhomogeneous dehazing~\cite{ancuti2020ntire}, perceptual extreme super-resolution~\cite{zhang2020ntire}, video quality mapping~\cite{fuoli2020ntire}, real image denoising~\cite{abdelhamed2020ntire}, real-world super-resolution~\cite{lugmayr2020ntire}, spectral reconstruction from RGB image~\cite{arad2020ntire} and demoireing~\cite{yuan2020demoireing}. 

\section{NTIRE 2020 Challenge}

The goals of the NTIRE 2020 Challenge on Real-World Image Super-Resolution is to (i) promote research into weak and unsupervised learning approaches for SR, that jointly enhance the image quality (ii) promote a benchmark protocol and dataset; and (iii) probe the current state-of-the-art in the field.
The challenge contains two tracks. Both tracks have the goal of upscaling with factor $4\times$. The competition was organized using the Codalab platform.

\subsection{Track 1: Image Processing Artifacts}
This track employs the benchmarking strategy described in~\cite{lugmayrICCVW2019}, which employs an artificial degradation operator to enable reference-based evaluation.

\parsection{Degradation operator}
We employ an undisclosed degradation operator which generates structured artifacts commonly produced by the kind of image processing pipelines found on very low-end devices. This type of degradation operator is very different from what has been used in previous challenges \cite{AIM2019RWSRchallenge}.
This operation is applied to all source domain images of train, validation and test.
According to the rules of the challenge, the participants were not permitted to try to reverse-engineer or with hand-crafted algorithms construct similar-looking degradation artifacts.
It was however allowed to try to \emph{learn} the degradation operator using generic techniques (such as deep networks) that can be applied to any other sort of degradations or source of natural images. The reason is that the method as a whole needs to generalize to different types of degradations and input domains.

\parsection{Data}
The dataset is constructed following the general strategy used for Track 2 in the previous edition of the challenge~\cite{AIM2019RWSRchallenge}.
We construct a dataset of source (\ie input) domain training images $\mathcal{X}_\text{train} = \{x_i\}$ by applying the degradation operation to the 2650 images of the Flickr2K~\cite{wang2018esrgan} dataset, without performing any downsampling. The target domain for training $\mathcal{Y}_\text{train} = \{y_j\}$ are the original 800 clean high-quality training images from DIV2K. For validation and testing, we employ the corresponding splits from the DIV2K~\cite{timofte2017ntire} dataset. The source domain images $\mathcal{X}_\text{val}$ and $\mathcal{X}_\text{test}$ are obtained by first downscaling the images followed by the degradation. The Ground Truth images for validation $\mathcal{Y}_\text{val}^\text{tr1}$ and test $\mathcal{Y}_\text{test}^\text{tr1}$ are the original DIV2K images. A visual example for source and target images are provided in Figure~\ref{fig:intro}.

\subsection{Track 2: Smartphone Images}
Here the task is to super-resolve real-world images obtained from a low-quality smartphone camera. The desired output quality is defined by set of clean high-quality images. We employ the iPhone3 images of the DPED~\cite{IgnatovKTVG17DPED} dataset as source domain $\mathcal{X}_\text{train}$.
For training and validation, we employ the corresponding predefined splits of DPED.
As a ground truth to super-resolved images above sensor size does not exist, we use crops of the validation set of DPED for a human perception study.
The target domain $\mathcal{Y}_\text{train}$ is the same as in Track~1.
A visual example for source and target images are provided in Figure~\ref{fig:intro}.

\subsection{Challenge phases} 
The challenge had three phases: (1) Development phase: the participants got training images and the LR images of the validation set. (2) Validation phase: the participants had the opportunity to measure performance using the PSNR and SSIM metrics by submitting their results on the server for Track~1. A validation leaderboard was also available. (3) Final test phase: the participants got access to the LR test images and had to submit their super-resolved images along with description, code and model weights for their methods.
\section{Challenge Results}

\begin{table*}[t]
	\centering%
		\begin{tabular}{lllll}
\toprule
             Team &        PSNR$\uparrow$ &        SSIM$\uparrow$ &       LPIPS$\downarrow$ &    MOS$\downarrow$ \\
\midrule
 Impressionism &  24.67$_{(16)}$ &  0.683$_{(13)}$ &  0.232$_{(1)}$ &  2.195$_{(1)}$ \\
 Samsung-SLSI-MSL &  25.59$_{(12)}$ &  0.727$_{(9)}$ &  0.252$_{(2)}$ &  2.425$_{(2)}$ \\
 BOE-IOT-AIBD &  26.71$_{(4)}$ &  0.761$_{(4)}$ &  0.280$_{(4)}$ &  2.495$_{(3)}$ \\
 MSMers &  23.20$_{(18)}$&  0.651$_{(17)}$&  0.272$_{(3)}$&  2.530$_{(4)}$ \\
 KU-ISPL &  26.23$_{(6)}$&  0.747$_{(7)}$&  0.327$_{(8)}$&  2.695$_{(5)}$ \\
 InnoPeak-SR &  26.54$_{(5)}$&  0.746$_{(8)}$&  0.302$_{(5)}$&  2.740$_{(6)}$ \\
 ITS425 &  27.08$_{(2)}$&  0.779$_{(1)}$&  0.325$_{(6)}$&  2.770$_{(7)}$ \\
 MLP-SR &  24.87$_{(15)}$&  0.681$_{(14)}$&  0.325$_{(7)}$&  2.905$_{(8)}$ \\
 Webbzhou &  26.10$_{(9)}$&  0.764$_{(3)}$&  0.341$_{(9)}$&  - \\
 SR-DL &  25.67$_{(11)}$&  0.718$_{(10)}$&  0.364$_{(10)}$&  - \\
 TeamAY &  27.09$_{(1)}$&  0.773$_{(2)}$&  0.369$_{(11)}$&  - \\
 BIGFEATURE-CAMERA &  26.18$_{(7)}$&  0.750$_{(6)}$&  0.372$_{(12)}$&  - \\
  BMIPL-UNIST-YH-1 &  26.73$_{(3)}$&  0.752$_{(5)}$&  0.379$_{(13)}$&  - \\
 SVNIT1-A &  21.22$_{(19)}$&  0.576$_{(19)}$&  0.397$_{(14)}$&  - \\
 KU-ISPL2 &  25.27$_{(14)}$&  0.680$_{(15)}$&  0.460$_{(15)}$&  - \\
 SuperT &  25.79$_{(10)}$&  0.699$_{(12)}$&  0.469$_{(16)}$&  - \\
 GDUT-wp &  26.11$_{(8)}$&  0.706$_{(11)}$&  0.496$_{(17)}$&  - \\
 SVNIT1-B &  24.21$_{(17)}$&  0.617$_{(18)}$&  0.562$_{(18)}$&  - \\
 SVNIT2 &  25.39$_{(13)}$&  0.674$_{(16)}$&  0.615$_{(19)}$&  - \\
 \midrule
 AITA-Noah-A &  24.65$_{(-)}$&  0.699$_{(-)}$&  0.222$_{(-)}$&  2.245$_{(-)}$ \\
 AITA-Noah-B &  25.72$_{(-)}$&  0.737$_{(-)}$&  0.223$_{(-)}$&  2.285$_{(-)}$ \\
 \midrule
 Bicubic &  25.48$_{(-)}$&  0.680$_{(-)}$&  0.612$_{(-)}$&  3.050$_{(-)}$ \\
 ESRGAN Supervised &  24.74$_{(-)}$&  0.695$_{(-)}$&  0.207$_{(-)}$&  2.300$_{(-)}$ \\
\bottomrule
\end{tabular}
		\vspace{2mm}
	\caption{Challenge results for \textbf{Track 1}. The top section in the table contains participating methods that are ranked in the challenge. The middle section contains participating approaches that deviated from the challenge rules, whose results are reported for reference but not ranked. The bottom section contains baseline approaches. Participating methods are ranked according to their Mean Opinion Score (MOS).}
	\label{tab:track1}
\end{table*}

\begin{table*}[t]
	\centering%
		\begin{tabular}{lllll|lr@{\hspace{6mm}}|r}
\toprule
                  Team & NIQE$\downarrow$ & BRISQUE$\downarrow$ & PIQE$\downarrow$ & NRQM$\uparrow$ & PI$\downarrow$ &  IQA-Rank$\downarrow$\hspace{-3mm} & MOR$\downarrow$ \\
\midrule
 Impressionism &  5.00$_{(1)}$ &  24.4$_{(1)}$ &  17.6$_{(2)}$ &  6.50$_{(1)}$ &  4.25$_{(1)}$ &  3.958 &  1.54$_{(1)}$ \\
 AITA-Noah-A &  5.63$_{(4)}$&  33.8$_{(5)}$&  29.7$_{(8)}$&  4.23$_{(8)}$&  5.70$_{(6)}$&  7.720 &  3.04$_{(2)}$\\
 ITS425 &  8.95$_{(18)}$&  52.5$_{(18)}$&  88.6$_{(18)}$&  3.08$_{(18)}$&  7.94$_{(18)}$&  14.984 &  3.30$_{(3)}$\\
 AITA-Noah-B &  8.18$_{(17)}$&  50.1$_{(12)}$&  88.0$_{(17)}$&  3.23$_{(15)}$&  7.47$_{(17)}$&  13.386 &  3.57$_{(4)}$\\
 Webbzhou &  7.88$_{(15)}$&  51.1$_{(15)}$&  87.8$_{(16)}$&  3.27$_{(14)}$&  7.30$_{(15)}$&  12.612 &  4.44$_{(5)}$\\
 Relbmag-Eht &  5.58$_{(3)}$&  33.1$_{(3)}$&  12.5$_{(1)}$&  6.22$_{(2)}$&  4.68$_{(2)}$&  4.060 &  - \\
 MSMers &  5.43$_{(2)}$&  38.2$_{(7)}$&  20.5$_{(3)}$&  5.22$_{(5)}$&  5.10$_{(3)}$&  5.420 &  - \\
 MLP-SR &  6.45$_{(8)}$&  30.6$_{(2)}$&  29.0$_{(6)}$&  6.12$_{(3)}$&  5.17$_{(4)}$&  5.926 &  - \\
 SR-DL &  6.11$_{(5)}$&  33.5$_{(4)}$&  29.4$_{(7)}$&  5.24$_{(4)}$&  5.43$_{(5)}$&  6.272 &  - \\
 InnoPeak-SR &  7.42$_{(13)}$&  39.3$_{(8)}$&  21.5$_{(4)}$&  5.12$_{(6)}$&  6.15$_{(9)}$&  7.716 &  - \\
 QCAM &  6.21$_{(6)}$&  44.2$_{(9)}$&  49.6$_{(9)}$&  4.10$_{(10)}$&  6.05$_{(8)}$&  8.304 &  - \\
 SuperT &  6.94$_{(10)}$&  50.2$_{(13)}$&  75.1$_{(11)}$&  4.23$_{(9)}$&  6.35$_{(10)}$&  9.612 &  - \\
 KU-ISPL &  6.79$_{(9)}$&  45.1$_{(10)}$&  61.6$_{(10)}$&  3.60$_{(13)}$&  6.59$_{(12)}$&  10.152 &  - \\
 BMIPL-UNIST-YH-1 &  7.03$_{(12)}$&  50.2$_{(14)}$&  81.5$_{(13)}$&  3.70$_{(12)}$&  6.66$_{(13)}$&  12.218 &  - \\
 BIGFEATURE-CAMERA &  7.45$_{(14)}$&  49.2$_{(11)}$&  87.1$_{(14)}$&  3.23$_{(16)}$&  7.11$_{(14)}$&  13.784 &  - \\
 \midrule
 Samsung-SLSI-MSL &  6.25$_{(7)}$&  37.3$_{(6)}$&  26.0$_{(5)}$&  4.31$_{(7)}$&  5.97$_{(7)}$&  6.662 &  - \\
 \midrule
 Bicubic &  7.97$_{(16)}$&  52.0$_{(17)}$&  87.2$_{(15)}$&  3.16$_{(17)}$&  7.40$_{(16)}$&  14.532 &  6.04$_{(6)}$\\
 RRDB &  7.01$_{(11)}$&  51.3$_{(16)}$&  76.0$_{(12)}$&  4.06$_{(11)}$&  6.48$_{(11)}$&  10.042 &  6.06$_{(7)}$\\
\bottomrule
\end{tabular}
		\vspace{2mm}
	\caption{Challenge results for \textbf{Track 2}. The top section in the table contains participating methods that are ranked in the challenge. The middle section contains participating approaches that deviated from the challenge rules, whose results are reported for reference but not ranked. The bottom section contains baseline approaches. Participating methods are ranked according to their Mean Opinion Rank (MOR).}
	\label{tab:track2}
\end{table*}

Before the end of the final test phase, participating teams were required to submit results, code/executables, and factsheets for their approaches.
From 292 registered participants in Track 1, 19 valid methods were submitted, stemming from 16 different teams. Track 2 had 251 registered participants, of which 15 valid methods were submitted from 14 different teams. Table~\ref{tab:track1} and~\ref{tab:track2} report the final results of Track 1 and 2 respectively, on the test data of the challenge. The methods of the teams that entered the final phase are described in Section~\ref{sec:methods} and the teams' members and affiliations are shown in Section~\ref{sec:affiliation}.

\subsection{Architectures and Main Ideas} 
Inspired by the results of the last challenge in AIM 2019~\cite{AIM2019RWSRchallenge} and on the success of recent approaches~\cite{lugmayrICCVW2019,manuelFS}, most top methods pursued a two step approach. The first step aims to learn a network that can transfer \emph{clean} images to the source domain. This network thus learns a degradation operator, adding the kind of noise and corruptions present in the source images. It is then used to generate paired training data for the second step, which involves learning the super resolution network itself. It is generally trained using pairs generated by first downscaling and then applying the learned degradation on images from the target domain set. Many works employed the DSGAN~\cite{manuelFS} framework from the winner of the AIM 2019 challenge~\cite{AIM2019RWSRchallenge} to learn the degradation operator in the first step.

Some of the top methods in this challenge proposed particularly notable alterations and extensions to the general idea described above for learning the degradation network. The AITA-Noah team (Sec.~\ref{sec:AITA-Noah}) employs an iterative approach for Track~1, alternating between learning the degradation and SR network. It also uses an explicit denoising algorithm and train a sharpening network to decrease the blurring effects from the former. Impressionism (Sec.~\ref{sec:Impressionism}) is the only team that aims to explicitly estimate the blur kernel in the image, for improved source data generation. For Track~2, it employs the KernelGAN~\cite{Bell19InternalGAN} for this purpose. It also aims to explicitly estimate the noise variance using source image patches. This approach led to superior sharpness and quality in the generated SR images for Track~2. There were also some alternative strategies proposed. In particular, the Samsung-SLSI-MSL team (Sec.~\ref{sec:Samsung-SLSI-MSL}) aim to train a robust SR network capable of handling different source domains by randomly sampling a variety of degradations during the training of the SR network.

For the Real-world Super-Resolution setting, the results in the challenge suggest that training strategy and careful degradation modelling is far more important than choice of SR architecture. For the latter, most top methods simply adopted popular architectures, such as the RRDB/ESRGAN~\cite{wang2018esrgan} and the RCAN~\cite{rcan}. Most methods also included adversarial and perceptual VGG losses, often based on the ESRGAN~\cite{wang2018esrgan} framework. Brief descriptions of the methods submitted from each team is given in Sec.~\ref{sec:methods}.

\subsection{Baselines}
We compare methods participating in the challenge with several baseline approaches.

\parsection{Bicubic}
Standard bicubic upsampling using MATLAB's \verb|imresize| function.

\parsection{RRDB PT}
The pre-trained RRDB \cite{wang2018esrgan}, using the network weights provided by the authors. The network was trained with clean images using bicubic down-sampling for supervision. The only objective is the PSNR oriented L1 loss.

\parsection{ESRGAN Supervised}
ESRGAN network~\cite{wang2018esrgan} that is fine-tuned in a fully supervised manner, by applying the synthetic degradation operation used in Track~1. The degradation was unknown for the participants. This method therefore serves as an upper bound in performance, allowing us to analyze the gap between supervised and unsupervised methods. We employ the source $\mathcal{X}_\text{train}$ and target $\mathcal{Y}_\text{train}$ domain train images respectively. Low-resolution training samples are constructed by first down-sampling the image using the bicubic method and then apply the synthetic degradation. The network is thus trained with real input and output data, which is otherwise inaccessible. As for previous baselines, the network is initialized with the pre-trained weights provided by the authors. Note that no supervised baseline is available for Track~2 since no ground-truth HR images exists.

\subsection{Evaluation Metrics} 
The aim of the challenge is to pursue good image quality as perceived by humans.
As communicated to the participants at the start of the challenge, the final ranking was therefore to be decided based on a human perceptual study. 

\parsection{Track~1}
For Track~1, the fidelity-based Peak Signal-to-Noise Ratio (PSNR) and the Structural Similarity index (SSIM) \cite{WangBSS04SSIM} was provided on the Codalab platform for quantitative feedback. These metrics are also reported here for the test set. Moreover, we report the LPIPS \cite{zhang2018unreasonable} distance, which is a learned reference-based image quality metric computed as the $L^2$ distance in a deep feature space. The network itself has been fine-tuned based on image quality annotations, to correlate better with human perceptual opinions. However, this metric needs to be used with great care since many methods employ feature-based losses using ImageNet pre-trained VGG networks, which in its design is very similar to LPIPS. Moreover, some methods directly use the LPIPS distance as a loss of for hyper-parameter tuning. We treat LPIPS as an indication of perceptual quality, but not as a metric to decide final rankings.

\begin{figure*}[t]
\centering
	\newcommand{\wid}{.19\linewidth}
	\newcommand{\im}[2]{
	\setlength\tabcolsep{-2pt}
    \begin{tabular}{c}
         \vspace{-4pt}
         \includegraphics[width=\wid]{figures/visuals/track1/0901/#1.jpg} \\
         \footnotesize{#2} 
    \end{tabular}}
    \im{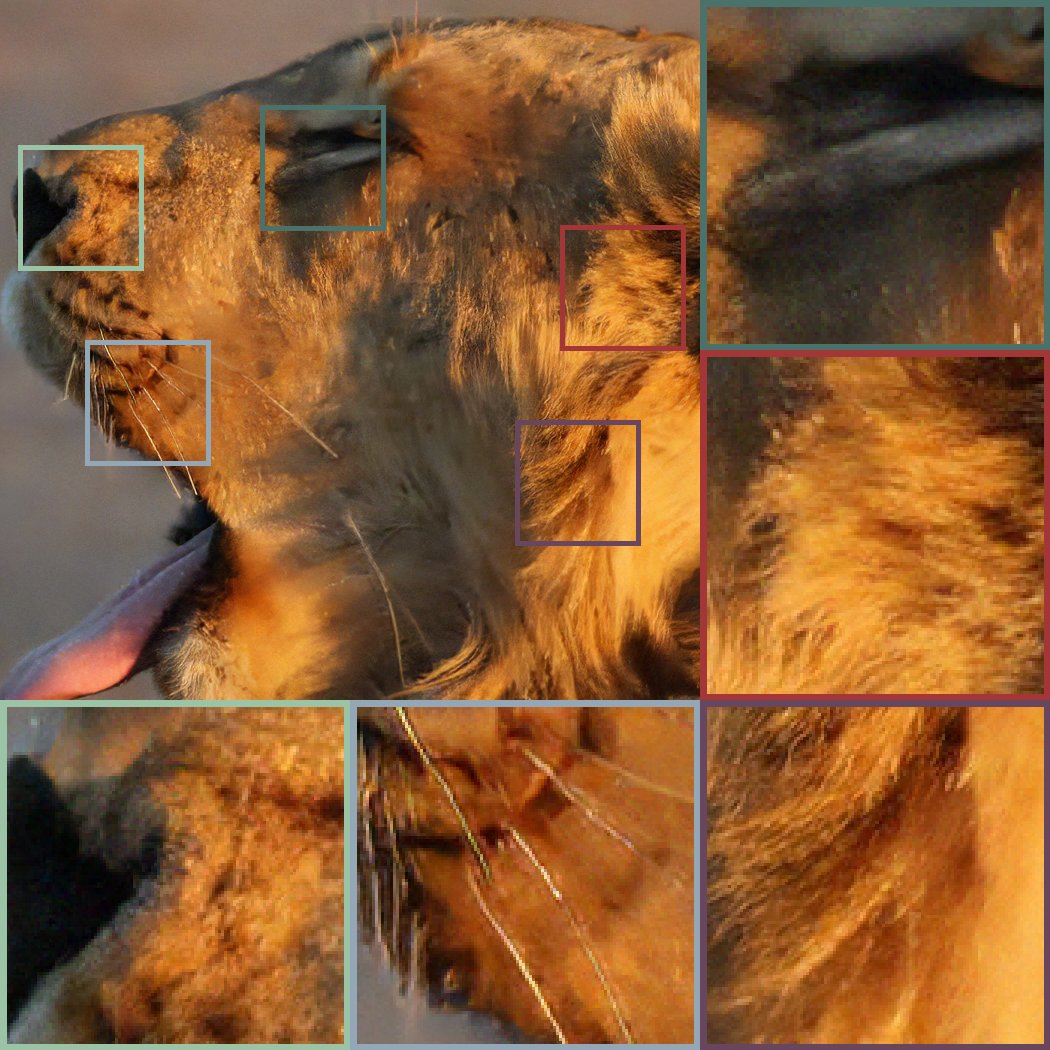}{Impressionism}
    \im{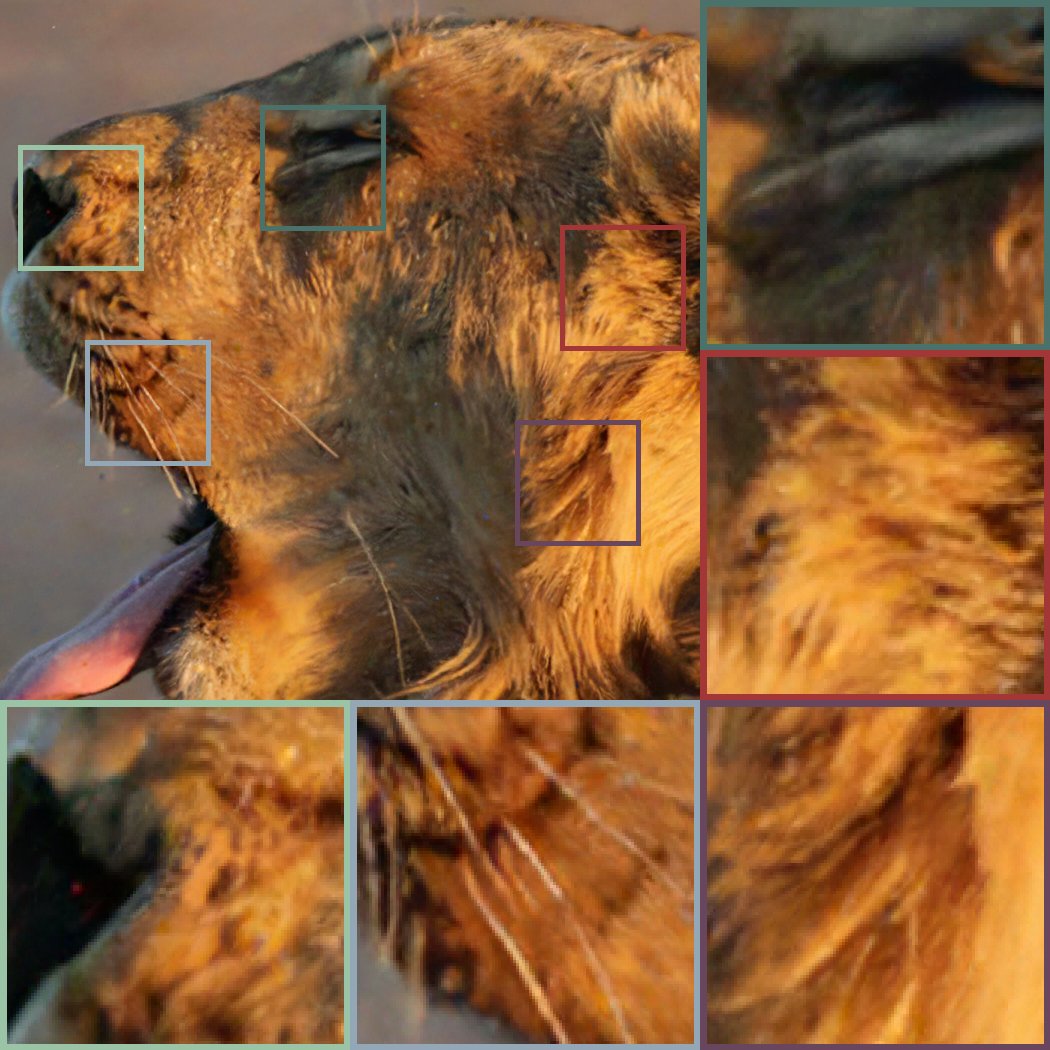}{Samsung-SLSI-MSL}
    \im{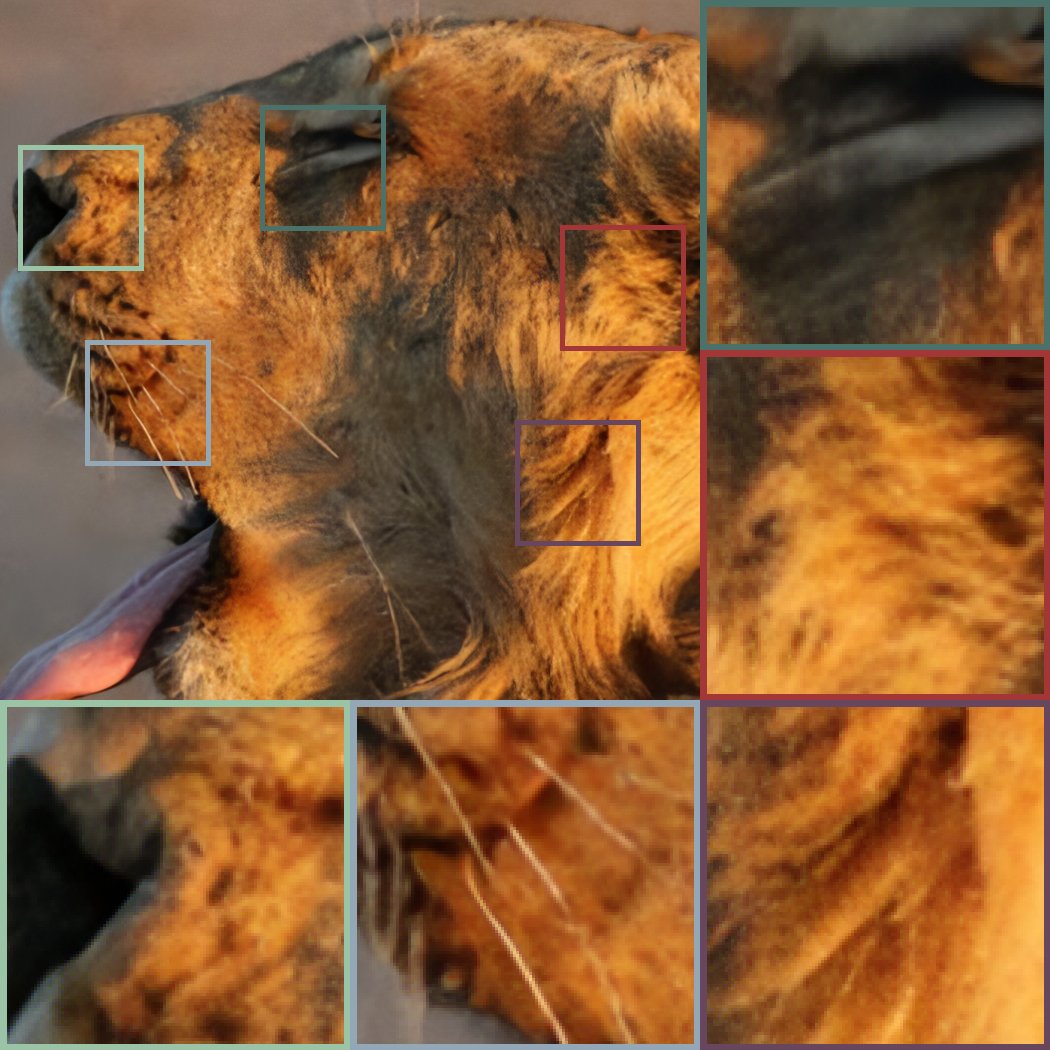}{BOE-IOT-AIBD}
    \im{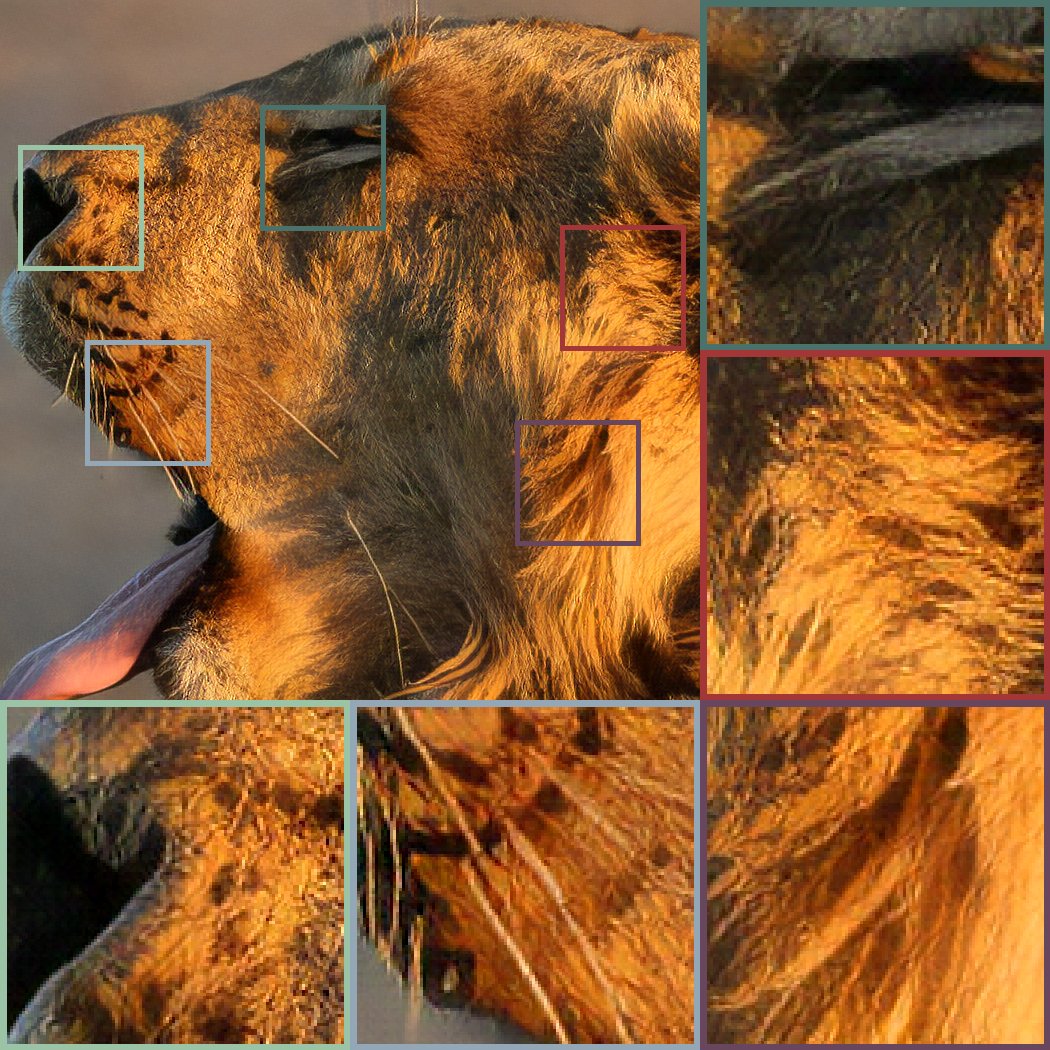}{MSMers}
    \im{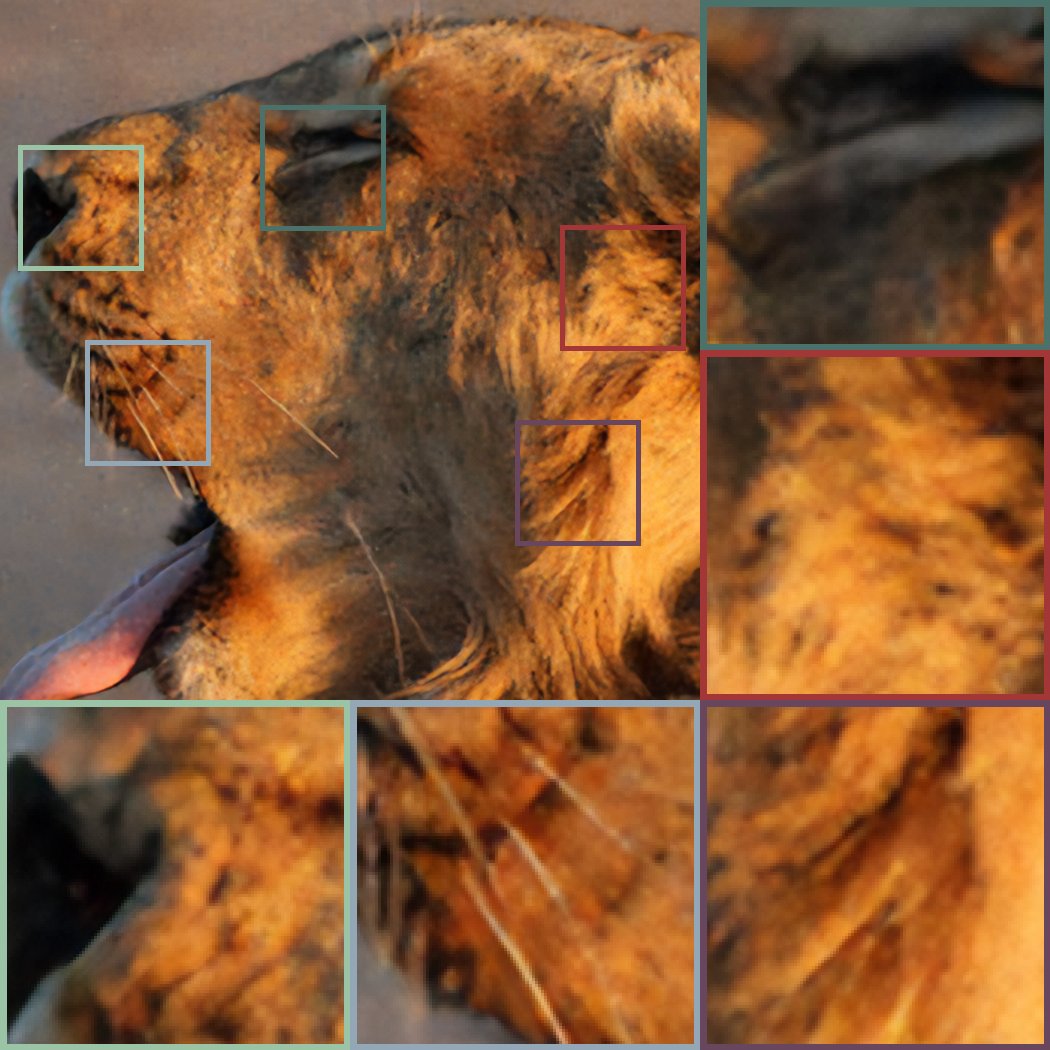}{KU ISPL}
    \im{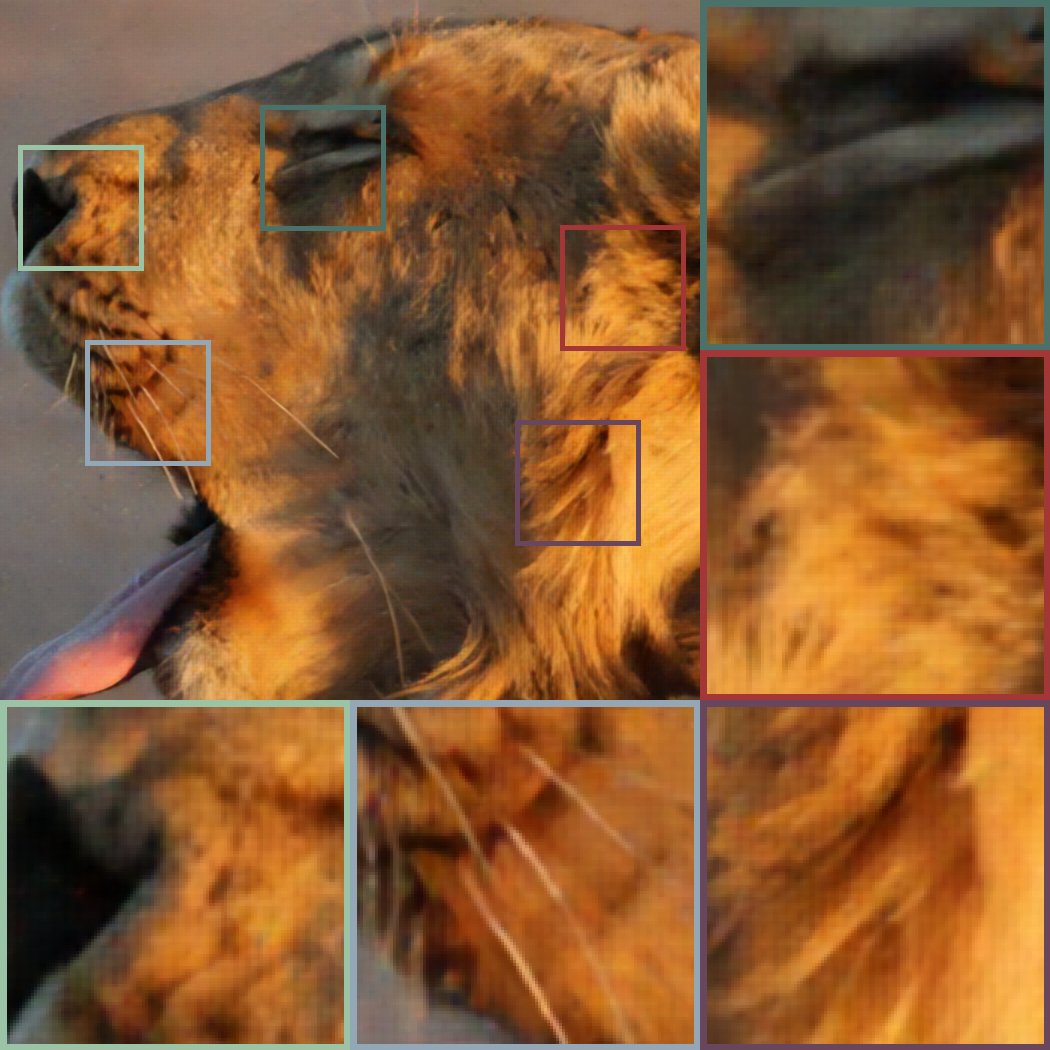}{InnoPeak-SR}
    \im{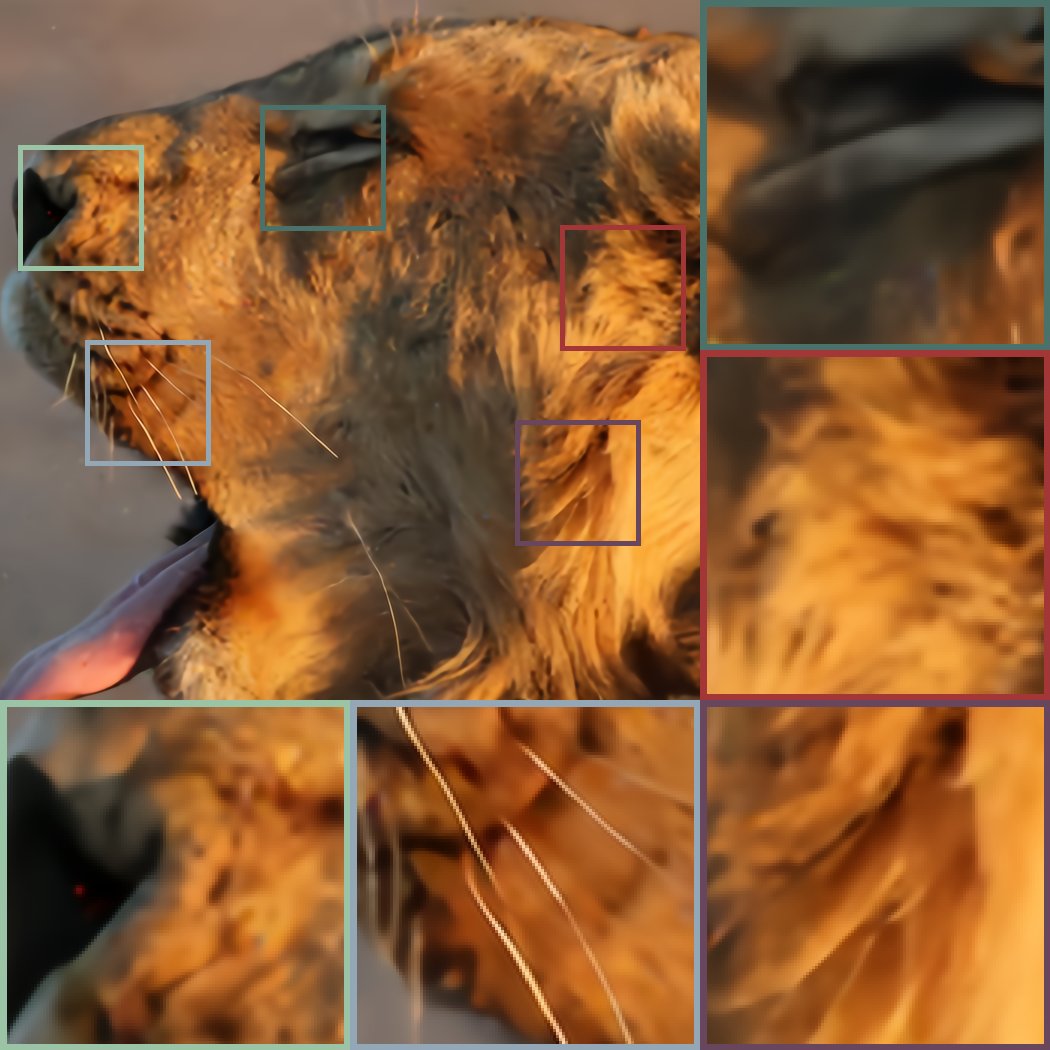}{ITS425}
    \im{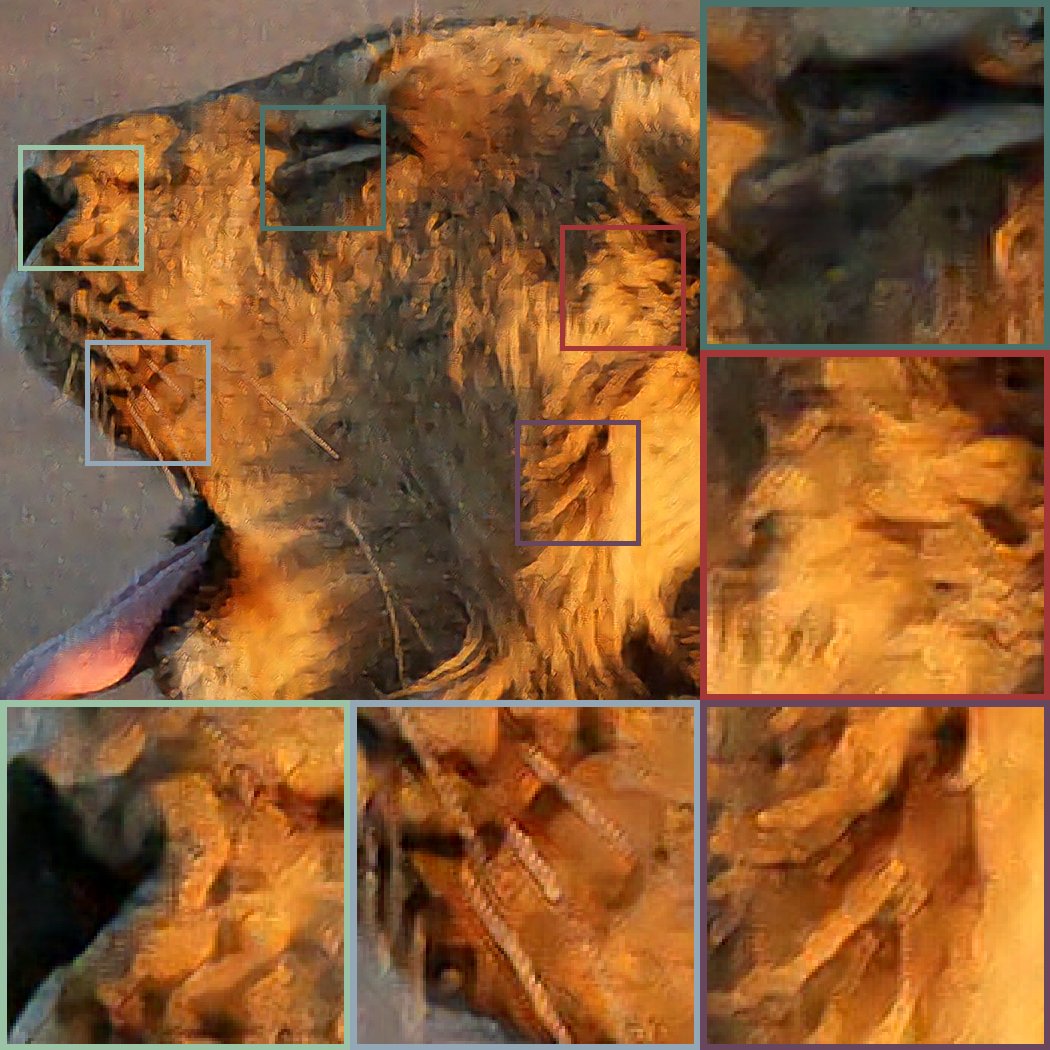}{MLP-SR}
    \im{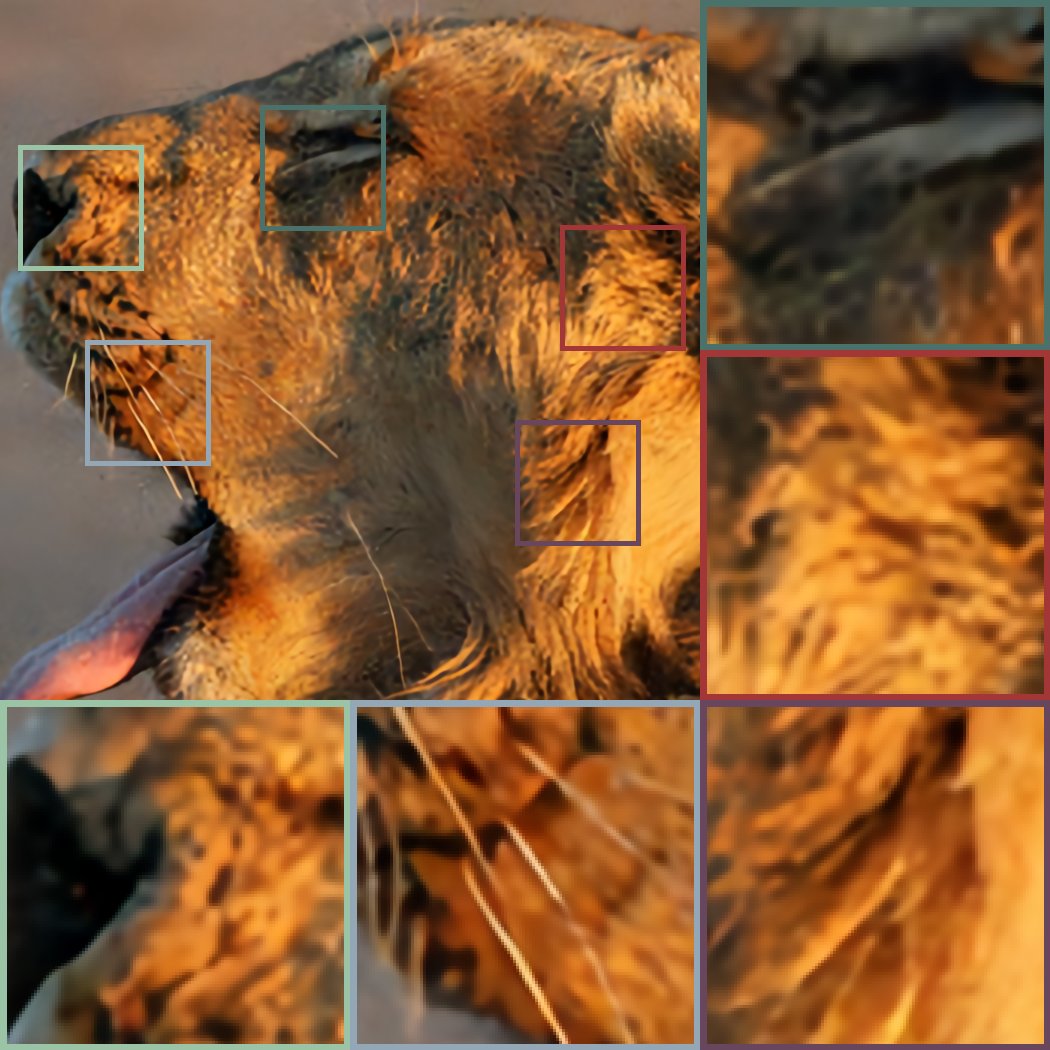}{Webbzhou}
    \im{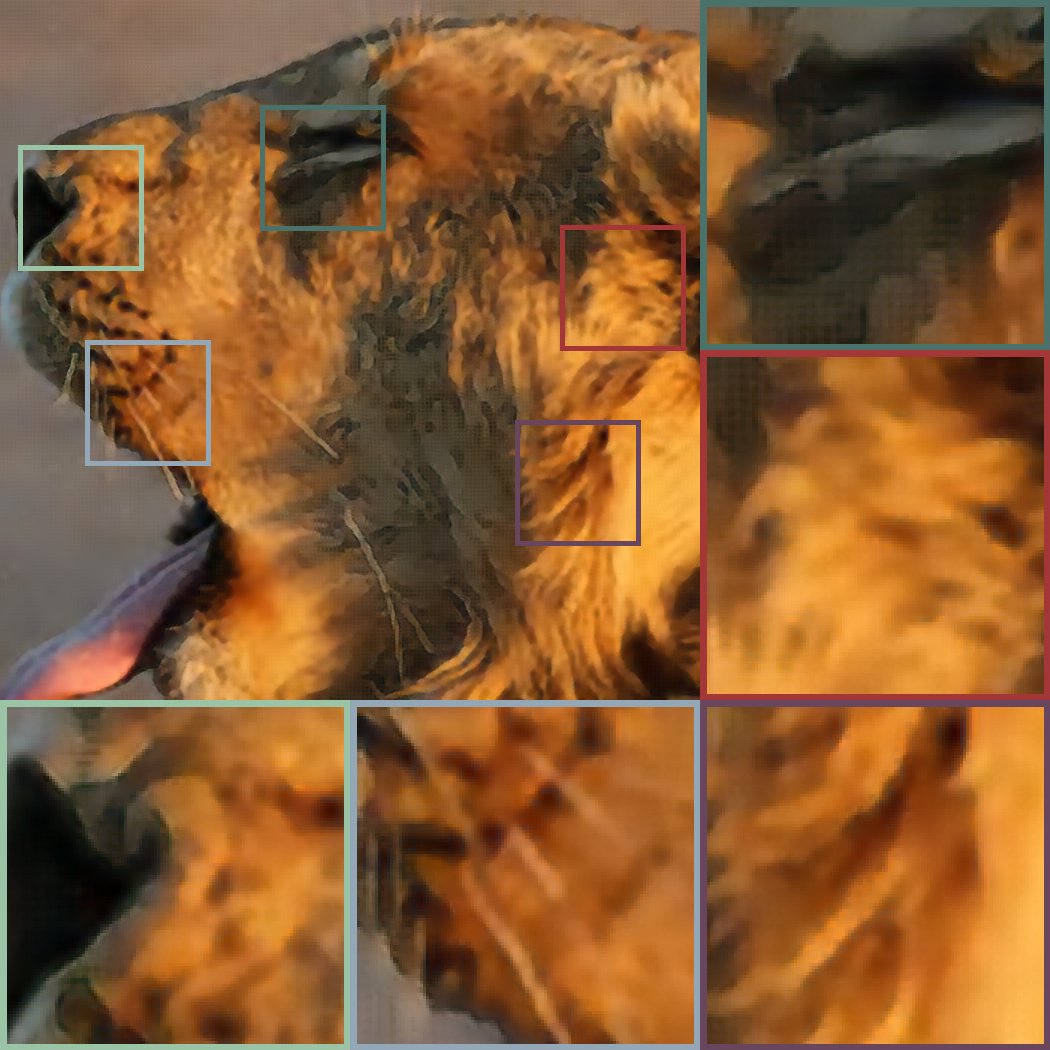}{SR-DL}
    \im{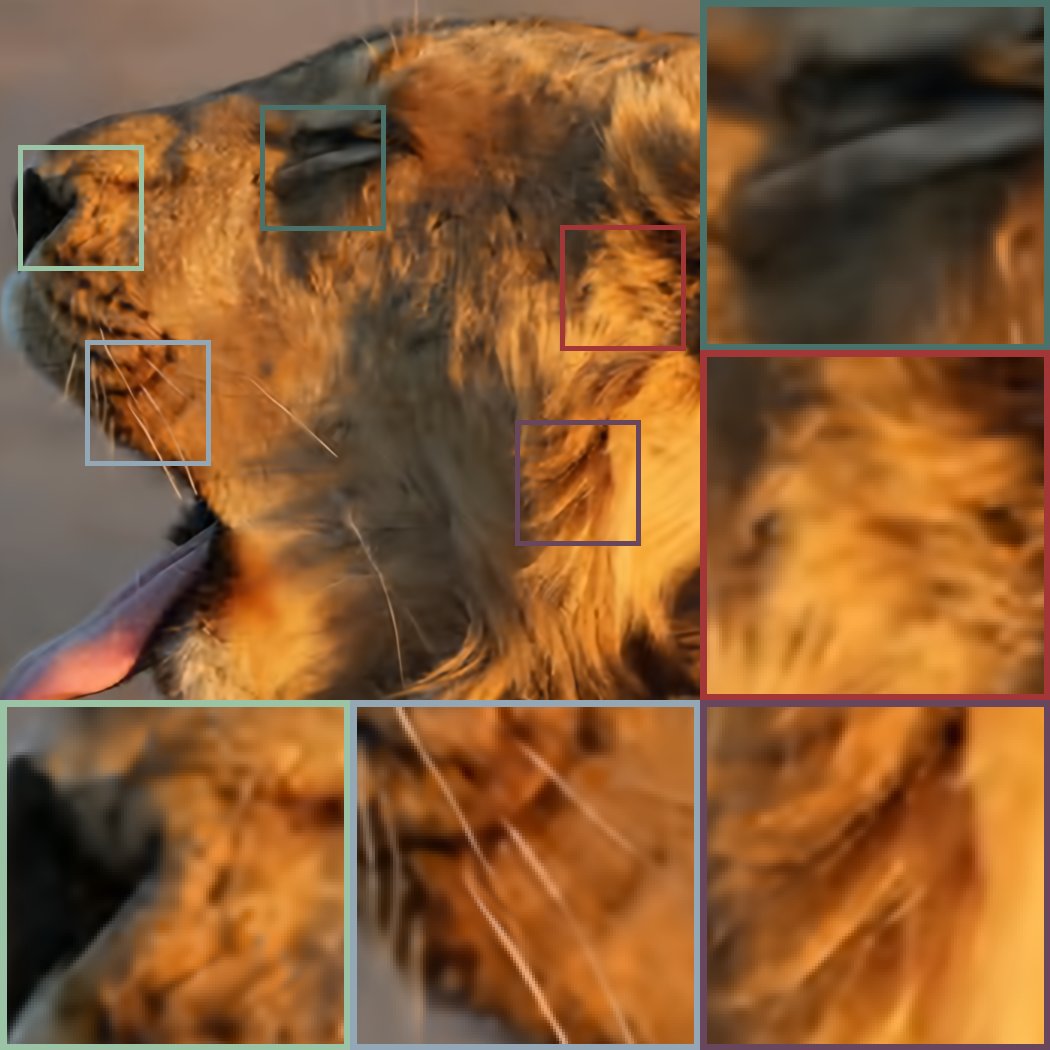}{TeamAY}
    \im{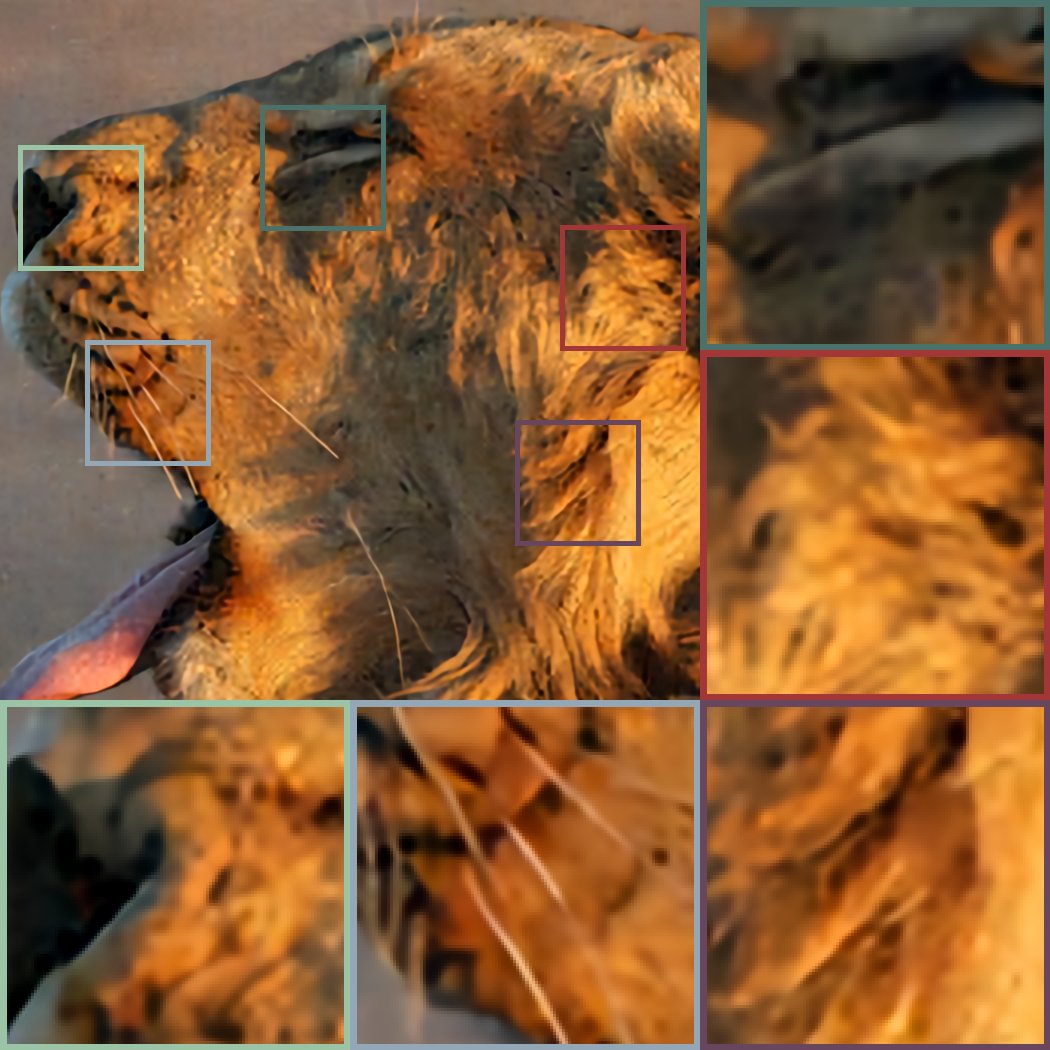}{BIGFEATURE-CAMERA}
    \im{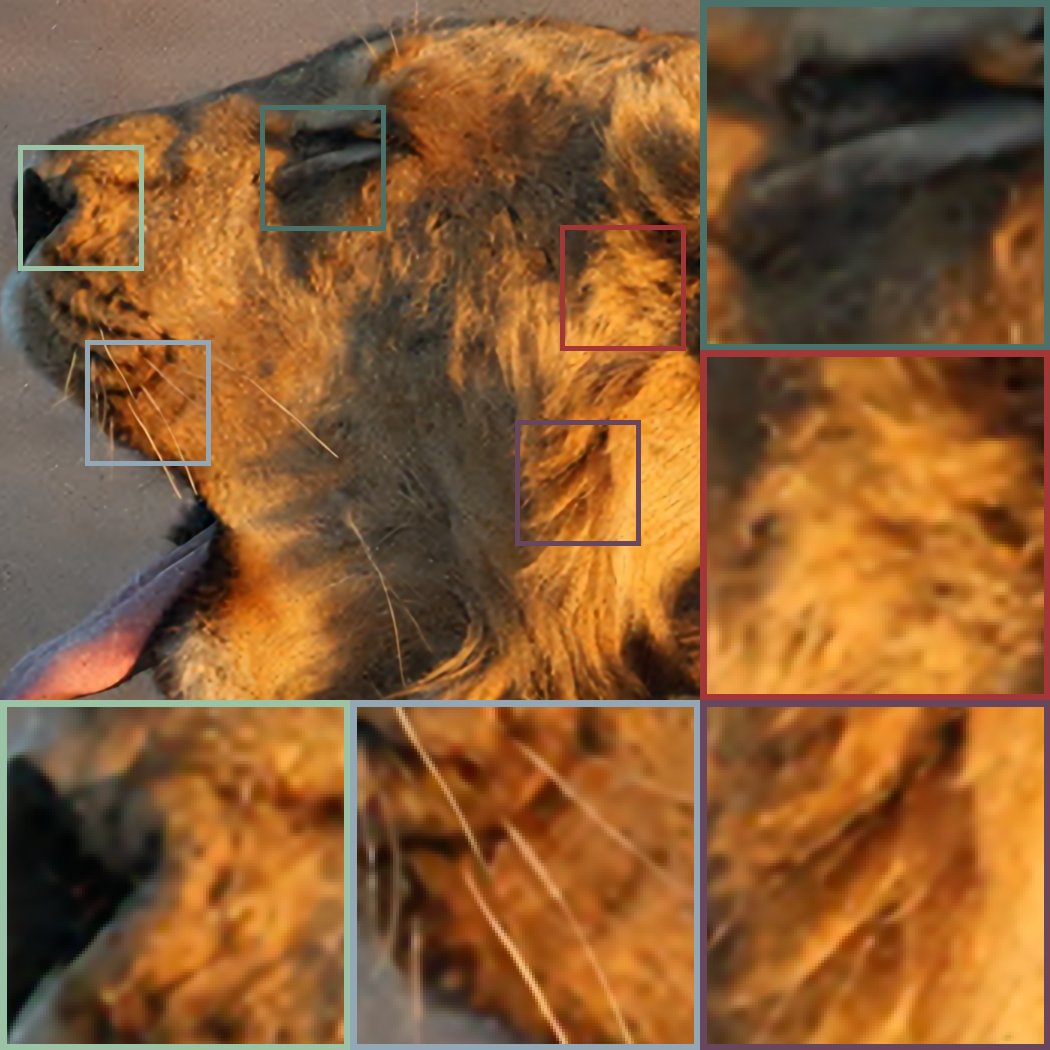}{BMIPL-UNIST-YH-1}
    \im{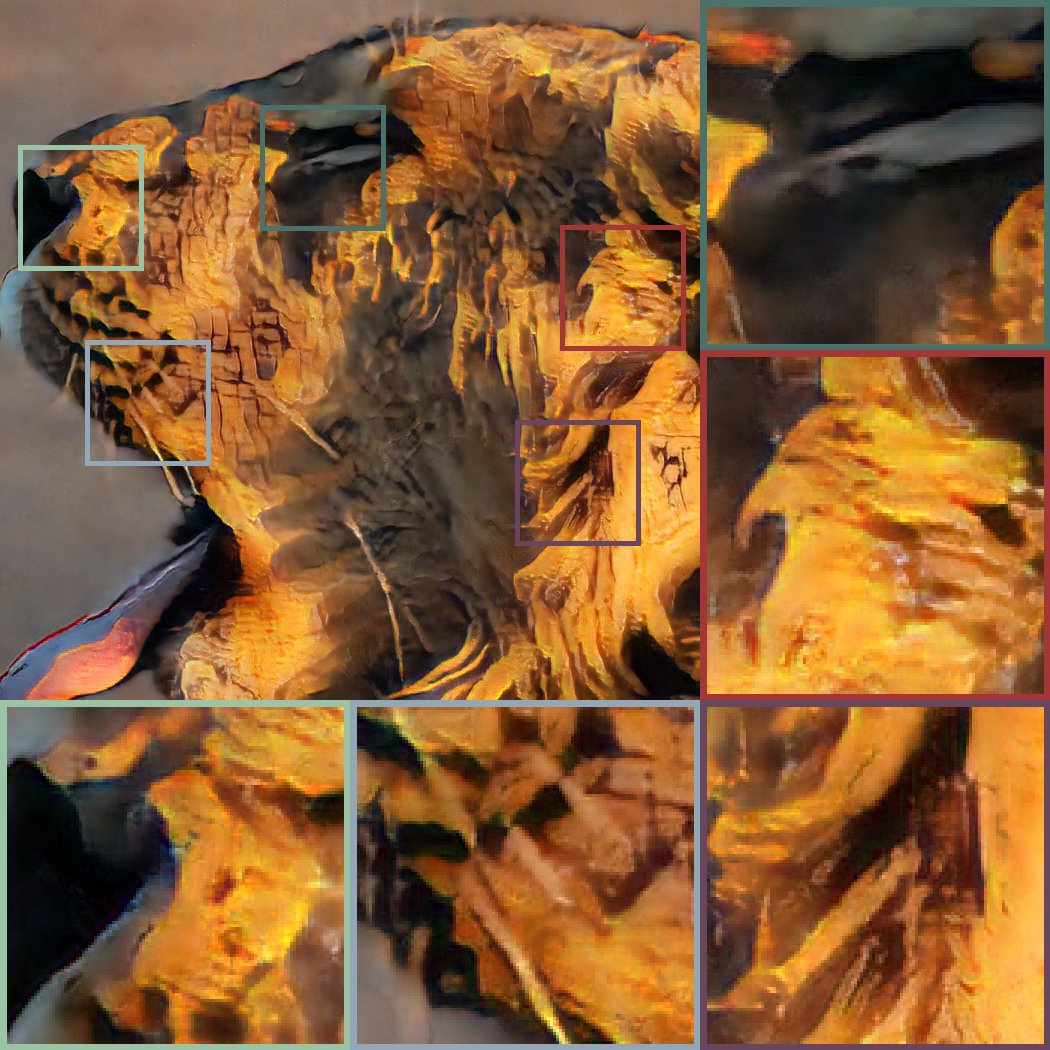}{SVNIT1-A}
    \im{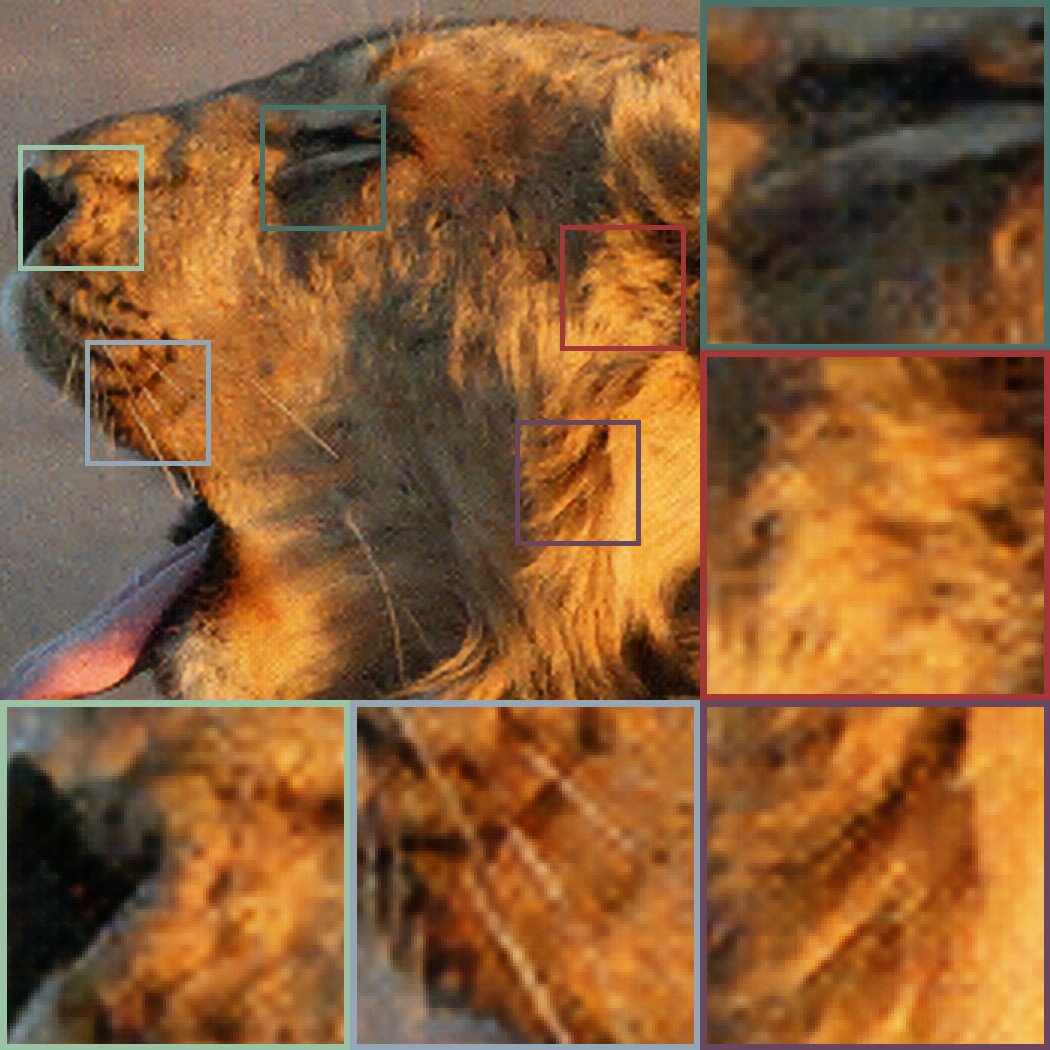}{KU-ISPL2}
    \im{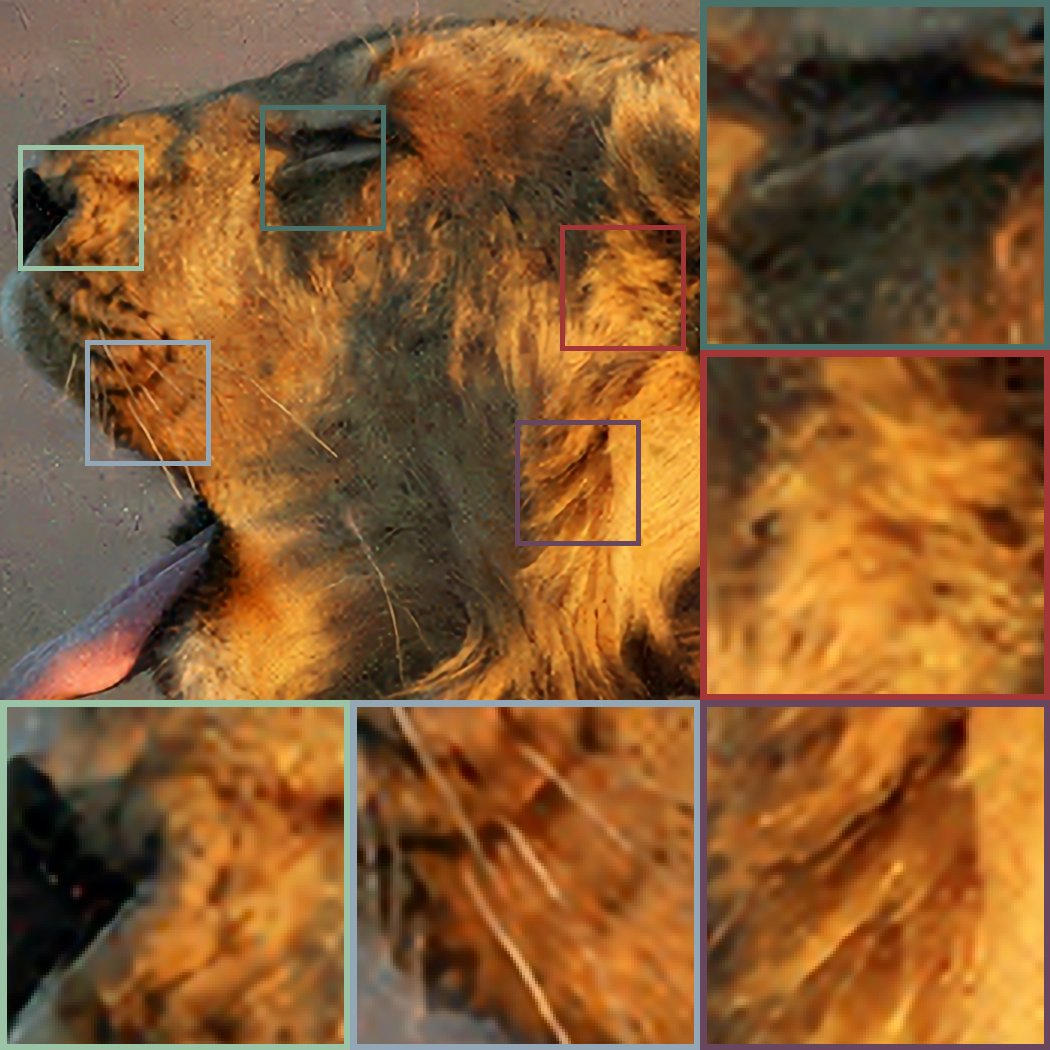}{SuperT}
    \im{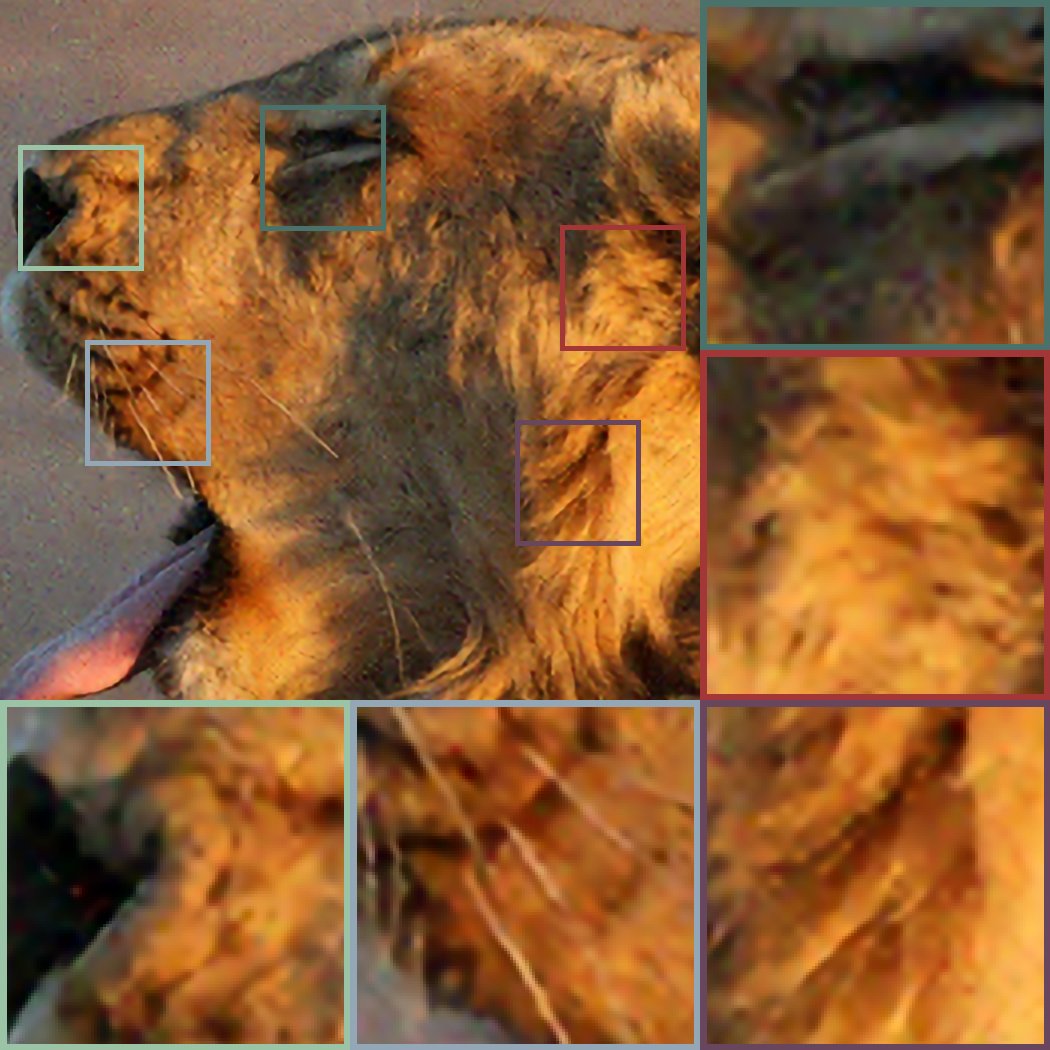}{GDUT-wp}
    \im{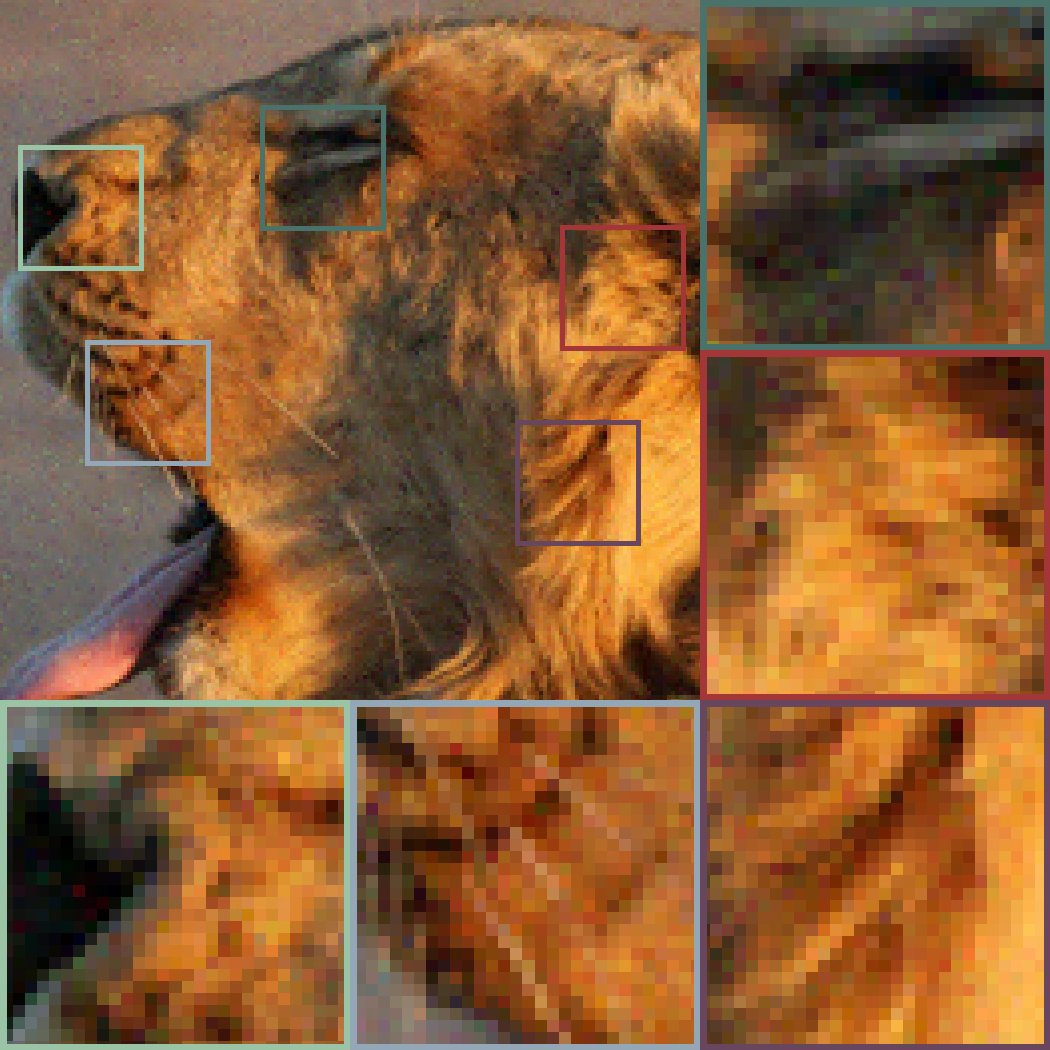}{SVNIT1-B}
    \im{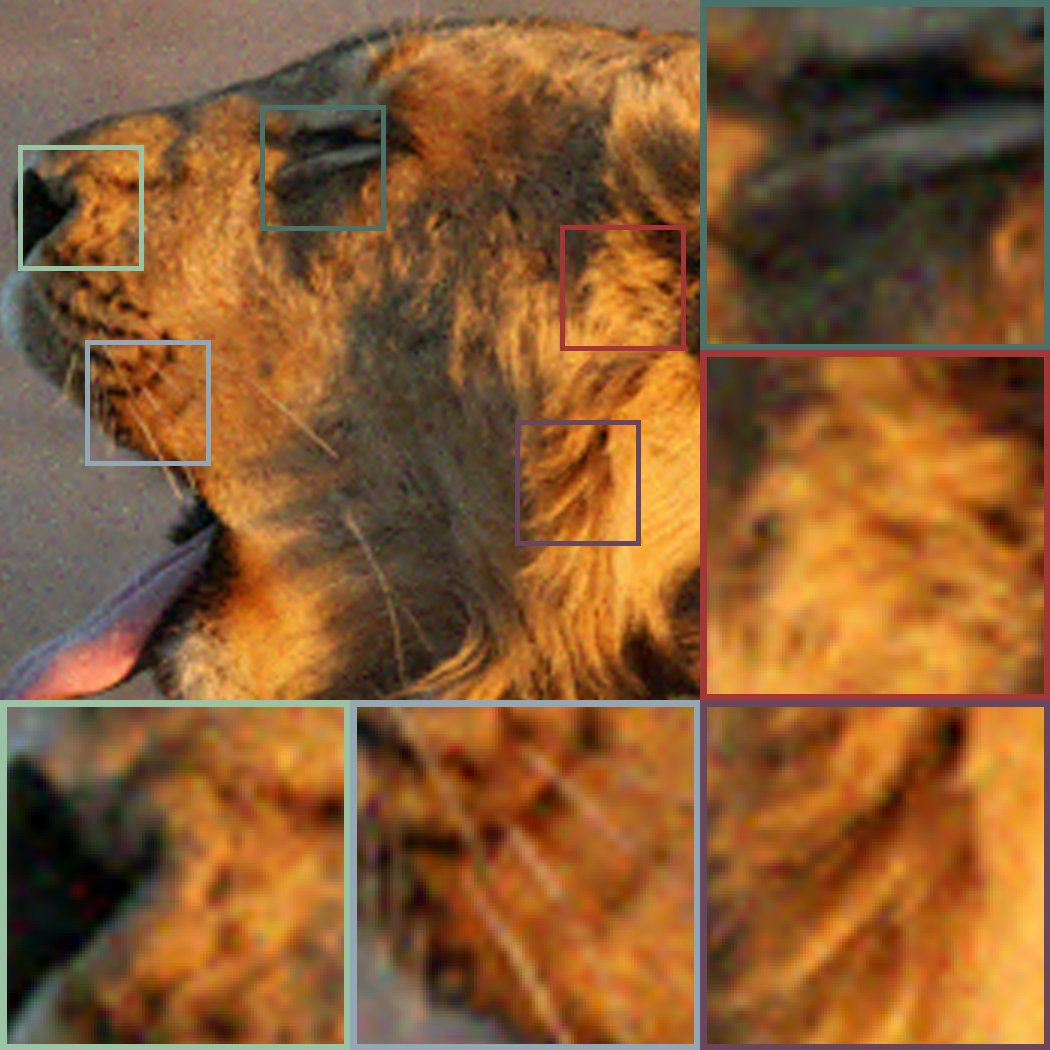}{SVNIT2}
    \im{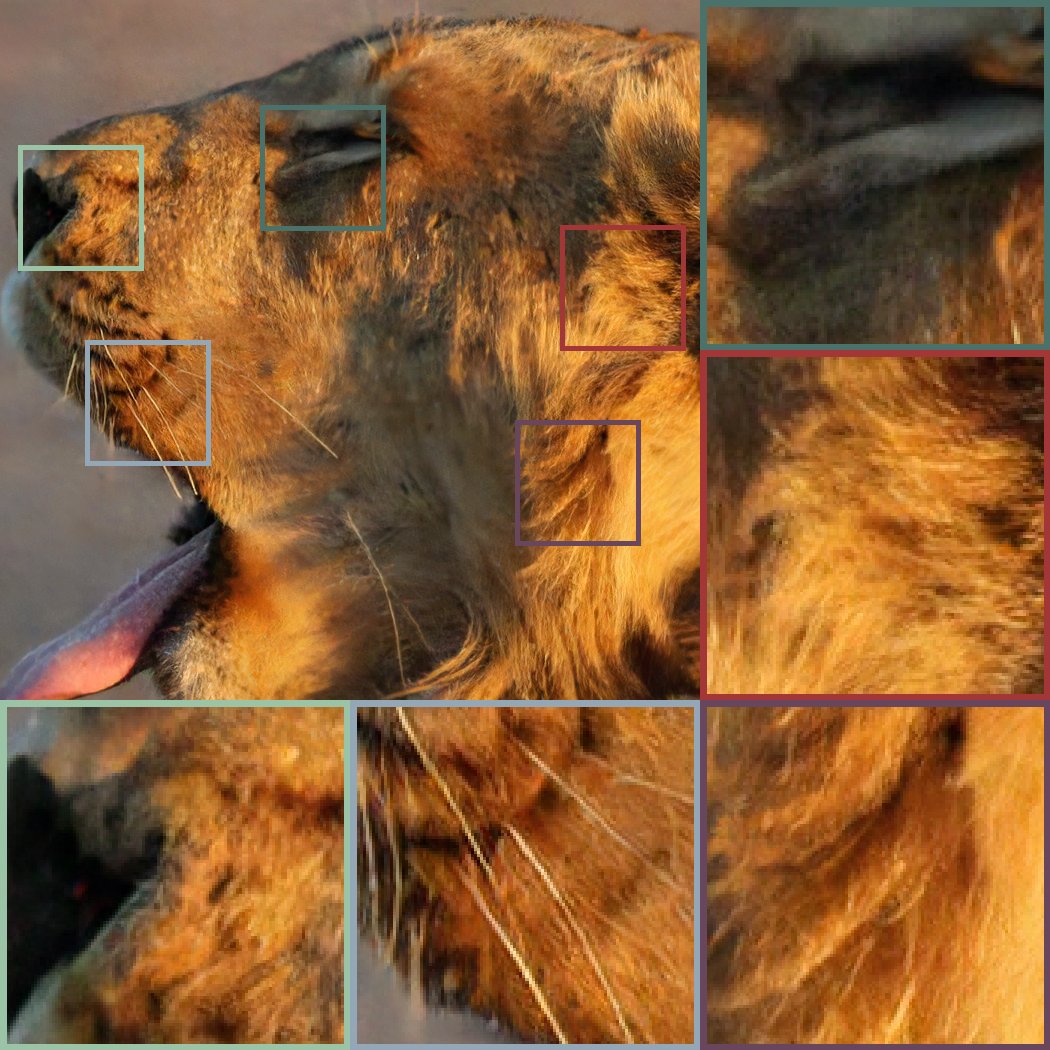}{AITA-Noah-A}
    \im{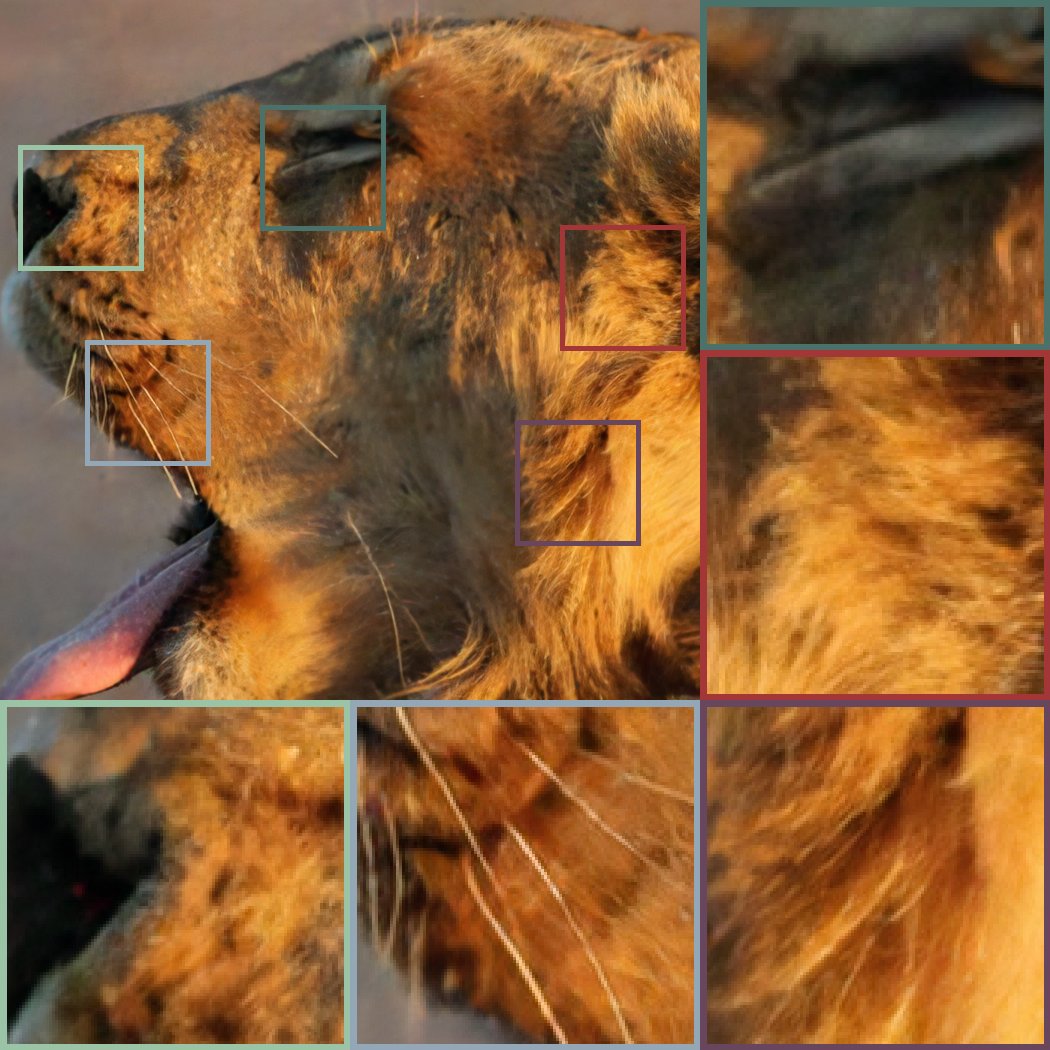}{AITA-Noah-B}
    \im{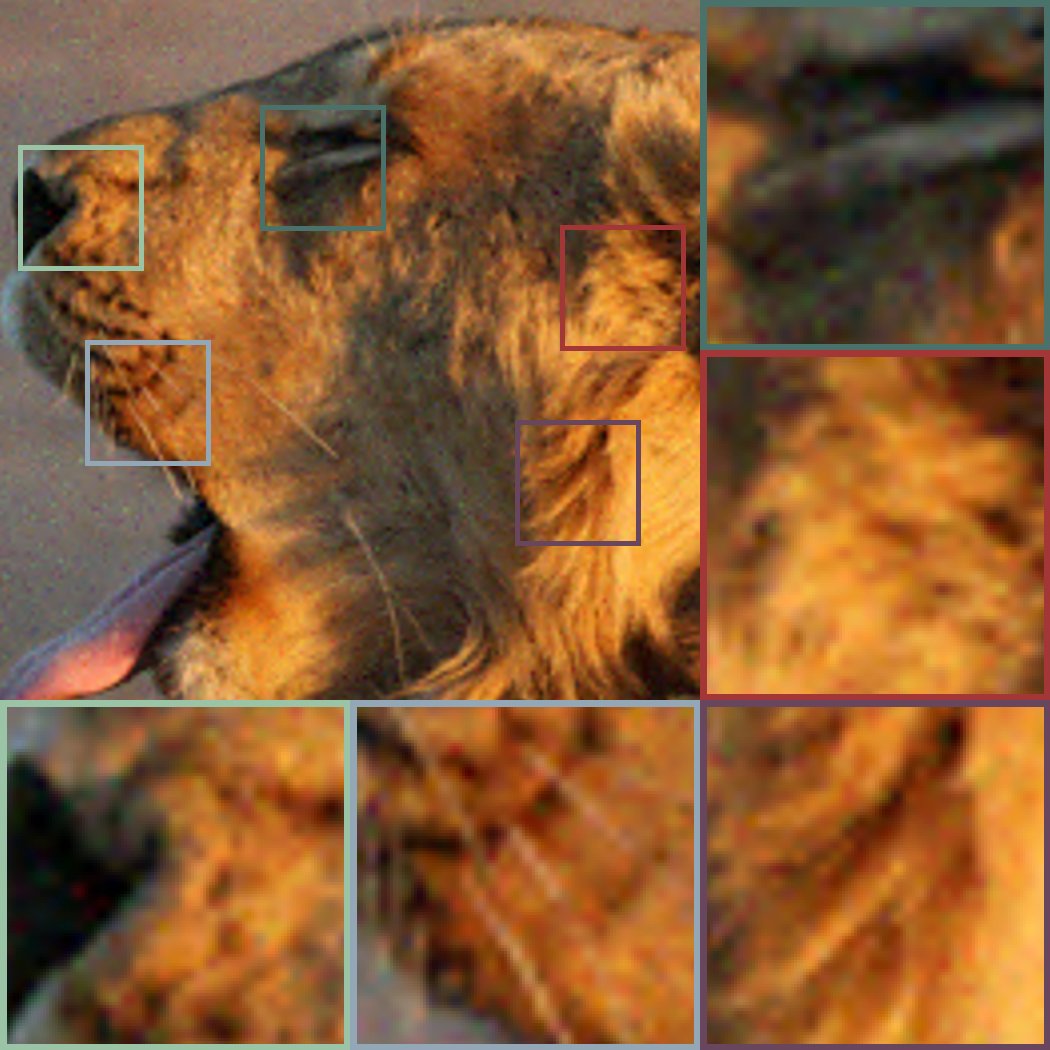}{Bicubic}
    \im{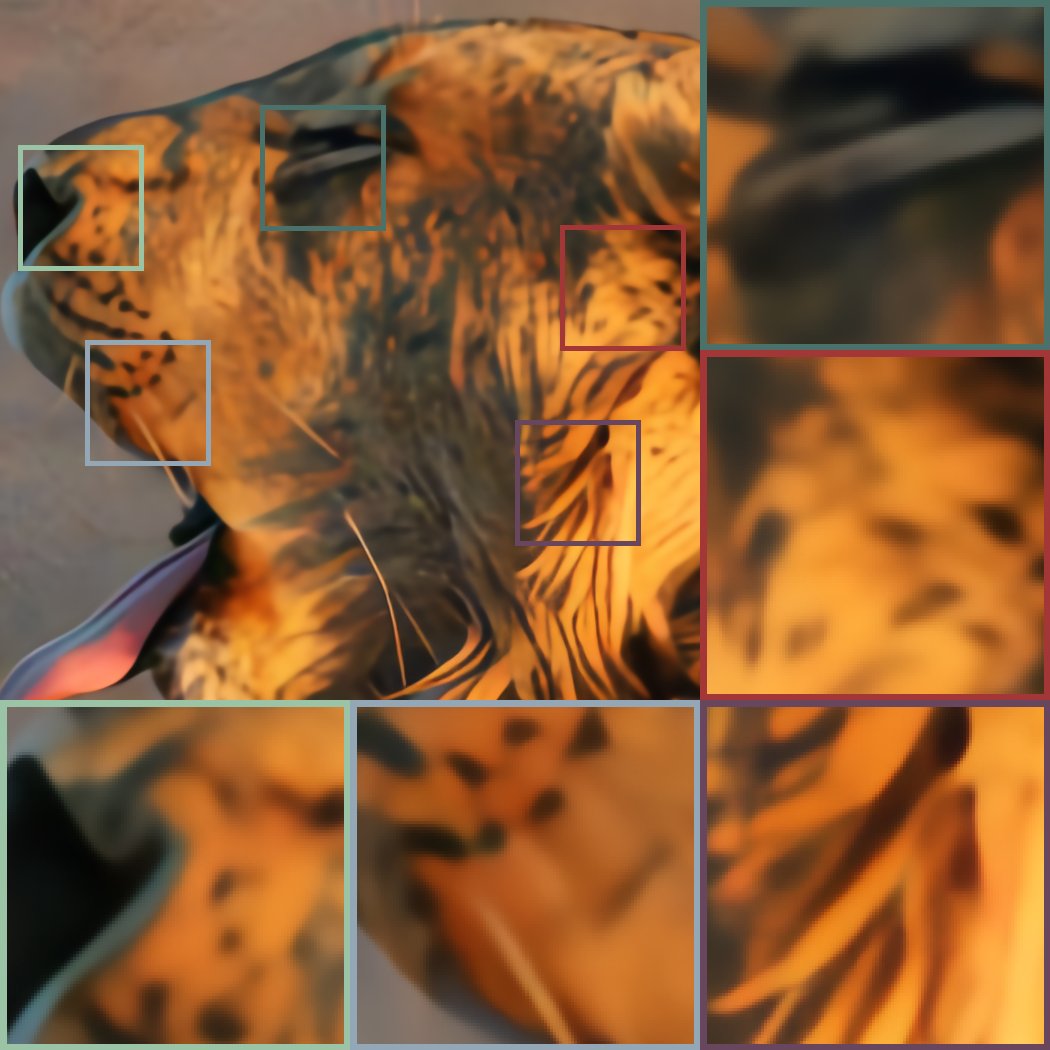}{RRDB Supervised}
    \im{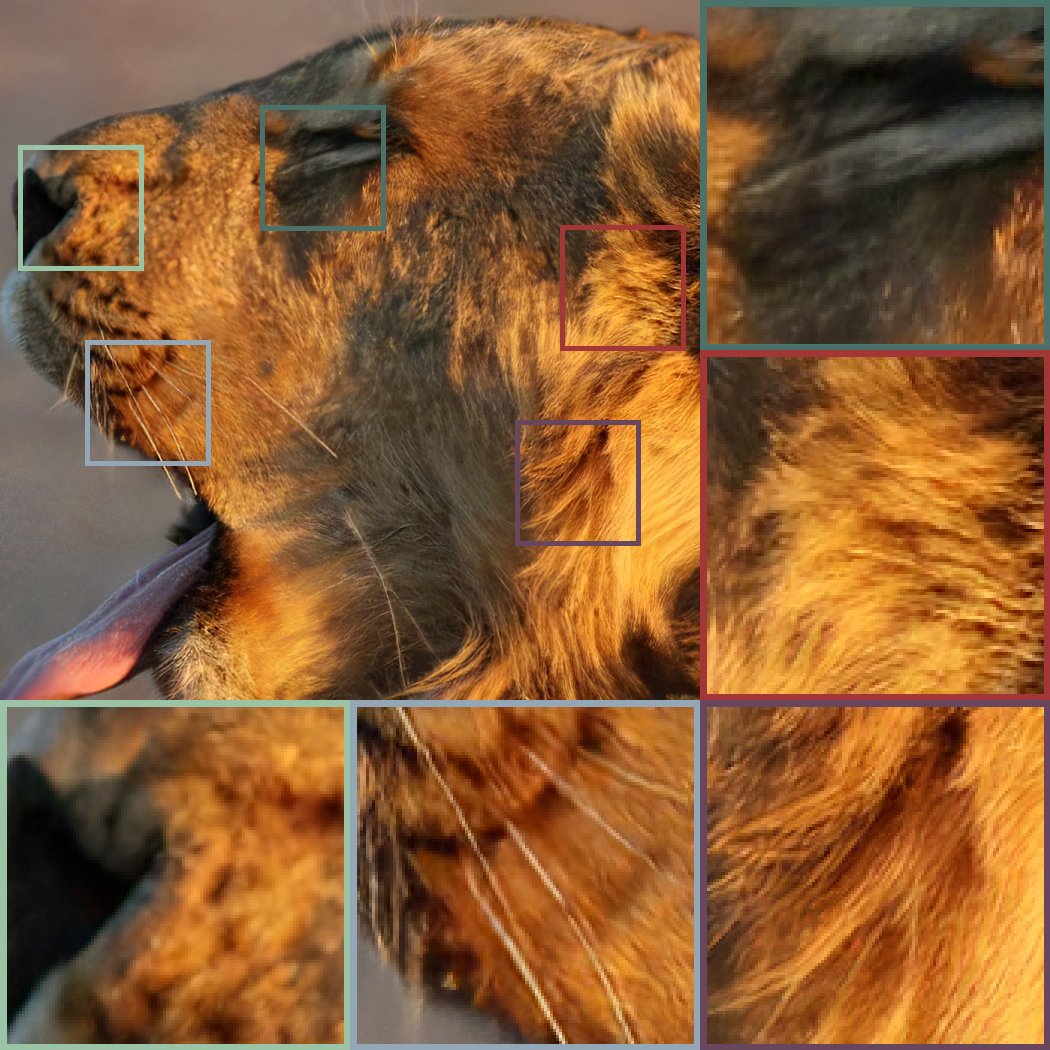}{ESRGAN Supervised}
    \im{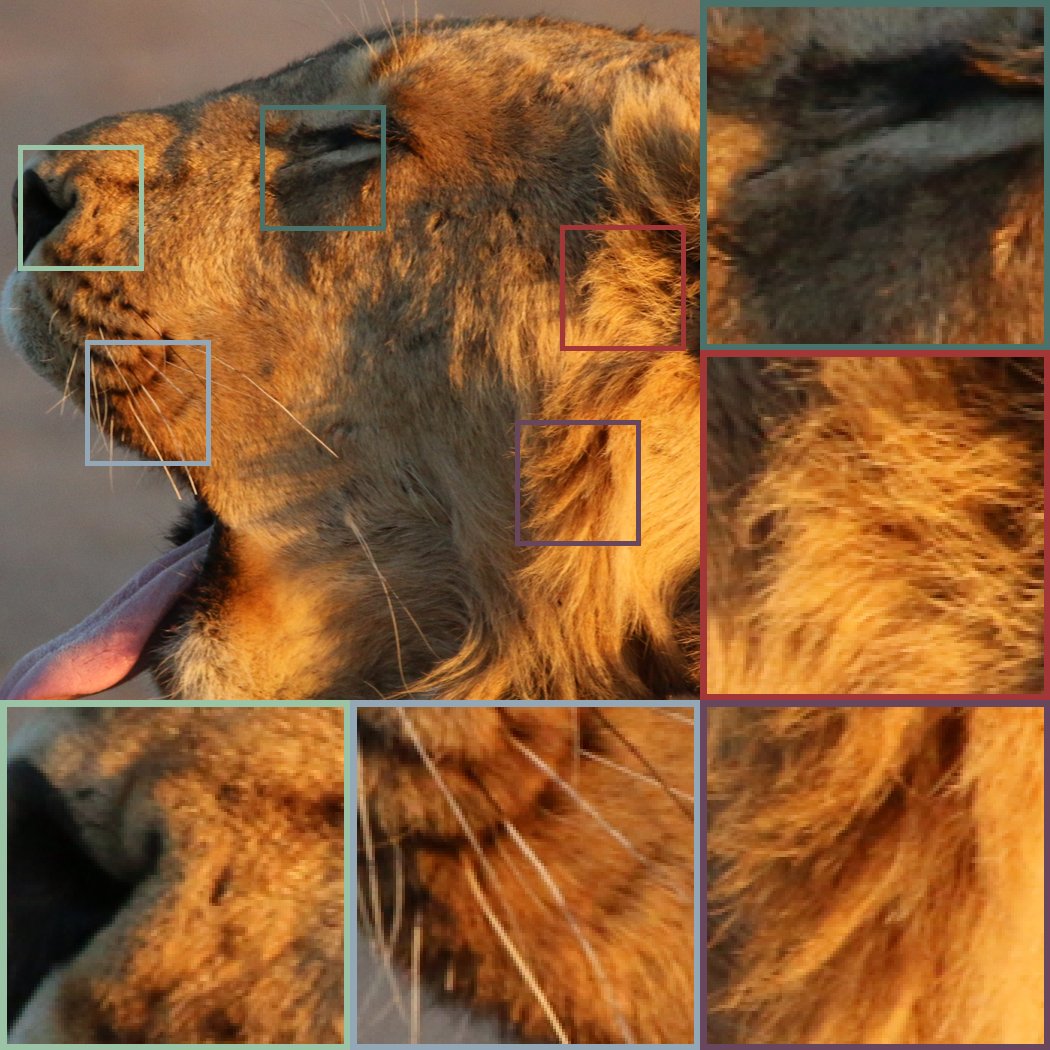}{Ground Truth}\vspace{1mm}
	\caption{Qualitative comparison between the participating approaches for Track 1.  ($4\times$ super-resolution)}\vspace{-3mm}
	\label{fig:visuals_track_1}
\end{figure*}

To obtain a final ranking of the methods, we performed a user study on Amazon Mechanical Turk. For Track~1, where reference images are available, we calculate the Mean Opinion Score (MOS) in the following manner. The test candidates were shown a side-by-side comparison of a sample prediction of a certain method and the corresponding reference ground-truth. They were then asked to evaluate the quality of the SR image \wrt the reference image using the 6-level scale defined as:  0 - 'Perfect',  1 - 'Almost Perfect',  2 - 'Slightly Worse',  3 - 'Worse',  4 - 'Much Worse',  5 - 'Terrible'. The images shown to the participants of the study were composed of zoomed crops, as shown in Figure~\ref{fig:visuals_track_1}. The human study was performed for the top 10 methods according to LPIPS distance, along with 4 baseline approaches.

\parsection{Track~2}
For Track 2, a ground truth reference does not exist due to the nature of the problem. Therefore we used several no-reference based image quality assessment (IQA) metrics. In particular, we report the NIQE~\cite{MittalSB13NIQE}, BRISQUE~\cite{mittal2011brisque} and PIQE~\cite{NDBCM15piqe}, using their corresponding MATLAB implementations. Moreover, we report the learned NRQM~\cite{MaYY017NRQM} IQA score. We also report two metrics that summarize the result of the computed IQA metrics. The Perceptual Index PI, previously employed in \cite{ignatov2018pirm}, is calculated as an adjusted mean of NIQE and NRQM. We also compute the mean IQA-Rank by taking the average image-wise rank achieved \wrt each of the four IQA metrics. In this case, taking the average rank is preferred over the average value, since the rank is not sensitive to the specific scaling or range of the particular metric.

Since no reference image exists in Track~2, the MOS score as defined for Track~1 cannot be computed. Instead, we compute the Mean Opinion Rank (MOR) by asking the study participants to rank the predictions of several methods in terms of image quality. For each question, the study participants were shown the SR results of all methods in the study for a particular image. These images were then ranked in terms of overall image quality. The MOR is then computed by averaging the assigned rank of each method, over all images and study participants. Since ranking too many entries at once is cumbersome and can lead to inaccurate results, we performed the human study on the top 5 approaches along with two baselines. As we did not find any of the IQA metrics previously discussed to correlate well with perceived image quality, the initial selection of top 5 methods was performed using a purely visual comparison performed by the challenge organizers. The top 5 methods were selected by assessing sharpness, noise, artifacts, and overall quality. The MOR scores were then computed using Amazon Mechanical Turk.

\begin{figure*}[t]
\centering
	\newcommand{\wid}{.155\linewidth}
	\newcommand{\im}[2]{
	\setlength\tabcolsep{-2pt}
    \begin{tabular}{c}
         \vspace{-4pt}
         \includegraphics[width=\wid]{figures/visuals/track2/0/#1.jpg} \\
         \scriptsize{#2} 
    \end{tabular}}
    \im{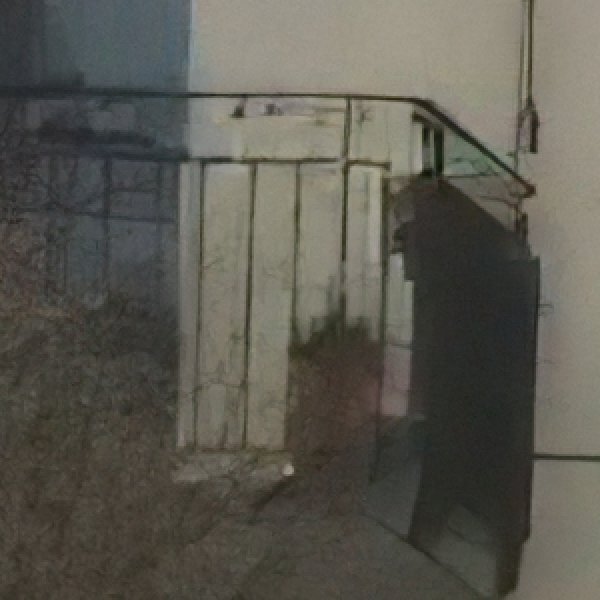}{Impressionism}
    \im{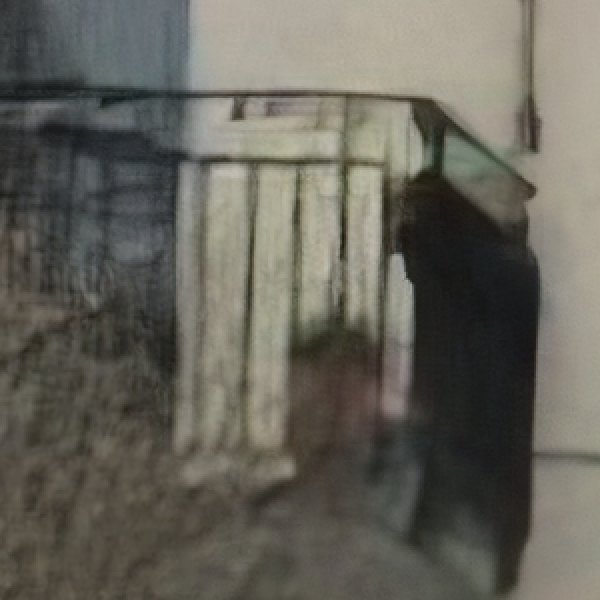}{AITA-Noah-A}
    \im{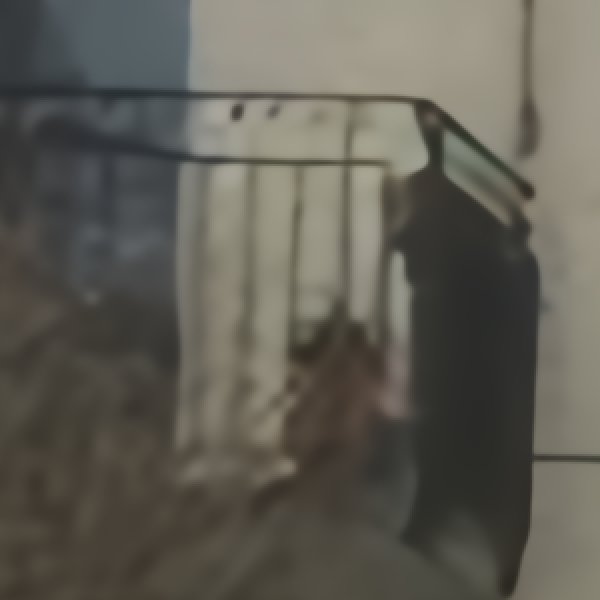}{ITS425}
    \im{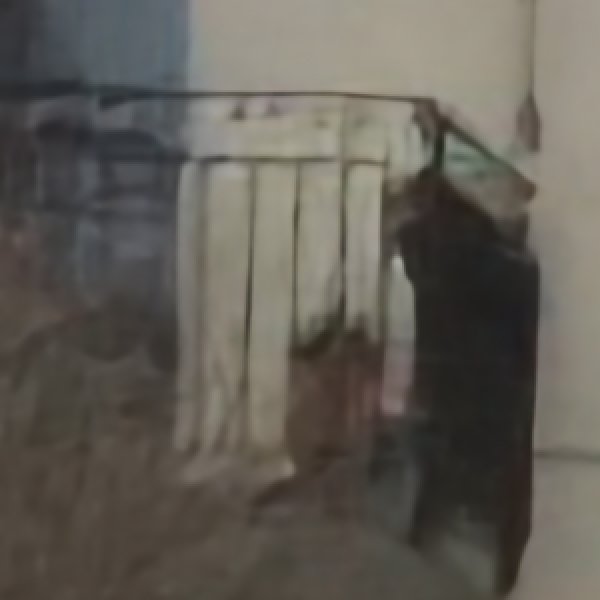}{AITA-Noah-B}
    \im{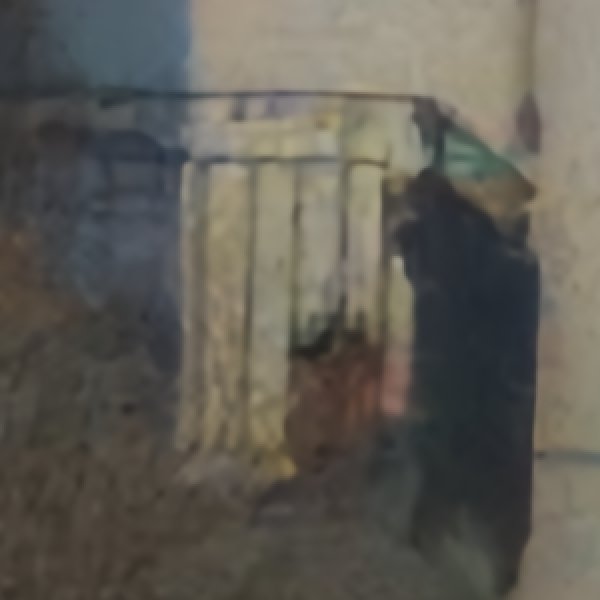}{Webbzhou}
    \im{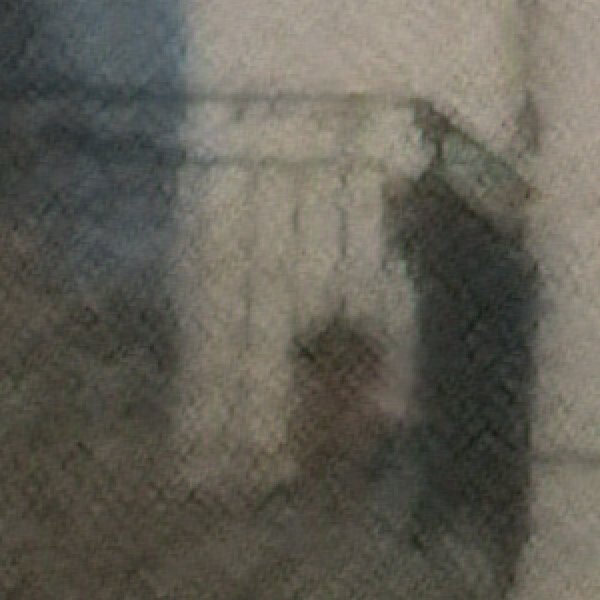}{Relbmag-Eht}
    \im{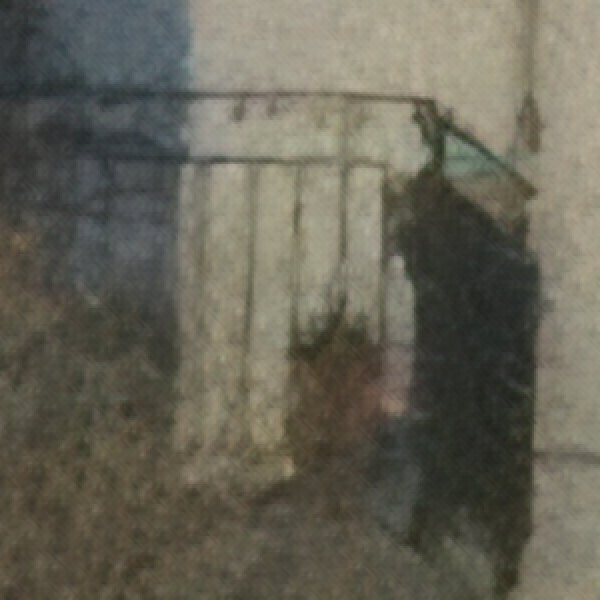}{MSMers}
    \im{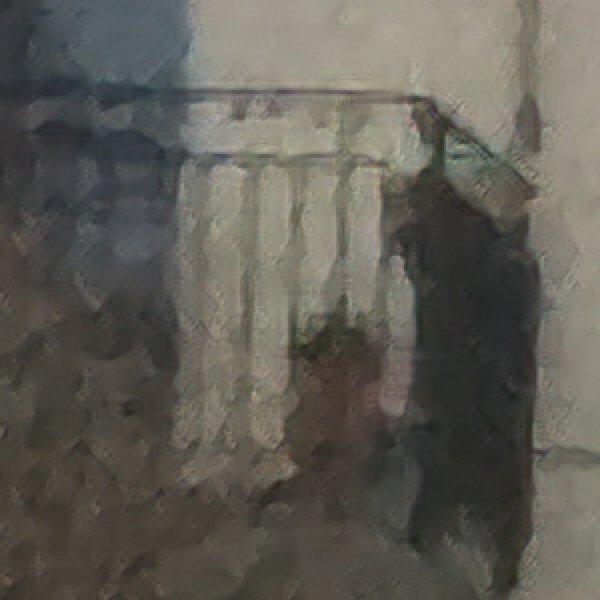}{MLP-SR}
    \im{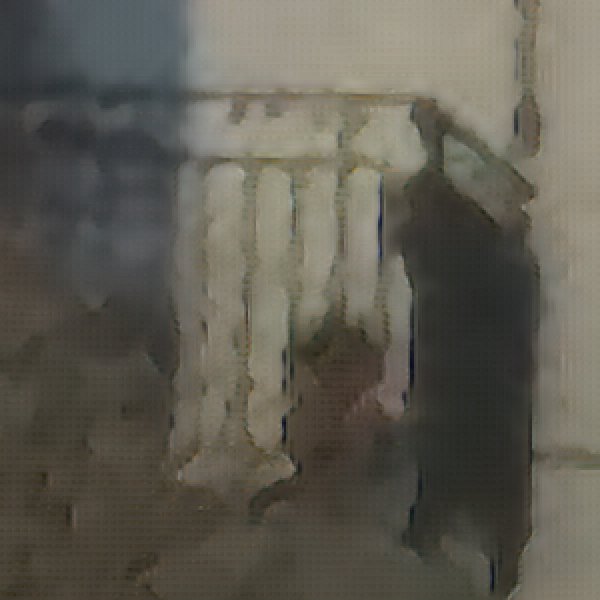}{SR-DL}
    \im{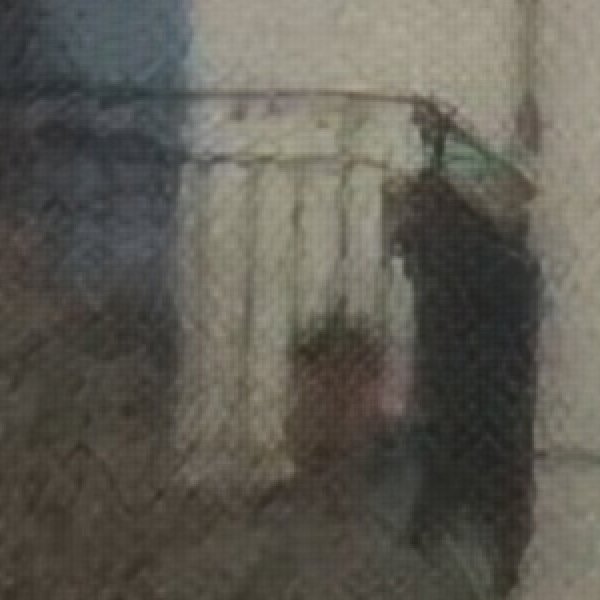}{InnoPeak-SR}
    \im{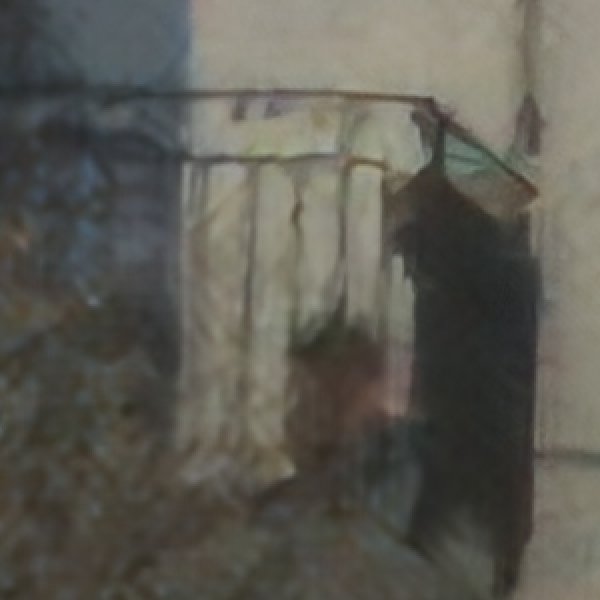}{QCAM}
    \im{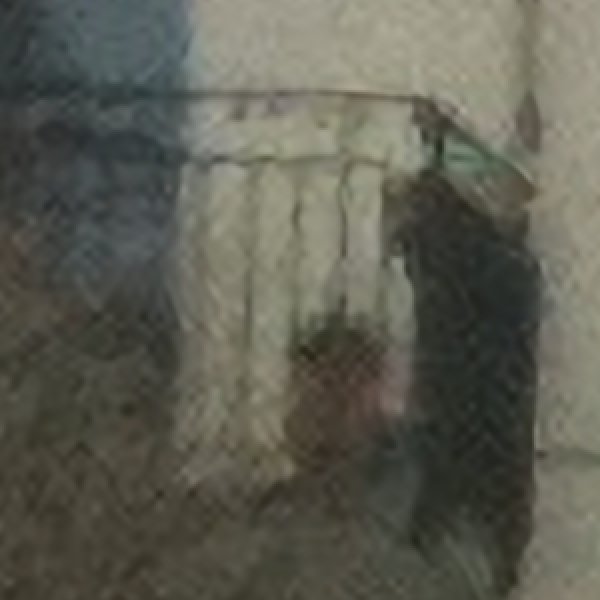}{SuperT}
    \im{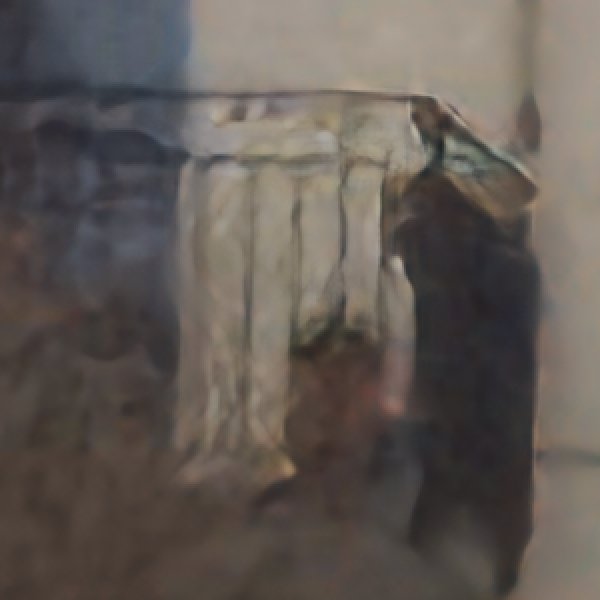}{KU-ISPL}
    \im{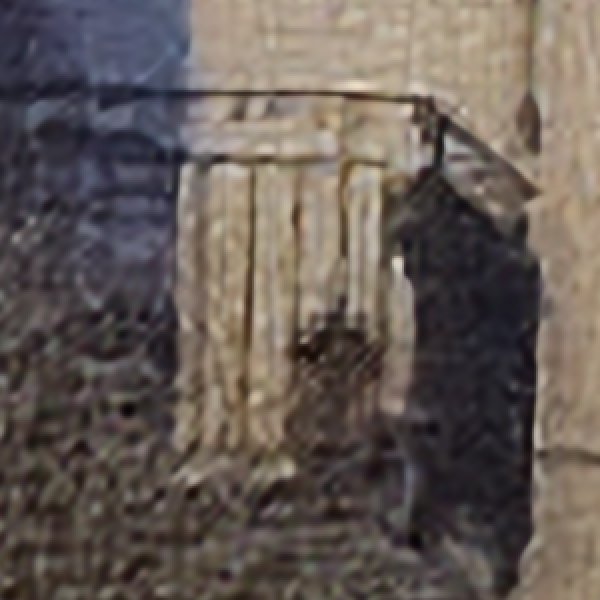}{BMIPL-UNIST-YH-1}
    \im{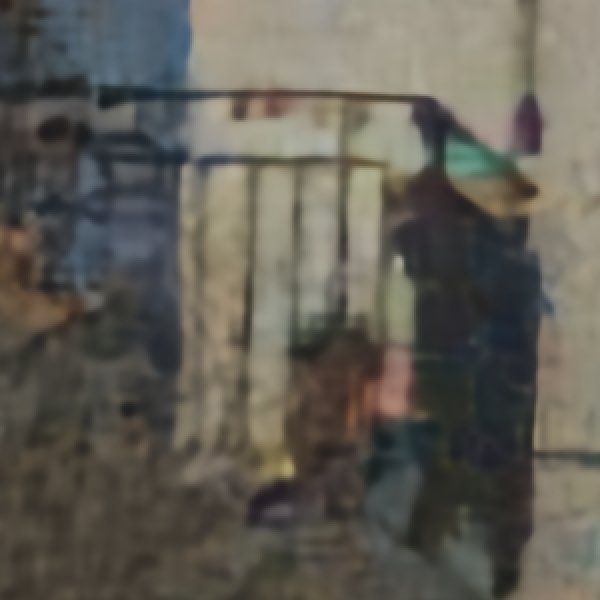}{BIGFEATURE-CAMERA}
    \im{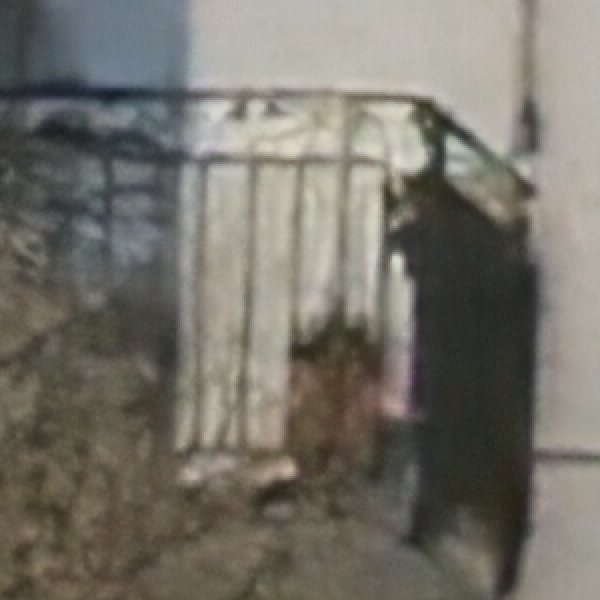}{Samsung-SLSI-MSL}
    \im{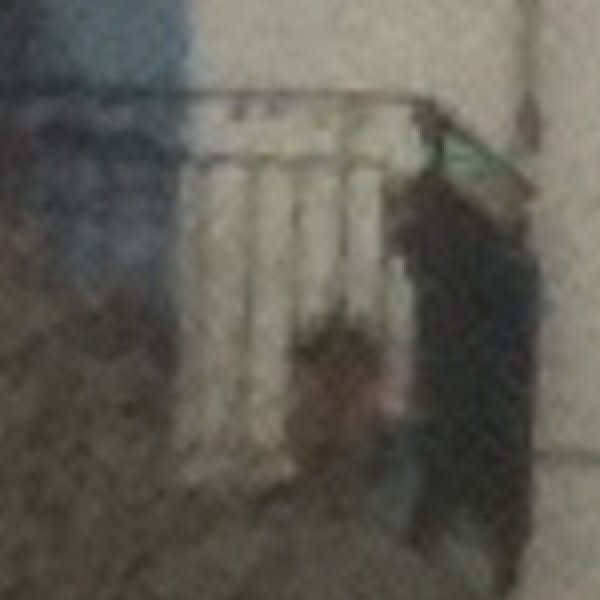}{Bicubic}
    \im{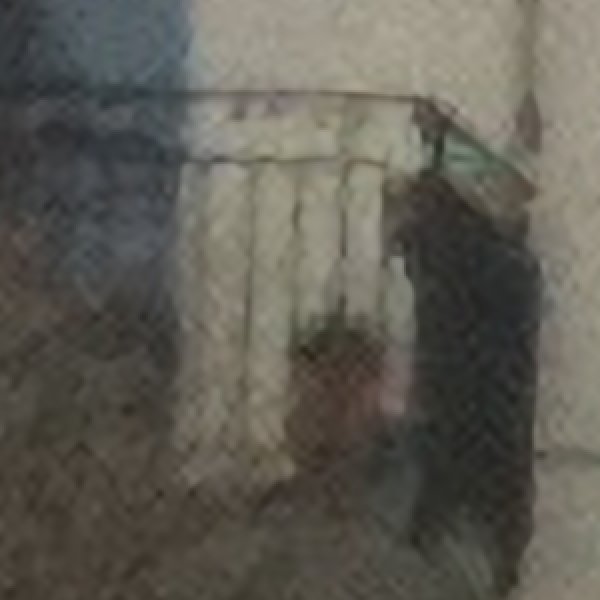}{RRDB Pre-trained}\vspace{3mm}
	\renewcommand{\im}[2]{
	\setlength\tabcolsep{-2pt}
    \begin{tabular}{c}
         \vspace{-4pt}
         \includegraphics[width=\wid]{figures/visuals/track2/3/#1.jpg} \\ 
         \scriptsize{#2} 
    \end{tabular}}
    \im{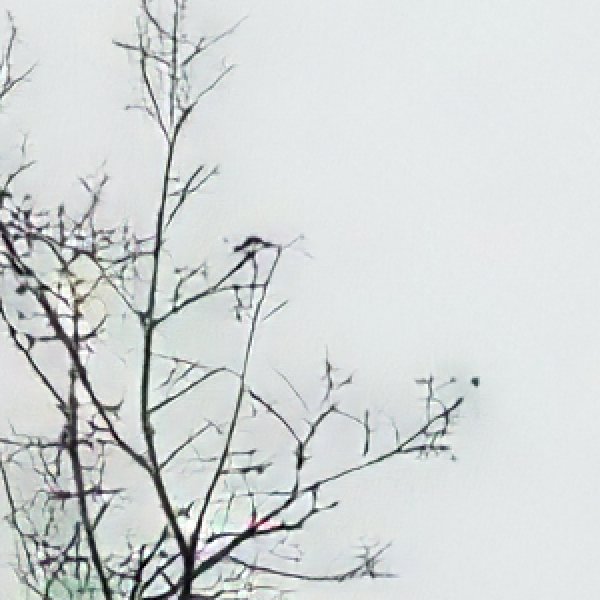}{Impressionism}
    \im{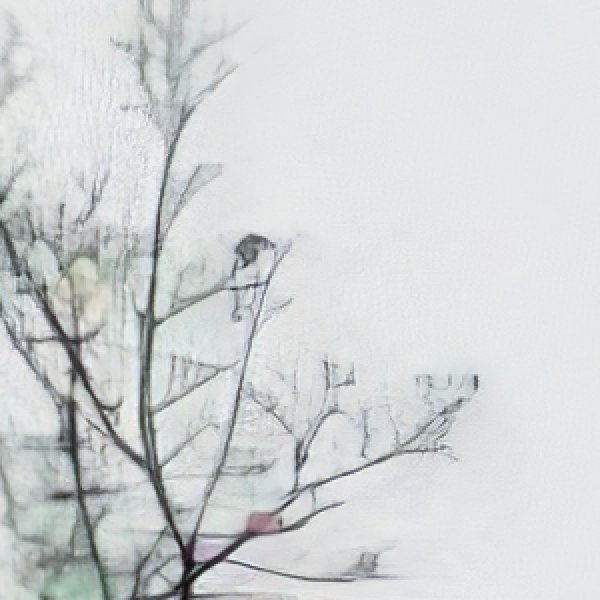}{AITA-Noah-A}
    \im{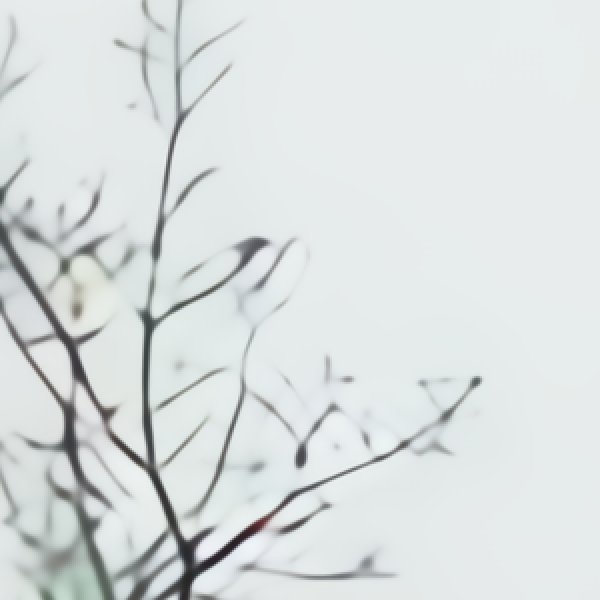}{ITS425}
    \im{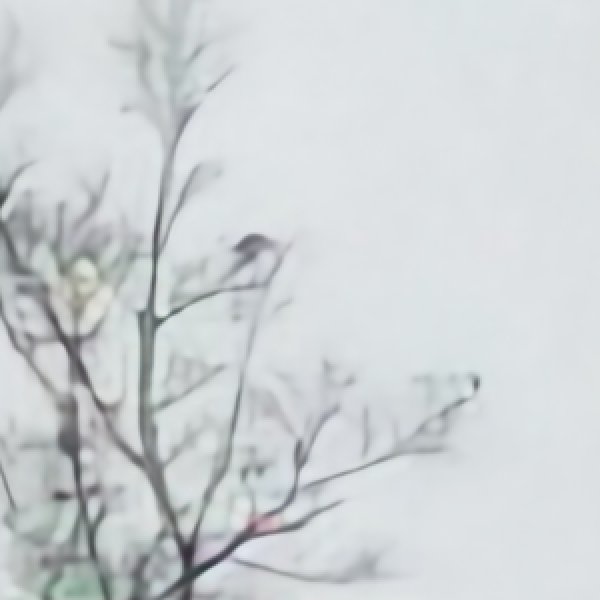}{AITA-Noah-B}
    \im{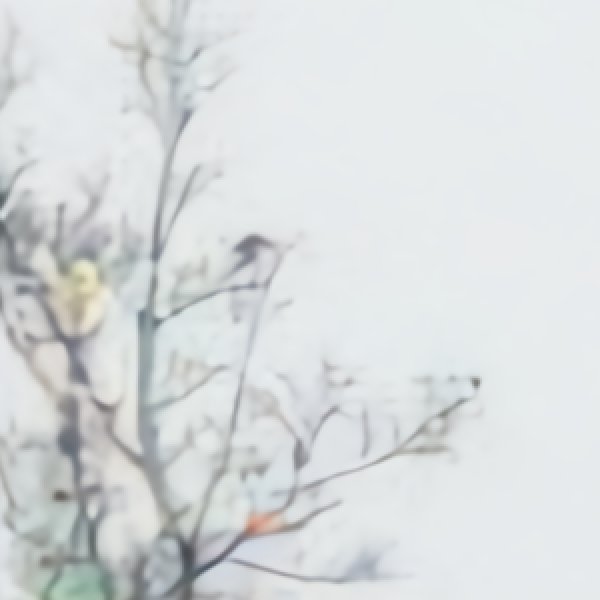}{Webbzhou}
    \im{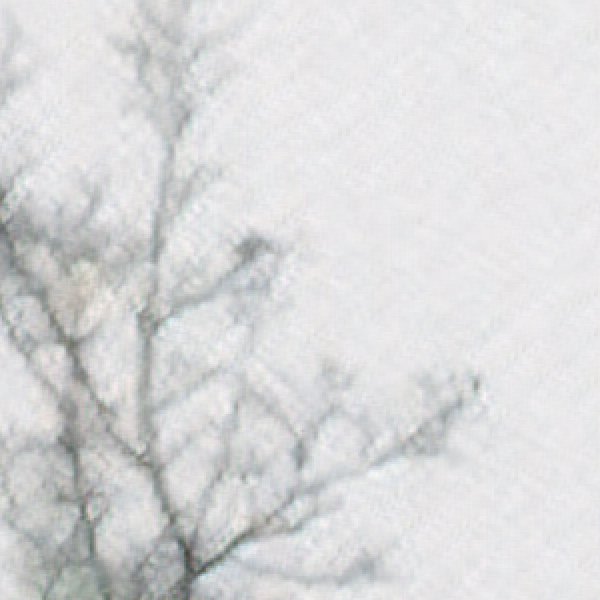}{Relbmag-Eht}
    \im{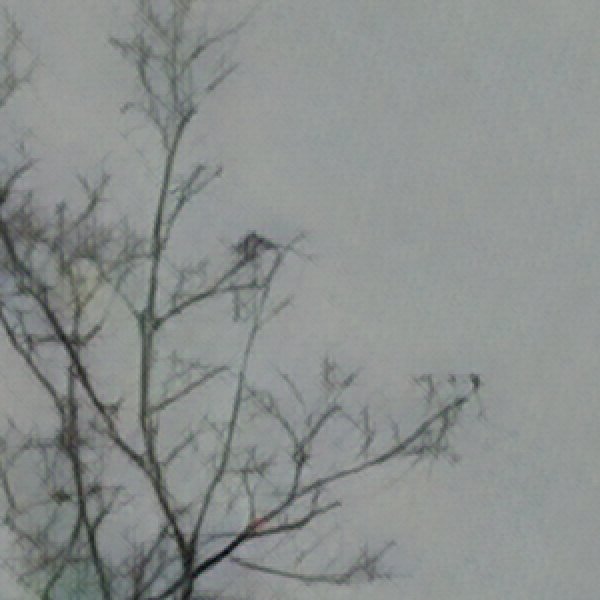}{MSMers}
    \im{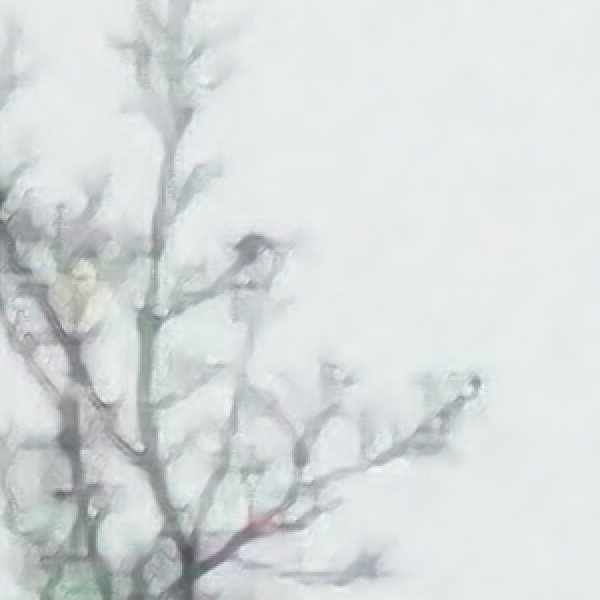}{MLP-SR}
    \im{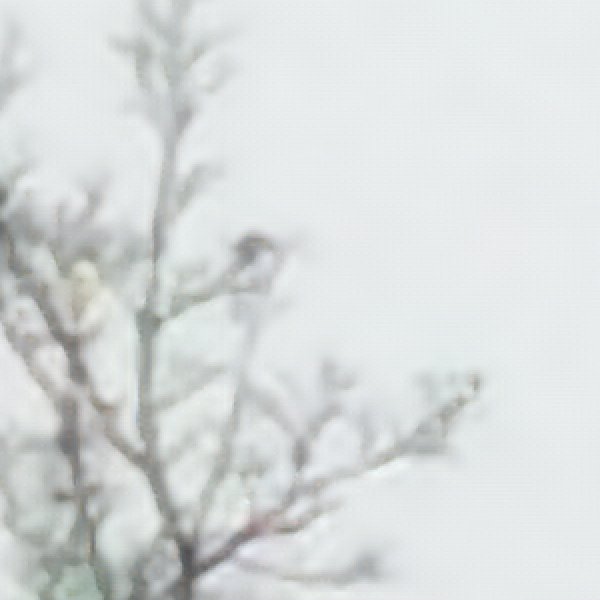}{SR-DL}
    \im{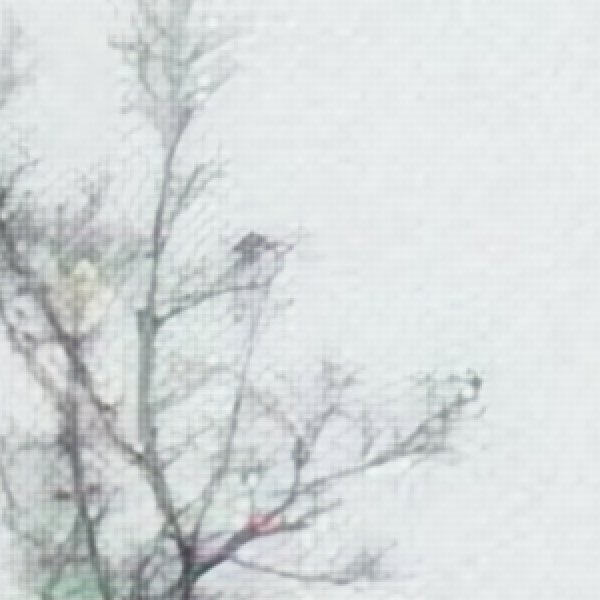}{InnoPeak-SR}
    \im{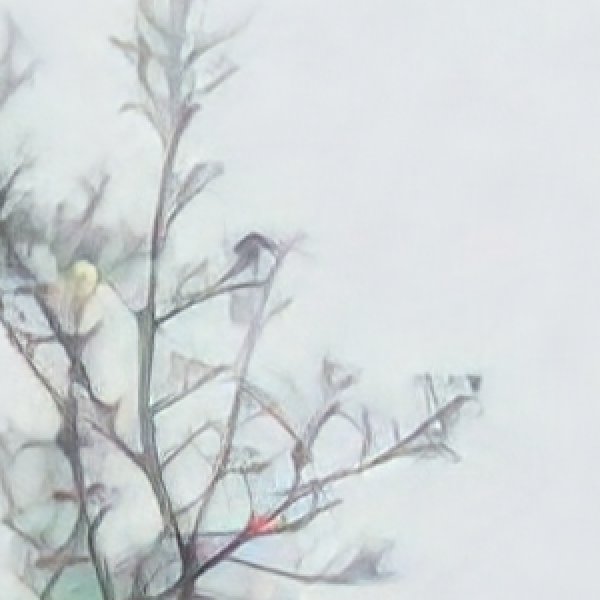}{QCAM}
    \im{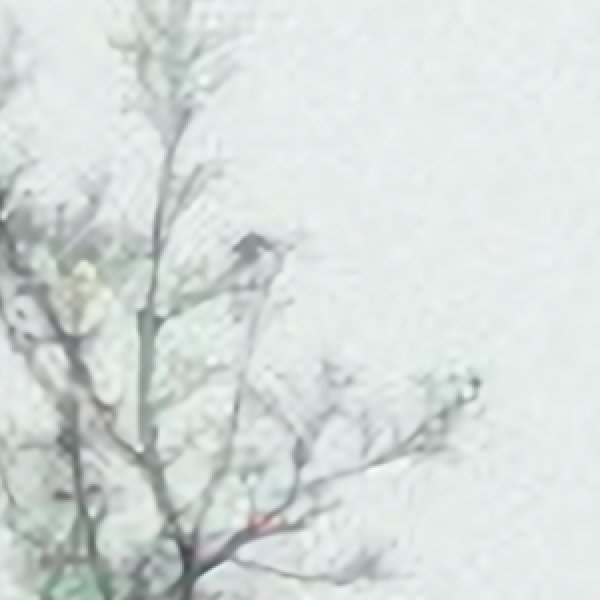}{SuperT}
    \im{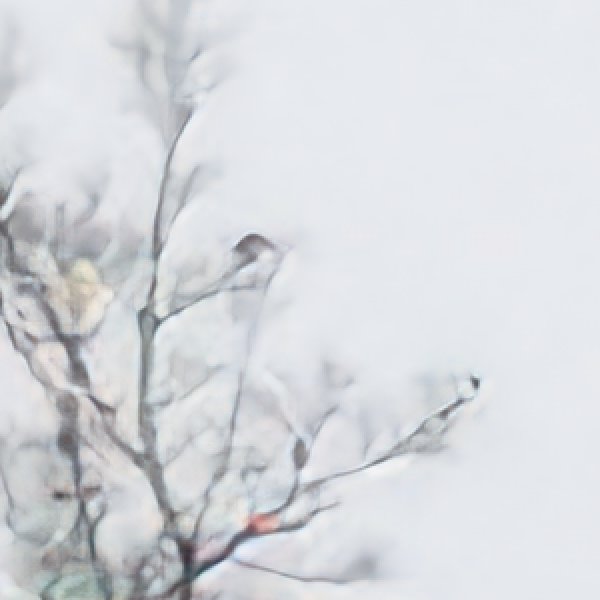}{KU-ISPL}
    \im{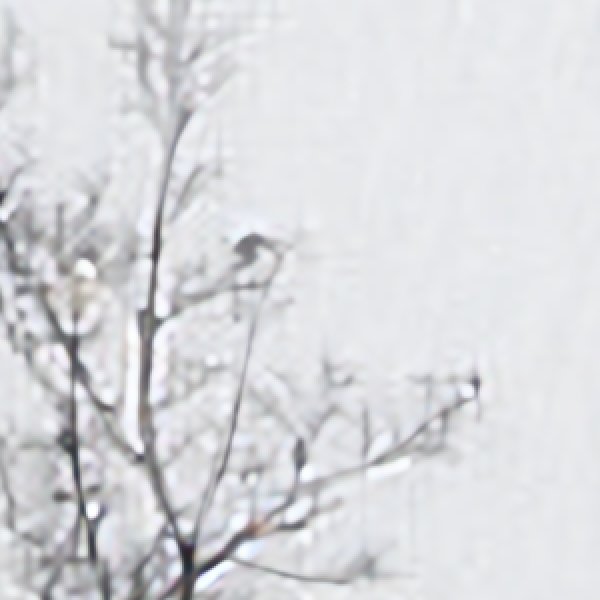}{BMIPL-UNIST-YH-1}
    \im{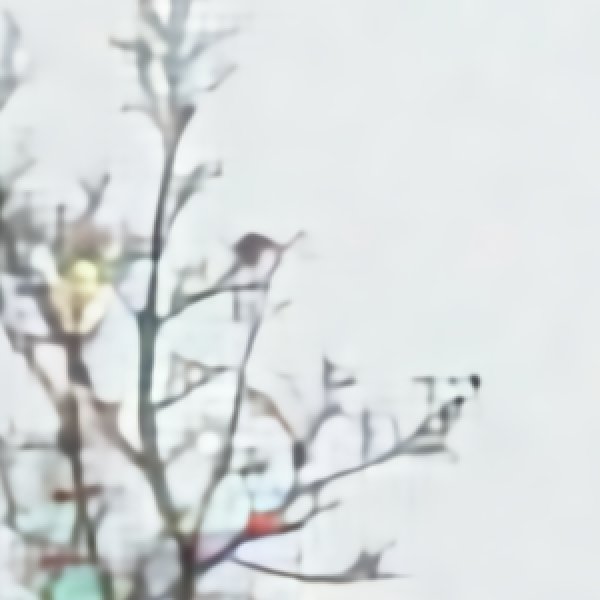}{BIGFEATURE-CAMERA}
    \im{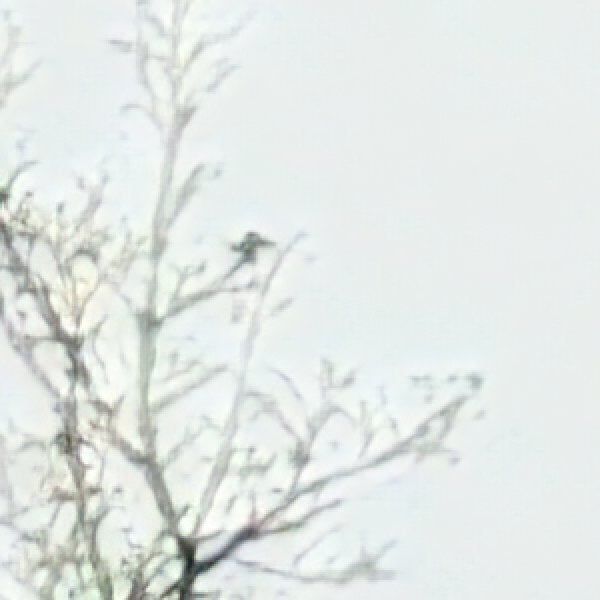}{Samsung-SLSI-MSL}
    \im{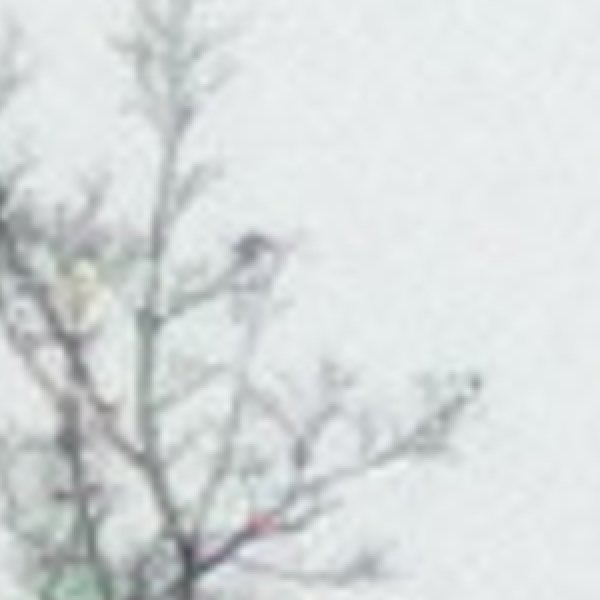}{Bicubic}
    \im{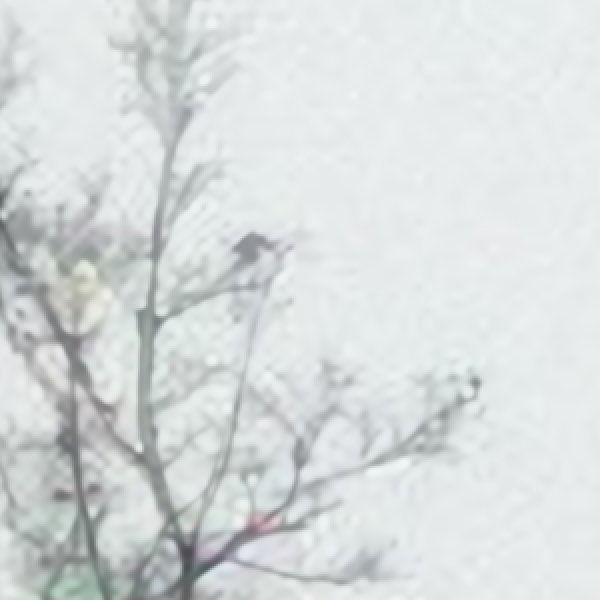}{RRDB Pre-trained}\vspace{1mm}
	\caption{Qualitative comparison between the participating approaches for Track 2. ($4\times$ super-resolution)}\vspace{-4mm}
	\label{fig:visuals_track_2}
\end{figure*}

\subsection{Track 1: Image Processing Artifacts}

Here we present the results for Track~1. All experiments presented were conducted on the test set. The results are shown in Table~\ref{tab:track1}. The Impressionism team achieves the best result, with a $9.5\%$ better MOS than the second entry, namely Samsung-SLSI-MSL. Both these teams take a more direct approach for simulating degradations for supervised SR learning. While Samsung-SLSI-MSL sample random noise distributions and down-scaling kernels, Impressionism aim to estimate the kernel and noise statistics. The following three approaches: BOE-IOT-AIBD, MSMers, and KU-ISPL, employ CycleGAN~\cite{chu2017cyclegan} or DSGAN~\cite{manuelFS} based methods to learn the degradation operator. Also the AITA-Noah team follows this general strategy, achieving impressive MOS results. However, their methods are not ranked in Track~1 since source domain images from the test set was used for training, which is against the rules of the challenge. Notable are also the results of ITS425, who achieve the second best PSNR and best SSIM, while preserving good perceptual quality. Also the third-ranked method BOE-IOT-AIBD achieves very impressive PSNR and SSIM.

When comparing with the previous edition of the challenge~\cite{AIM2019RWSRchallenge}, the performance of the proposed method has improved substantially. In~\cite{AIM2019RWSRchallenge}, most method achieved similar or worse results than simple Bicubic interpolation. Here, all top-10 approaches achieved better MOS than the Bicubic baseline. Moreover, while a large gap to supervised methods was reported in~\cite{AIM2019RWSRchallenge}, in this year challenge, the winning Impressionism method even beats the ESRGAN baseline, which is trained with full supervision. While this can also be partly explained by other modifications and hyper-parameter settings, it clearly demonstrates that the performance gap to supervised SR methods is significantly narrower.
Visual results for all methods in Figure~\ref{fig:visuals_track_1}. 

\subsection{Track 2: Smartphone Images}
Quantitative results for Track~2 are reported in Table~\ref{tab:track2}. In this track, the Impressionism method outperforms other approaches by a large margin in the human study (MOR). This is also confirmed in the visual examples shown in Figure~\ref{fig:visuals_track_2}. The generated images are superior in sharpness compared to those of other approaches. Moreover, the SR images contain almost no noise and few artifacts. While AITA-Noah and ITS425 also generate clean images, they lack the sharpness and detail of Impressionism. We believe this to be largely due to the kernel estimation performed in the latter approach, employing KernelGAN for this purpose. This allows the SR network to take the pointspread function of the specific camera sensor into account.

We observe that Impressionism also achieves the best average IQA-Rank. However, note that while the Relbmag-Eht team achieves a similar IQA-Rank, their result severely suffers from a structured noise pattern. This suggests that standard IQA metrics are not well suited as evaluation criteria in this setting and data. Interestingly, the Samsung-SLSI-MSL team employed the paired DSLR images provided by~\cite{IgnatovKTVG17DPED}. This approach is therefore not ranked in this track. However, this approach does still not achieve close to the same level of sharpness as Impressionism. 

Despite being the first challenge of its kind, the top participating teams achieved very impressive results in this difficult real-world setting, where no reference data is available. In particular, the Impressionism team achieves not only a higher resolution image, but also substantially better image quality than the source image taken by the camera.

\clearpage

\section{Challenge Methods and Teams}
\label{sec:methods}

This sections give brief descriptions of the participating methods. A summary of all participants is given in table~\ref{tab:teams}.

\newcommand{\xtrain}{\mathcal{X}}
\newcommand{\ytrain}{\mathcal{Y}}

\newcommand{\teamsection}[1]{\subsection{#1}\label{sec:#1}}

\teamsection{Impressionism}

The team Impressionism proposes a novel framework, introduced in \cite{Impressionism}, to improve the robustness of the super-resolution model on real images, which usually fails when trained on bicubic downsampled data. 
To generate more realistic LR images, they design a real-world degradation process that maintains important original attributes. Specifically, they focus on two aspects: 1) The blurry LR image is obtained by downsampling High-Resolution (HR) images with estimated kernels from real blurry images. 2) The real noise distribution is restored by injecting collected noise patches from real noisy images.
From the real-world (source domain) dataset $\mathcal{X}$  and the clean HR (target domain) dataset  $\mathcal{Y}$, the team thus aims to construct domain-consistent data $\{{\mathbf I}_\text{LR},{\mathbf I}_\text{HR}\} \in \{\mathcal{X},\mathcal{Y}\}$.

\parsection{Clean-up}
Since bicubic downsampling can remove high-frequency noise, they directly do a bicubic downsampling on the image from $\mathcal{X}$ to obtain more HR images. Let ${\mathbf I}_\text{src} \in \mathcal{X}$, and ${\mathbf k_\text{bic}}$ be the ideal bicubic kernel. Then the image is downsampled with a clean-up scale factor $s$ as ${\mathbf I}_\text{HR} = ({\mathbf I}_\text{src} * {\mathbf k_\text{bic}}){\downarrow}_{s}$.
Then the images after downsampling are regarded as clean HR images, that is ${\mathbf I}_\text{HR} \in \mathcal{Y}$. 

\parsection{Downsampling}
The team performs downsampling on the clean HR images using the estimated kernels by KernelGAN \cite{Bell19InternalGAN}. The downsampling process is a cross-correlation operation followed by sampling with stride $s$,
\begin{equation}
{\mathbf I}_{D} = ({\mathbf I}_{HR} * {\mathbf k}_i) {\downarrow}_s \,,\; i \in \{1,2, \ldots, m\},
\end{equation}
where ${\mathbf I}_{D}$ denotes the downsampled image, and ${\mathbf k}_i$ refers to the specific blur kernel. 

\parsection{Noise Injection}
Mere estimation of the blurry kernel cannot accurately model the degradation process of $\mathcal{X}$. By observing the real data, they find that the noise is usually combined with content of the image. In order to decouple noise and content, they design a filtering rule to collect noise patches $\{{\mathbf n}_i, i \in \{1,2 \cdots l\} \}$ with their variance in a certain range $\sigma^2 ({\mathbf n}_i) < v$,
where $\sigma^2(\cdot)$ denotes the variance, and $v$ is the maximum value of variance. 
Then these patches will be added to ${\mathbf I}_{D}$ as,
\begin{equation}
{\mathbf I}_{LR} = {\mathbf I}_{D} + {\mathbf n}_i\,,\; i \in \{1,2 \cdots l\}\,.
\end{equation}
After downsampling HR images with the estimated kernels and injecting collected noise, they obtain ${\mathbf I}_{LR} \in \mathcal{X}$.  

\parsection{Network Details}
Based on ESRGAN \cite{wang2018esrgan}, they train a super-resolution model on constructed paired data $\{{\mathbf I}_{LR},{\mathbf I}_{HR}\} \in \{\mathcal{X},\mathcal{Y}\}$.  Three losses are applied to training including pixel loss $L_{1}$, perceptual loss $L_{per}$, and adversarial loss $L_{adv}$. Different from default setting, they use patch discriminator \cite{isola2017image} instead. Overall, the final training loss is as follows:
\begin{equation}
L_{total} = {\lambda}_{1} * L_{1} + {\lambda}_\text{per} * L_\text{per} + {\lambda}_\text{adv} * L_\text{adv},
\end{equation}
where ${\lambda}_{1}$, ${\lambda}_\text{per}$, and ${\lambda}_\text{adv}$ are set as 0.01, 1, and 0.005 empirically.

\begin{figure}[t]
	\centering%
	\includegraphics[width=\columnwidth]{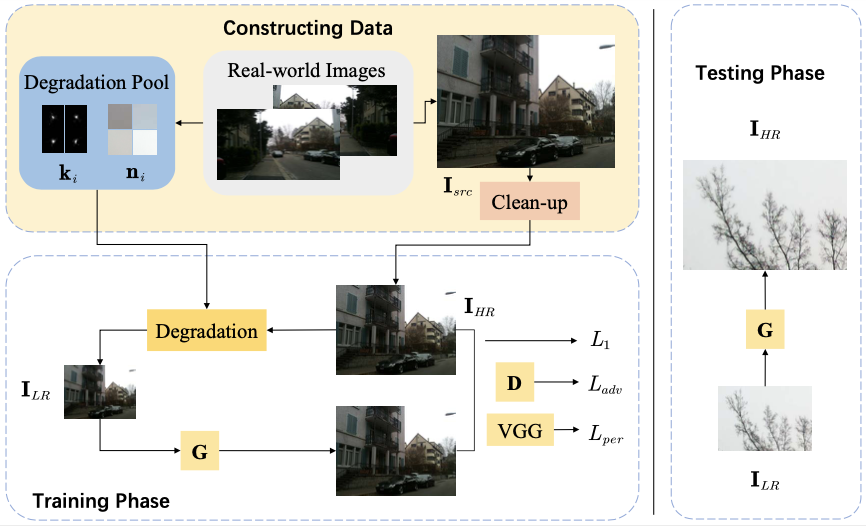}
	\caption{Overview of the method by the \textbf{Impressionism} team.}
	\label{fig:Impressionism}
\end{figure}

\begin{table*}[t]
	\centering%
	\resizebox{1.01\textwidth}{!}{%
		\newcommand{\specialcell}[2][l]{%
  \begin{tabular}[#1]{@{}c@{}}#2\end{tabular}}
\begin{tabular}{lllcccc}
\toprule
&&&\multicolumn{2}{c}{Track 1}&\multicolumn{2}{c}{Track 2}\\
Team Name &  Username in Codalab &                                                                    Additional Data & Traintime [h] & Runtime [sec] & Traintime [h] & Runtime [sec] \\
\midrule
 AITA-Noah-A &  AITA &  \specialcell{Track 1 and 2: AIM-2019 pretrained model. \\ Track 2: only use external 400 div8k images.} &   8 &  0.5 &   8 &  0.5 \\
 AITA-Noah-B &  Noah\_TerminalVision &  AIM-2019 pretrained ESRGAN-FS model.  &  8 &  5 &  8 &  0.3 \\
 BIGFEATURE\_CAMERA &  conson0214 &  DSGAN for LR-HR pairs, DF2K to pre-train SR model.  &  22 &  0.25  &  22 &  0.25 \\
 BMIPL\_UNIST\_YH\_1 &  syh &  RCAN Super Resolution model &   32 &  40 &   12 &  40 \\
 BOE-IOT-AIBD &  eastworld &  739 pexels.com images, downsized to 2K &  264 &  38.20 &  no &  no \\
 GDUT-wp &  HouseLee &  - &   10 &  0.85 &  no &  no \\
 ITS425 &  Ziyao\_Zong &  - &   24 &  1.34 &  24 &  1.24 \\
 Impressionism &  xiaozhongji &  RRDB\_PSNR\_x4.pth released by the ESRGAN authors &   12 &  1.3 &   32 &  0.9 \\
 InnoPeak\_SR &  qiuzhangTiTi &  10,000 collected images &  12 &  0.15 &  12 &  0.15 \\
 KU-ISPL2 &  Kanghyu Lee &  VGG19 was used for VGG loss &  2 &  0.02 &  no &  no  \\
 KU\_ISPL &  gtkim &  VGG-19 model for perceptual loss &  168 &  6.48 &   168 &  4.11 \\
 MLP\_SR &  raoumer &  - &   28.57 &  1.289 &   0 &  967 \\
 MSMers &  huayan &  CycleGan, RCAN &    72 &  0.483 &   63 &  0.343 \\
 QCAM &  tkhu &  AIM2019 &  no &  no &   15 &  0.21 \\
 Relbmag Eht &  Timothy\_Cilered &  - &  no &  no &  8.9 &  1.09 \\
 SR\_DL &  ZhiSong\_Liu &  - &  15 &  4 &   15 &  1 \\
 SVNIT1-A &  kalpesh\_svnit &  - &  50 &  1.09 &  no &  no \\
 SVNIT1-B &  Kishor &  - &   50 &  0.85 &  no &  no \\
 SVNIT2 &  vishalchudasama &  - &  50 &  0.92 &  no & no \\
 Samsung\_SLSI\_MSL &  Samsung\_SLSI\_MSL &  Flickr2K for Track 1, DPED for Track 2. &   72 &  1 &   24 &  1 \\
 SuperT &  tongtong &  DIV2K &    48 &  0.64 &   48 &  0.64 \\
 TeamAY &  nmhkahn &  - &   100 &  20 &  no &  no \\
 Webbzhou &  Webbzhou &  - &   60 &  0.5 &   60 &  0.5 \\
\bottomrule
\end{tabular}}
	\label{tab:teams}
	\caption{Information about the participating teams in the challenge.}
\end{table*}

\teamsection{AITA-Noah}

This method, which is detailed in~\cite{AITA-Noah}, adopts the idea of learning the degradation operator in order to synthetically generate paired training data for SR network.
For Track~1, an approach termed \emph{Iterative Domain Adaptation} is developed. The source training data $\mathcal{X}_{tr}$ and downsampled target training data $\mathcal{Y}_{tr\downarrow}$ are first processed with a denoising algorithm (Non-local Means), denoted $D$. The sets $D(\mathcal{Y}_{tr\downarrow})$ and $\mathcal{Y}_{tr\downarrow}$ are then used to train a sharpening network $S$, in a fully supervised manner. When applied to the source data, $S(D(x_{tr}))$ generates images that are clean and sharp. This set can then be used to train a degradation operator $G$, using pairs from $S(D(\mathcal{X}_{tr}))$ and $\mathcal{X}_{tr}$. This is then used to train a super-resolution network $SR$ using pairs generated by $G(\mathcal{Y}_{tr\downarrow})$ and $\mathcal{Y}_{tr}$. The approach then proceeds by iteratively improving the degradation model $G$ using pairs $f(\mathcal{X}_{tr})$ generated by the current SR model $f$ and $\mathcal{X}_{tr}$, and improving the super-resolution model $f$ using pairs $G(\mathcal{Y}_{tr})$ generated by the current degradation operator $G$ and $\mathcal{Y}_{tr}$. In practice, the team used the 100 source validation images and 100 source test images as $\mathcal{X}_{tr}$. The team is not ranked in track 1, since according to the challenge rules, the test data should not be used during training, even in unpaired form.

\begin{figure}[t]
	\centering%
	\includegraphics[width=\columnwidth]{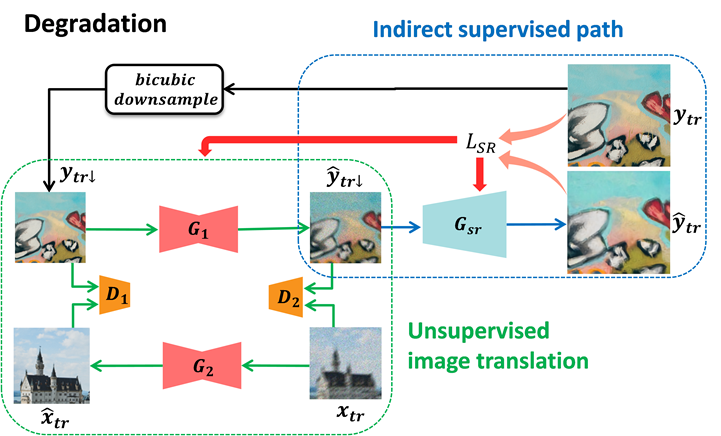}
	\caption{Overview of the CycleSR method used by \textbf{AITA-Noah} to learn the degradation operation for Track~2.}
	\label{fig:AITA-Noah}
\end{figure}

For Track~2, the team adopts the CycleSR framework~\cite{UISRhan2019,lugmayrICCVW2019} to generate degrade images. As illustrated in Fig.~\ref{fig:AITA-Noah}, this framework is composed of two stages: 1) unsupervised image translation between real LR images $\mathcal{X}_{tr}$ and synthetic LR images, \ie, $4\times$ bicubic downsampled HR images $\mathcal{Y}_{tr}$, denoted by $\mathcal{Y}_{tr\downarrow}$; 2) supervised super-resolution from degraded LR images $\hat{\mathcal{Y}}_{tr\downarrow}$ to get $\hat{\mathcal{Y}}_{tr}$. In detail, the approach first takes the unsupervised image translation model CycleGAN~\cite{zhu2017unpaired} for mapping between domain $\mathcal{X}_{tr}$ and $\mathcal{Y}_{tr\downarrow}$. An SR module SRResNet is employed after CycleGAN to super-resolve $\hat{\mathcal{Y}}_{tr\downarrow}$ to get $\hat{\mathcal{Y}}_{tr}$ and compute the loss $L_\text{SR}$ with ground truth $\mathcal{Y}_{tr}$. Hence, with an image translation model and an SR module together and a joint training strategy, we are able to train a model that super-resolves real LR images to HR images with an indirect supervised path. Compared with degradation directly using original CycleGAN, benefiting from the pixel-wise feedback of the SR module, CycleSR can alleviate color and brightness changes during degradation.

In both tracks, the same super-resolution architecture, based on the ESRGAN is used. The team furthermore use an LR-conditional frequency-separation discriminator to train the model and employ AutoML to tune the loss weights, employing LPIPS~\cite{zhang2018unreasonable} and NIQE~\cite{MittalSB13NIQE} as objective.
Two versions of this approach was submitted, with the significant differences as follows:

\parsection{AITA-Noah-A} 
For Track~1, this version uses the method described above. For Track~2, it includes an extra 400 images selected from DIV8K~\cite{div8k} in the target domain set $\mathcal{Y}_{tr}$ to improve data diversity.

\parsection{AITA-Noah-B} 
For Track~1, this approach additionally uses an ensemble fusion strategy (\ie running inference on the vertical flipped/horizontal flipped/transposed images of the original input, and then average the results), in addition to above. For Track~2, no extra data was used and no adversarial loss was used during training the ESRGAN model (\ie only RRDBNet was used).

\teamsection{Samsung-SLSI-MSL}

\begin{figure}[t]
	\centering%
	\includegraphics[width=\columnwidth]{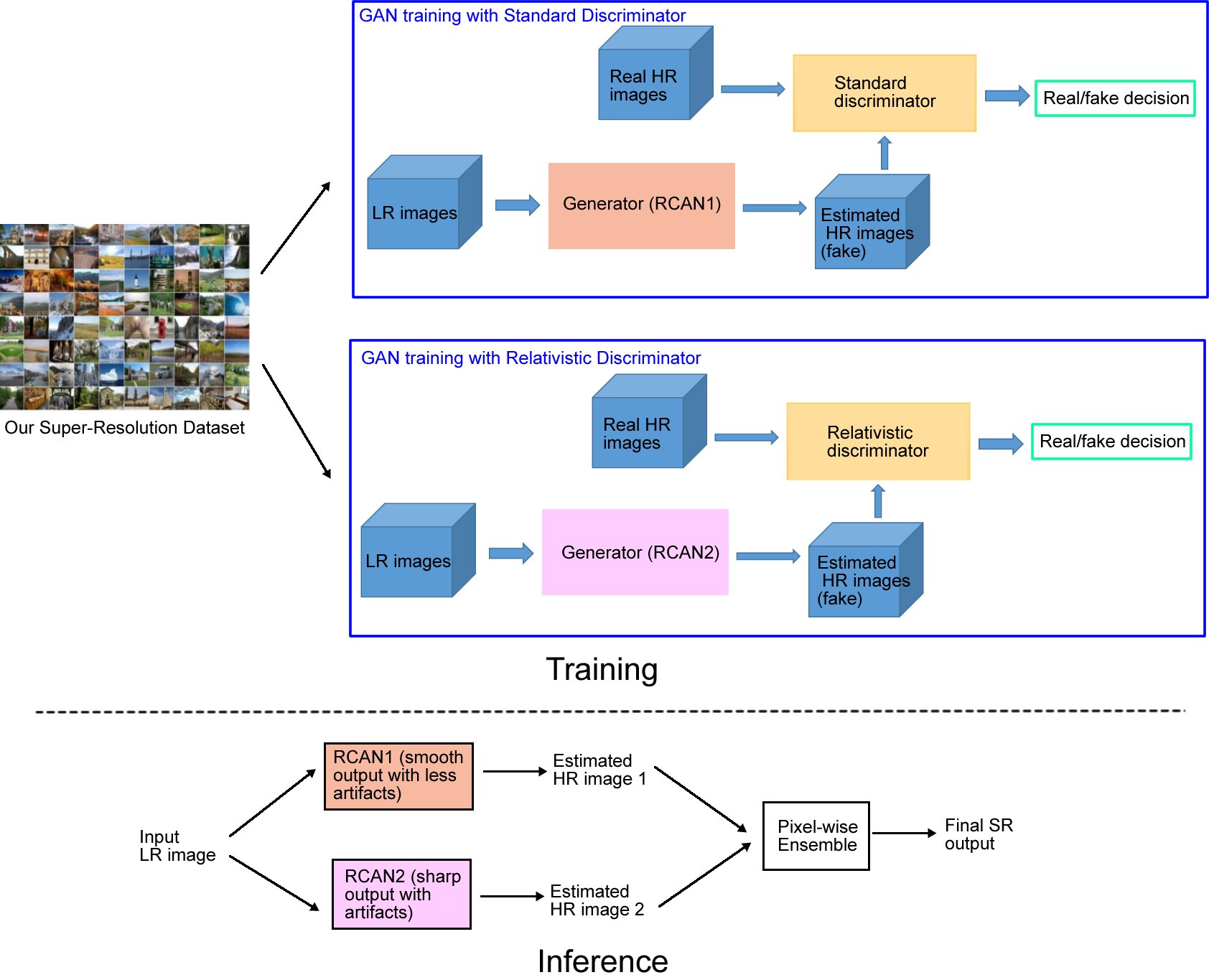}
	\caption{Overview of the SR method used by \textbf{Samsung-SLSI-MSL}.}
	\label{fig:Samsung-SLSI-MSL}
\end{figure}

For Track~1, this team aims to train a generic SR model that is robust to various image degradations, which can therefore be applied in real-world scenarios without knowledge of the specific degradation operator. This is performed by sampling diverse degradation types during training. The strategy proposed in the blind denoising method~\cite{RenEL18} is extended by adding downscaling and blur. The training set is generated by sampling different downscaling (\eg bilinear, nearest neighbor or bicubic), blur kernels (Gaussian kernel with different sigma), and noise distributions (additive Gaussian, Poisson, Poisson-Gaussian with randomly sampled parameters). The SR model consists of the RCAN~\cite{rcan} architecture, which is trained with a GAN loss while emphasizing the perceptual losses. To further improve the perceptual quality, they deploy an ensemble of two different GANs, and  use cues from the image luminance and adjust to generate better HR images at low-illumination. The workflow is given in Fig. \ref{fig:Samsung-SLSI-MSL}.

For Track~2, real world SR on images captured by mobile devices, the same GANs are trained by weak supervision on a mobile SR training set that they constructed to have LR-HR image pairs, from the DPED dataset which provides registered mobile-DSLR images at the same scale~\cite{IgnatovKTVG17DPED}. They use the mobile images as LR,  and apply the track 1 generic SR model on the paired DSLR  images to create super resolved HR images with good perceptual quality. This method is considered as a kind of \emph{Supervised} approach, and does not compete with the other participants in Track~2.
Details about the proposed method can be found in~\cite{samsung}.

\teamsection{MSMers}

\begin{figure}[t]
	\centering%
	\includegraphics[width=\columnwidth]{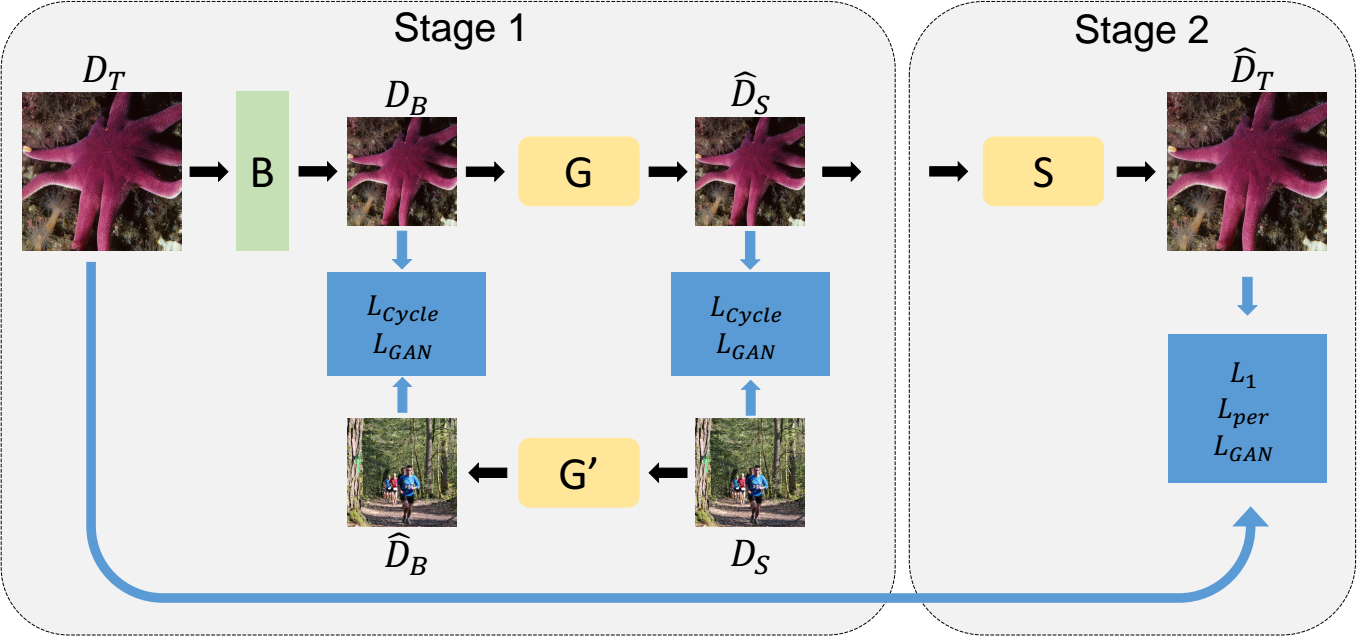}
	\caption{Overview of the method by the \textbf{MSMers} team.}
	\label{fig:MSMers}
\end{figure}

This method takes inspiration from \cite{lugmayrICCVW2019}, developing a two-stage approach. First, a degradation operator is learned in an unsupervised manner. This is then used to generate paired data for the second stage, in which the SR network is learned. Specifically, CycleGAN \cite{zhu2017unpaired} is
adopted in the first stage to learn a mapping from bicubic downsampled
HR to real LR. To keep the color consistent, the weight of the identity loss is increased in the setting. 
As for the second stage, RCAN \cite{rcan} is used to super-resolve the
LR image, which is first trained on L1 loss. On top of that, perceptual loss
and adversarial loss are added for better perceptual quality. Specifically,
we use features of VGG19 relu5-1 layer to compute a perceptual loss and
the WGAN-GP \cite{WGAN-GP} as adversarial loss. The method is visualized in Figure~\ref{fig:MSMers}.

\teamsection{BOE-IOT-AIBD}

\begin{figure}[b]
	\centering%
	\includegraphics[width=\columnwidth]{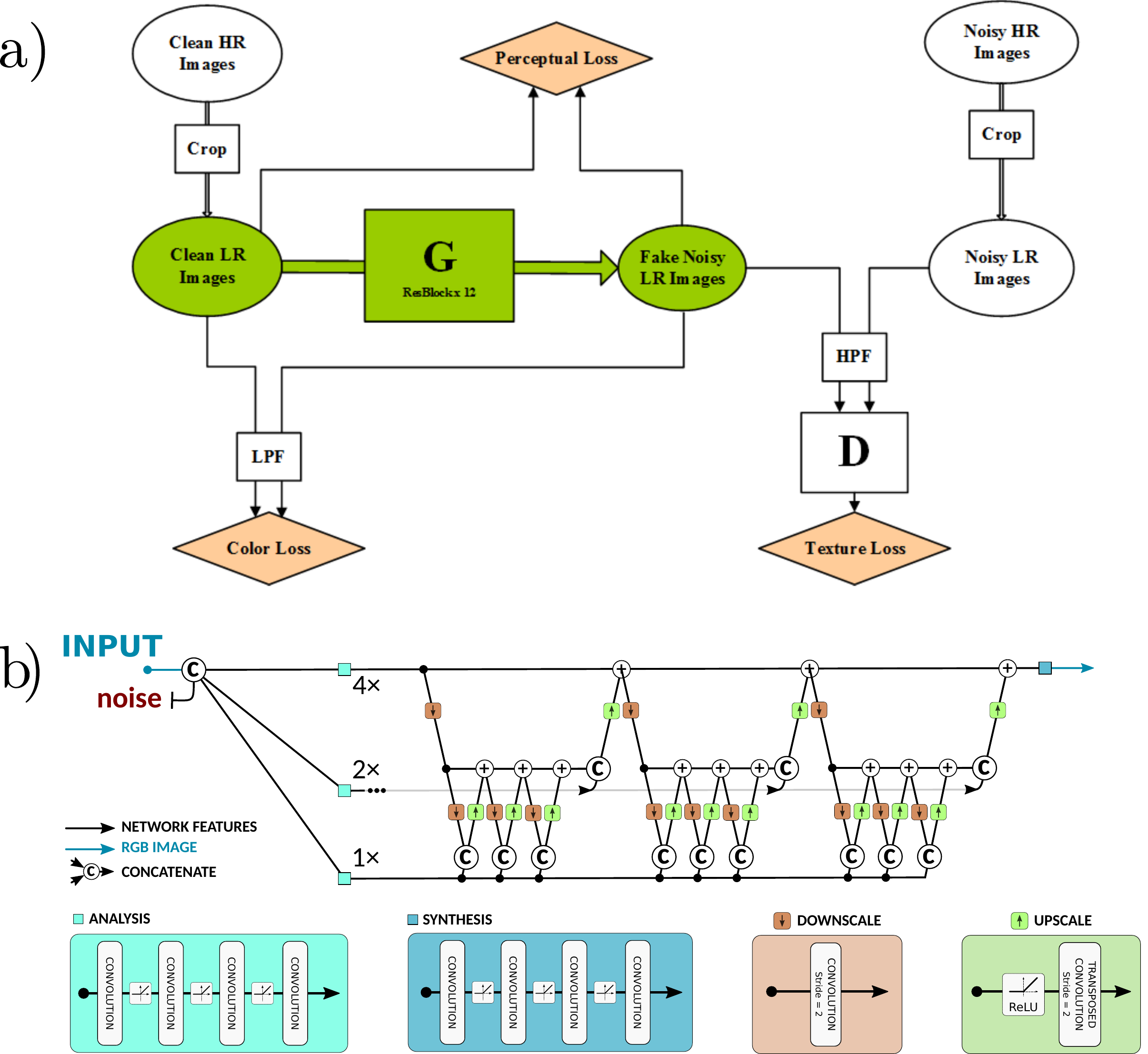}
	\caption{Overview of the DSGAN \cite{manuelFS} based method used by the \textbf{BOE-IOT-AIBD} team to learn the degradation operation.}
	\label{fig:BOE-IOT-AIBD}
\end{figure}

This team aims to learn the degradation operator in order to generate paired SR training samples. To this end, it employs solution provided by DSGAN \cite{manuelFS} to artificially generate LR images, as shown in Figure~\ref{fig:BOE-IOT-AIBD}a. These are then used to train an SR model. For this, it uses the modified MGBPv2~\cite{MGBPv2} network, proposed in the winning solution of the AIM ExtremeSR challenge \cite{AIM2019ESR}. It is adapted to $4\times$ upscaling by using a triple--V cycle (instead of the W--cycle) and adding multi--scale denoising modules as shown in Figure~\ref{fig:BOE-IOT-AIBD}b. During inference, an overlapping patch approach is used to further allow upscaling of large images. The training strategy employs a multiscale loss, combining distortion and perception losses on the output images. Model selection was performed by selecting low NIQE results on validation set and human tests based on ITU--T P.910. An additional set of 739 collected images for training. The team only participated in Track~1.

\teamsection{InnoPeak-SR}

\begin{figure}[t]
	\centering%
	\includegraphics[width=\columnwidth]{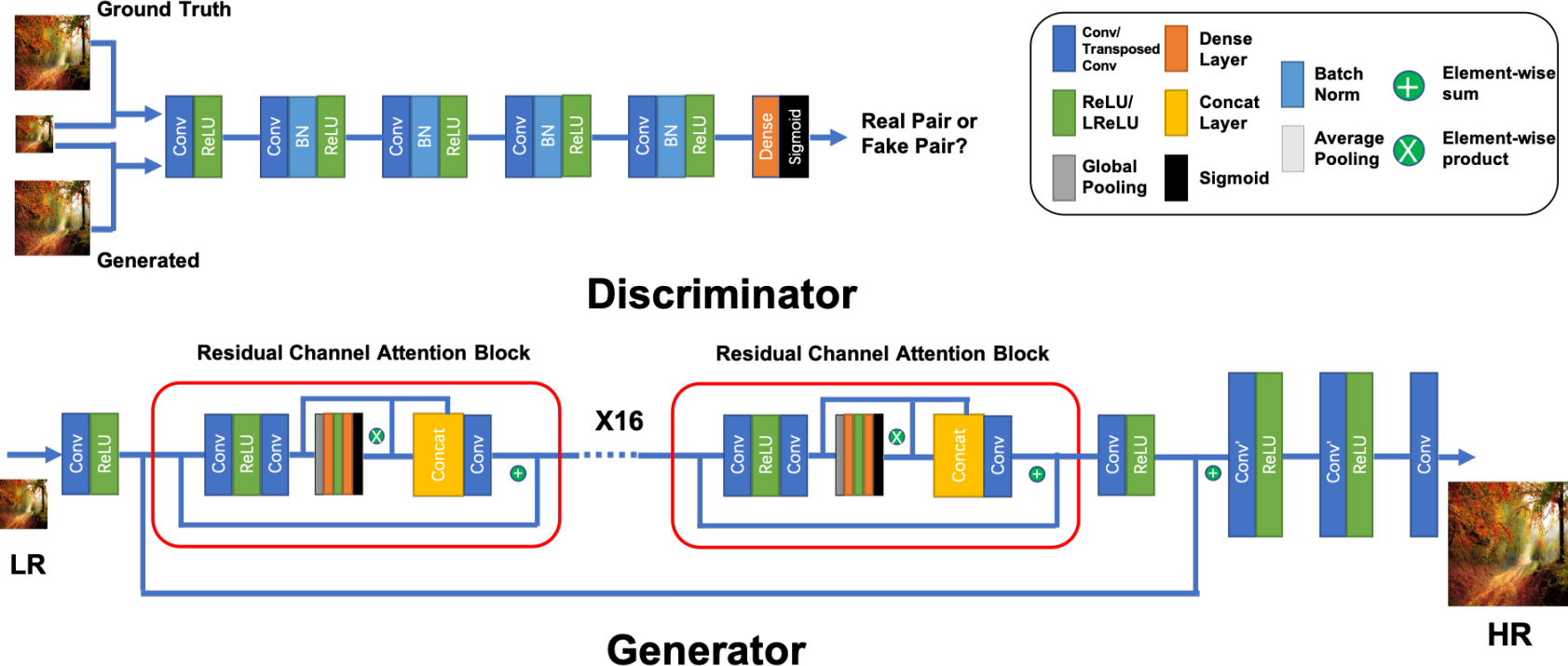}
	\caption{Architectures employed by the \textbf{InnoPeak-SR} team.}
	\label{fig:InnoPeak-SR}
\end{figure}

This approach does not directly address the unavailability of paired training data. Instead, it aims to develop a robust architecture capable of generalizing to the degradations present in the real-world setting, while trained using standard strategies. The SR network consists of a residual channel attention generator, visualized in Figure~\ref{fig:InnoPeak-SR}. It mainly consists of four parts: shallow feature extraction, residual
channel attention feature extraction, upscale module, and reconstruction. 
The discriminator network is implemented using four repeated $4\times 4$ convolution layers, followed by BatchNorm and ReLU. The networks are trained in a standard GAN  fashion. The generator additionally uses $L_1$, VGG, SSIM and gradient losses. The authors additionally used 10000 images from the ImageNet dataset for training.

Details about the proposed method can be found in~\cite{cai2020residual}

\teamsection{ITS425}

This team focus on improving the SR network architecture. The image degradation operator is first learned using an improved version of the DSGAN~\cite{manuelFS}, by using a smaller generator model than that of the original work. Unlike other methods, this team also aims to improve the quality of the target domain images. This is performed by training denoising and detail enhancement models to improve the target domain HR training images. The SR models is based on the RDN architecture~\cite{RDN}. It is modified by using the add operation instead of concatenate, which not only reduces
the amount of calculations of the model but also reduces the high-level
 information that is passed back to the final layers.

\teamsection{MLP-SR}

\begin{figure}[t]
	\centering%
	\includegraphics[width=\columnwidth]{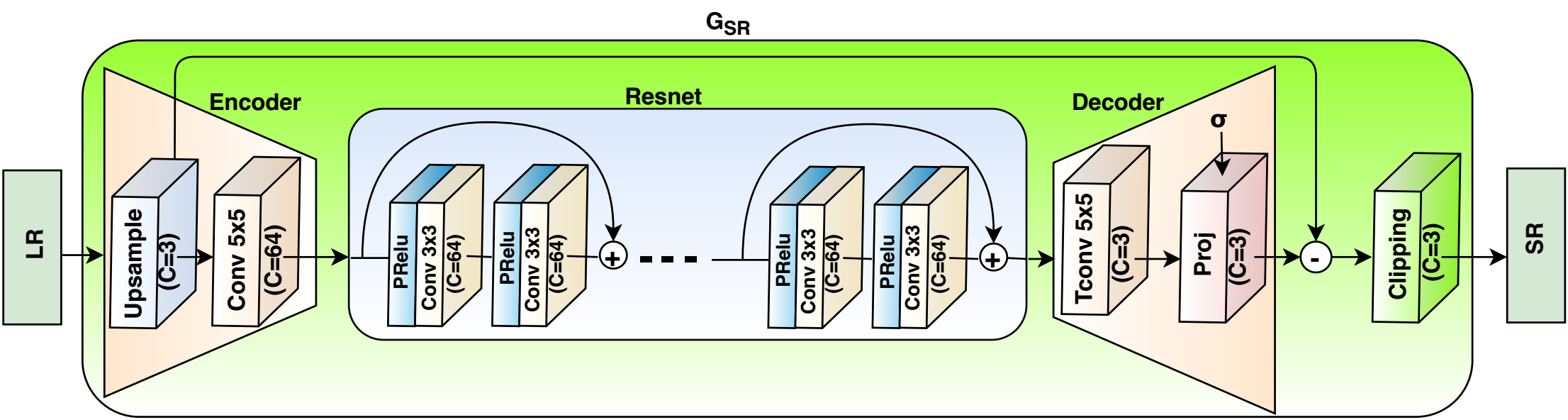}
	\caption{SR architecture employed by the \textbf{MLP-SR} team.}
	\label{fig:MLP-SR}
\end{figure}

This team follow a two stage approach. First, a DSGAN~\cite{manuelFS} (winner of the AIM2019 RWSR challenge) network and training strategy is employed to learn the image degradation mapping. This is then used to generate paired SR training data for the second stage. The team proposes a SR architecture, shown in Figure~\ref{fig:MLP-SR}, inspired by a physical image formation model. It uses a encoder-decoder structure. The inner ResNet
consists of 5 residual blocks with two pre-activation Conv layers. The pre-activation is the parametrized rectified linear unit (PReLU). The trainable projection layer \cite{UDN} inside Decoder computes the proximal map with the estimated
noise standard deviation and handles the data fidelity and prior terms. The
noise realization is estimated in the intermediate ResNet that is sandwiched
between Encoder and Decoder. The estimated residual image after Decoder is
subtracted from the LR input image. Reflection
padding is also used before all Conv layers to ensure slowly-varying changes at
the boundaries of the input images. The generator structure can also be described as the generalization of one stage TNRD \cite{Chen17} and UDNet \cite{UDN} that have
good reconstruction performance for image denoising problem.
For the discriminator, it employs the architecture used in SRGAN~\cite{ledig2017photo}, with the relativistic loss used in ESRGAN~\cite{wang2018esrgan}. In addition, $L_1$, Total-Variation and VGG losses are used.

\teamsection{KU-ISPL}

\begin{figure}[t]
	\centering%
	\includegraphics[width=\columnwidth]{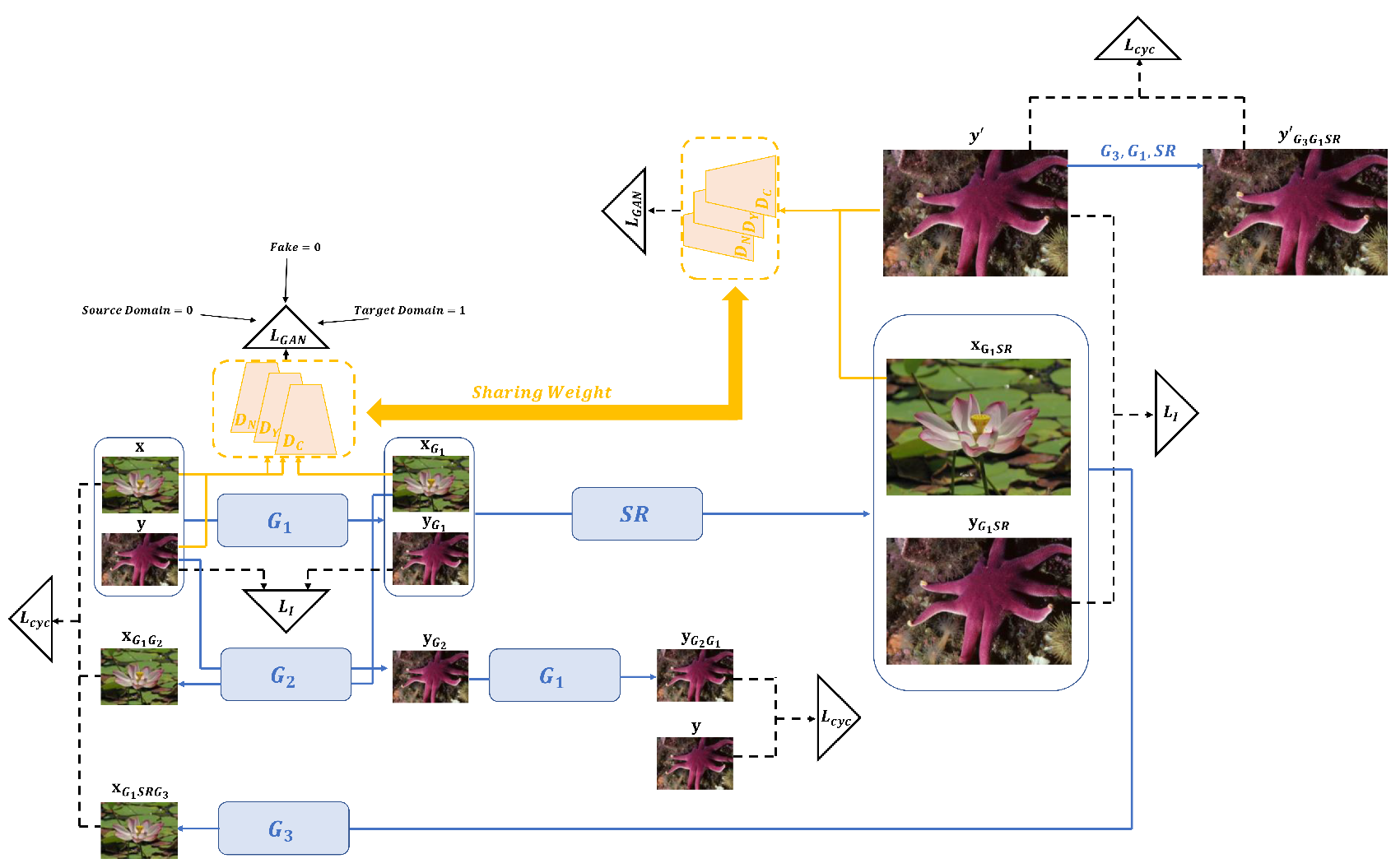}
	\caption{Overview of the learning approach proposed by the \textbf{KU-ISPL} team.}
	\label{fig:KU-ISPL}
\end{figure}

This team propose an un-paired GAN-based framework~\cite{kim2020unsupervised}. It consists of three generators, one
SR model and three discriminators. The overall architecture
is visualized in Figure~\ref{fig:KU-ISPL}.
The generators $G_1$, $G_2$, and $G_3$ constitute a modified CinCGAN \cite{yuan2018unsupervised}. Residual networks are used for these architectures. $G_3$ further downsamples the image by a factor of $4$.
The SR model is based on ESRGAN \cite{wang2018esrgan}. Bilinear upsampling is introduced into the architecture to preserve details and avoid checkerboard patterns induced by the transposed convolution module. 
The three discriminators $D_N$, $D_C$, and $D_Y$ are trained with different losses: adversarial noise loss, adversarial color loss, and adversarial texture
loss respectively. The $D_N$ uses a raw image, which contains noise signal. The $D_C$ and
$D_Y$ employ a Gaussian blurred image and a grayscale image, respectively,
as in WESPE~\cite{WESPE}. To improve performance of the discriminators,
source domain images are used when the discriminators are trained. Instead of classifying real or fake, the discriminator
distinguishes between source and target domain images. The generator
is trained to make target domain-like fake images and the discriminator
is trained to classify fake images as a source domain image.
The cycle consistency and identity loss each consist of three
losses: a pixel-wise $L_1$ loss, a VGG perceptual loss, and an SSIM loss. 

\teamsection{Webbzhou}

This team aims to first learn the degradation process in order to generate data for a second-stage SR network training. The degradation learning is based on the frequency separation in DSGAN~\cite{manuelFS}. Furthermore, in order to alleviate the color shift in degradation process, the team proposed a generator based on Color Attention Residual Block (CARB)~\cite{Zhou2020GuideFS}. In addition, the team modified the discrimnator of ESRGAN \cite{wang2018esrgan} which treats high frequency and low frequency separately. Finally, an EdgeLoss with Canny operator is constructed to further enhance details of edge.

\teamsection{SR-DL}

The team propose a joint image denoising and super-resolution model by using
generative Variational AutoEncoder (dSRVAE)~\cite{liu2020unsupervised}. It includes two parts: a Denoising AutoEncoder (DAE) and a Super-Resolution Sub-Network (SRSN).
With the absence of target images, a simple discriminator is trained together with the autoencoder to encourage the SR images to pick up the
desired visual pattern from the reference images.
During the training, Denoising AutoEncoder (DAE) is trained first by using source image training set. Then the Super-Resolution Sub-Network
(SRSN) is attached as a small head to the DAE which forms the proposed dSRVAE to output super-resolved images. Together with dSRVAE,
a simple convolutional neural network is used as a discriminator to distinguish whether generated SR images are close to the original input images.

\begin{figure}[t]
	\centering%
	\includegraphics[width=\columnwidth]{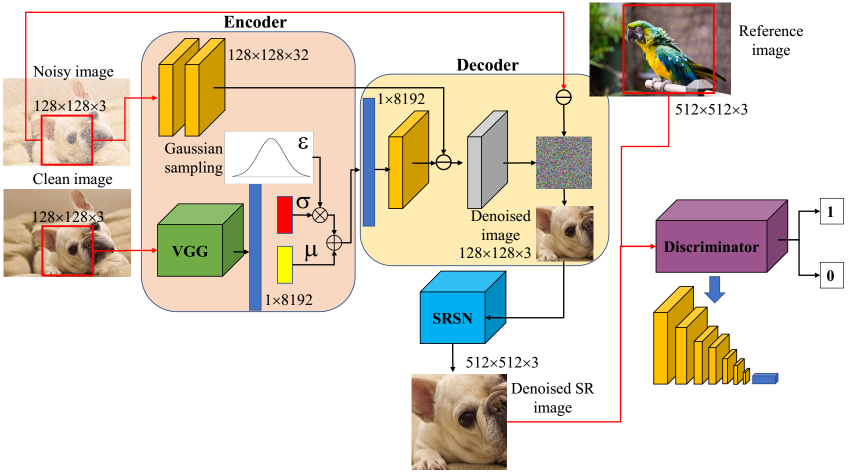}
	\caption{Overview of the method proposed by the \textbf{SR-DL} team.}
	\label{fig:SR-DL}
\end{figure}

The method is visualized in Figure~\ref{fig:SR-DL}. The proposed dSRVAE network first uses the encoder
to learn the latent vector of the clean image. A Gaussian model randomly
samples from the latent vector to the decoder. The input noisy LR image
is also included as a conditional constraint to supervise the reconstruction
of the decoder. Combining both noisy image features and latent features,
the decoder learn the noise pattern. Finally, the estimated clean image is
obtained by subtracting the estimated noise from the input noisy image.
At the second stage, Super-Resolution Sub-Network (SRSN) is added to
the end of the Denoising AutoEncoder to take both bicubic interpolated
original clean and estimated denoised images as input to generation superresolution result. Since there is no ground truth of super-resolved images,
a discriminator is trained to distinguish the super-resolution results and
cropped reference image. The balance is achieved when the discriminator
cannot distinguish between reference and denoised SR image.

\teamsection{TeamAY}
This team proposes a simple but strong method for unsupervised SR (SimUSR). Their approach is based on the zero-shot super-resolution (ZSSR)~\cite{shocher2018zssr} which trains the image-specific network at runtime using only a single given test image $\mathbf{I}_{LR}$. The ZSSR enables to optimize the model even if high-resolution images are not accessible. However, ZSSR suffers from high runtime latency and inferior performance compared to the supervised SR methods. To mitigate such issues, this team first slightly relax the constraint of ZSSR and assumes that it is relatively easy to collect the LR images, $\{\mathbf{I}_{LR_1},\dots,\mathbf{I}_{LR_N}\}$. Thanks to this assumption, they can convert fully unsupervised SR into the supervised learning regime by generating  multiple pseudo-pairs $\{(\mathbf{I}'_{LR_1}, \mathbf{I}'_{HR_1}),\dots(\mathbf{I}'_{LR_N}, \mathbf{I}'_{HR_N})\}$ by
$$(\mathbf{I}'_{LR_k}, \mathbf{I}'_{HR_k}) = (\mathbf{I}_{LR_k}^{son}, \mathbf{I}_{LR_k}^{father}), \;\; \text{for}\;k = 1\dots N.$$ where $\mathbf{I}_{LR}^{son} = \mathbf{I}_{LR}\downarrow_{s, k}$ and $\mathbf{I}_{LR}^{father} = \mathbf{I}_{LR}$.

Though this is a very simple correction, their modification brings several benefits: It allows their framework to exploit every benefit of supervised learning. For instance, unlike ZSSR, their SimUSR can utilize recently developed network architectures and training techniques that provide huge performance gains. In addition, since the online (runtime) training is not necessary, SimUSR can significantly reduce its runtime latency. For the NTIRE 2020 challenge, they use pretrained RCAN~\cite{rcan} (on bicubic $\times$4 scale) as a backbone model of SimUSR. Also, they attach ad-hoc denoiser (BM3D~\cite{HouZYC11BM3D}) before train the SimUSR method.
Details about the proposed method can be found in~\cite{TeamAY}.

\teamsection{Bigfeature-Camera}

This method use DSGAN~\cite{manuelFS} to learn the degradation, used for generating paired training data. In the second stage a RNAN~\cite{rnan} based SR network is trained. It is modified to handle multiple scales and by adding a contrast channel attention layers~\cite{rcan} along with local attention blocks.

\teamsection{BMIPL-UNIST-YH}

This method focus on how to train on unpaired data. Similar to \cite{lugmayrICCVW2019}, a CycleGAN is used to learn the degradation. In the second stage, and RCAN~\cite{rcan} SR architecture is trained on generated data.

\teamsection{SVNIT1}

This team combines self- and unsupervised strategies to train the SR network without supervision. For the self-supervised part, the LR input is upsampled bicubically and used for a pixel-wise loss. The unsupervised losses consist of a Total-Variation loss and a deep image quality loss. For the latter loss, a pre-trained quality assessment network was used. Details about the proposed method can be found in~\cite{SVNIT1}. Two versions of this approach was submitted:

\parsection{SVNIT1-A} 
In addition to above, this version employs an adversarial loss on the SR output. The discriminator architecture is inspired by~\cite{Radford16}.

\parsection{SVNIT1-B} 
Instead of a descriminator, this variant Variational Encoder which follows the architectural guidelines in~\cite{Radford16}.

\teamsection{SVNIT2} 

This method uses cyclic consistency between an SR network and a downscaling network. Two generator are trained: the SR generator going from LR to HR and the downscaling generator going from HR to LR. In addition to cycle consistency, the VGG loss, GAN loss, and a learned image quality loss is employed.

\teamsection{KU-ISPL2}

This team base their approach on SRGAN~\cite{ledig2017photo}. This is extended with a multi-scale convolutional block, that combines the results of convolutions with different kernel sizes.

\teamsection{SuperT}

This method uses a balanced Laplacian pyramid network~\cite{LaiHA017DeepLapPyr} for progressive image super-resolution. For training, both degraded and clean images are used with standard downsampling them for training data generation.

\teamsection{GDUT-wp}

This method uses an ensemble of SRResNets trained on bicubic downsampled data. The idea is that by selecting the best from an ensemble, the effect of random artifacts can be reduced.

\teamsection{MLP-SR}

This method is based on the DSGAN~\cite{manuelFS} approach. The loss of the super-resolution method consists of a VGG, GAN, TV and L1 loss. To improve the fidelity, they further used a ensemble method at test time~\cite{Timofte16SevenWays}. Details about the proposed method can be found in~\cite{MLP-SR}.

\teamsection{Relbmag-Eht}

Instead of
generating `fake' natural image as DSGAN~\cite{manuelFS}, this team aims to improve this method to aggregate this paring procedure into the super-resolution model. To supervise this matching from HR or bicubic images
to natural images, a module with discriminators both in the LR
and HR phase is proposed. It allows the downsampling model to learn from upsampling results. The ESRGAN~\cite{wang2018esrgan} is used as SR model.

\teamsection{QCAM}

This work fine-tunes a pretrained SR model on real data using only supervision in the low-resolution. That is, it aims to minimize the loss $\min_\theta \|D(f_\theta(x)) - x\|^2$ for source images $x$. Here, $f_\theta$ is the SR model with parameters $\theta$ and $D$ is the bicubic downsampling operation.

\section{Conclusions}
This paper presents the setup and results of the NTIRE 2020 challenge on real world super-resolution. Contrary to conventional super-resolution, this challenge addresses the real world setting, where paired true high and low-resolution images are unavailable. For training, only one set of unpaired source and target input images were provided to the participants. The source images have unknown degradations, while the target images are clean, high quality images. The challenge contains two tracks, where the goal was to super-resolve images with Image Processing artifacts (Track 1) or low-quality smart-phone images (Track 2). The challenge had in total 22 teams competing in the final step. Most of the participating were influenced AIM 2019 and demonstrated interesting and innovative solutions. Our goal is that this challenge stimulates future research in the area of unsupervised learning for image super-resolution and other similar tasks, by serving as a standard benchmark and by the establishment of new baseline methods.

\section*{Acknowledgements}
We thank the NTIRE 2020 sponsors: Huawei, Oppo, Voyage81, MediaTek, DisneyResearch$\mid$Studios, and Computer Vision Lab (CVL) ETH Zurich.

\setcounter{section}{0}
\renewcommand{\thesection}{Appendix~\Alph{section}}

\section{Teams and affiliations}
\label{sec:affiliation}

\subsection*{AIM2019 organizers}
\noindent\textbf{Members:}\\
Andreas Lugmayr (andreas.lugmayr@vision.ee.ethz.ch)\\
Martin Danelljan (martin.danelljan@vision.ee.ethz.ch)\\
Radu Timofte (radu.timofte@vision.ee.ethz.ch)

\noindent\textbf{Affiliation:} Computer Vision Lab, ETH Zurich

\subsection*{SR-DL} \noindent\textbf{Title:} Generative Variational AutoEncoder for Real Image Super-Resolution

\noindent\textbf{Team Leader:}\\
Zhi-Song Liu (zhi-song.liu@inria.fr)

\noindent\textbf{Members:}\\
Zhi-Song Liu, LIX - Computer science laboratory at the École polytechnique [Palaiseau]\\
Li-Wen Wang, Center of Multimedia Signal Processing, The Hong Kong Polytechnic University\\
Marie-Paule Cani, LIX - Computer science laboratory at the École polytechnique [Palaiseau]\\
Wan-Chi Siu, Center of Multimedia Signal Processing, The Hong Kong Polytechnic University

\subsection*{MSMers} \noindent\textbf{Title:} Cycle-based Residual Channel Attention Network for Real-World Super-Resolution

\noindent\textbf{Team Leader:}\\
Fuzhi Yang (yfzcopy0702@sjtu.edu.cn)

\noindent\textbf{Members:}\\
Fuzhi Yang, Shanghai Jiao Tong University,\\
Huan Yang, Microsoft Research Beijing, P.R. China,\\
Jianlong Fu, Microsoft Research Beijing, P.R. China,\\

\subsection*{GDUT-wp} \noindent\textbf{Title:} Ensemble of ResNets for Image Restoration

\noindent\textbf{Team Leader:}\\
Hao Li (2111903004@mail2.gdut.edu.cn)

\noindent\textbf{Members:}\\
Hao Li, Guangdong University of Technology\\
Yukai Shi, Guangdong University of Technology\\
Junyang Chen, Guangdong University of Technology

\subsection*{KU-ISPL} \noindent\textbf{Title:} Unsupervised Real-World Super Resolution with Cycle-in-Cycle Generative Adversarial Network and Domain-Transfer Discriminator.

\noindent\textbf{Team Leader:}\\
Gwantae Kim (gtkim@ispl.korea.ac.kr)

\noindent\textbf{Members:}\\
Gwantae, Kim, Intelligent Signal Processing Laboratory, Korea University\\
Kanghyu, Lee,  Intelligent Signal Processing Laboratory, Korea University\\
Jaihyun, Park, Intelligent Signal Processing Laboratory, Korea University\\
Junyeop, Lee, Intelligent Signal Processing Laboratory, Korea University\\
Jeongki, Min, Intelligent Signal Processing Laboratory, Korea University\\
Bokyeung, Lee, Intelligent Signal Processing Laboratory, Korea University\\
Hanseok, Ko, Intelligent Signal Processing Laboratory, Korea University
    
\subsection*{TeamAY} \noindent\textbf{Title:} SimUSR: A Simple but Strong Baseline for Unsupervised Image Super-resolution

\noindent\textbf{Team Leader:}\\
Namhyuk Ahn (aa0dfg@ajou.ac.kr)

\noindent\textbf{Members:}\\
Namhyuk, Ahn, Ajou University\\
Jaejun, Yoo, EPFL\\
Kyung-Ah, Sohn, Ajou University    

\subsection*{MLP-SR} \noindent\textbf{Title:} Deep Generative Adversarial Residual Convolutional Networks for Real-World Super-Resolution

\noindent\textbf{Team Leader:}\\
Rao Muhammad Umer (engr.raoumer943@gmail.com)

\noindent\textbf{Members:}\\
Rao Muhammad Umer,  University of Udine, Italy.\\
Christian Micheloni,  University of Udine, Italy.

\subsection*{BOE-IOT-AIBD} \noindent\textbf{Title:} DSGAN and Triple-V MGBPv2 for Real Super-Resolution

\noindent\textbf{Team Leader:}\\
Pablo Navarrete Michelini (pnavarre@boe.com.cn)

\noindent\textbf{Members:}\\
Pablo, Navarrete Michelini, BOE Technology Group Co. Ltd.\\
Fengshuo, Hu, BOE Technology Group Co. Ltd.\\
Yanhong, Wang, BOE Technology Group Co. Ltd.\\
Yunhua, Lu, BOE Technology Group Co. Ltd.

\subsection*{SuperT} \noindent\textbf{Title:} Fast and Balanced Laplacian Pyramid Networks for Progressive image super-resolution

\noindent\textbf{Team Leader:}\\
Tongtong Zhao (daitoutiere@gmail.com)

\noindent\textbf{Members:}\\
Jinjia,Peng, Dalian Maritime University\\
Huibing,Wang, Dalian Maritime University

\subsection*{BIGFEATURE-CAMERA} \noindent\textbf{Title:} Deep Residual Mix Attention Network for Image Super-Resolution

\noindent\textbf{Team Leader:}\\
Kaihua Cheng (consonwm0909@gmail.com)

\noindent\textbf{Members:}\\
Kaihua Cheng, Guangdong OPPO Mobile Telecommunications Corp., Ltd\\
Haijie Zhuo, Guangdong OPPO Mobile Telecommunications Corp., Ltd

\subsection*{KU-ISPL2} \noindent\textbf{Title:} Modular generative adversarial network based super-resolution

\noindent\textbf{Team Leader:}\\
Kanghyu Lee (khlee@ispl.korea.ac.kr)

\noindent\textbf{Members:}\\
Gwantae Kim is with Department of Video Information Processing, Korea University\\
Junyeop Lee is with School of Electrical Engineering, Korea University\\
Jeongki Min is with School of Electrical Engineering, Korea University\\
Bokyeung Lee is with School of Electrical Engineering, Korea University\\
Jaihyun Park is with School of Electrical Engineering, Korea University\\
Hanseok Ko is with School of Electrical Engineering, Korea University

\subsection*{Impressionism} \noindent\textbf{Title:} Real World Super-Resolution via Kernel Estimation and Noise Injection

\noindent\textbf{Team Leader:}\\
Xiaozhong Ji (shawn\_ji@163.com)

\noindent\textbf{Members:}\\
Xiaozhong Ji, Tencent Youtu Lab
Yun Cao, Tencent Youtu Lab
Ying Tai, Tencent Youtu Lab 
Chengjie Wang, Tencent Youtu Lab 
Jilin Li, Tencent Youtu Lab     
Feiyue Huang, Tencent Youtu Lab 

\subsection*{Relbmag Eht} \noindent\textbf{Title:} Network of Aggregated Downsampler-Upsampler with Dual Resolution Domain Matching

\noindent\textbf{Team Leader:}\\
Timothy Haoning Wu (1700012826@pku.edu.cn)

\noindent\textbf{Members:}\\
Haoning Wu, Peking University

\subsection*{ITS425} \noindent\textbf{Title:} Addptive Residual dense block network for real image super-resolution

\noindent\textbf{Team Leader:}\\
Ziyao Zong (824924664@qq.com)

\noindent\textbf{Members:}\\
Ziyao Zong, North China University of Technology\\
Shuai Liu, North China University of Technology\\
Biao Yang, North China University of Technology

\subsection*{AITA-Noah} \noindent\textbf{Title:} Real World Super-Resolution with Iterative Domain Adaptation, CycleSR and Conditional Freqency Separation GAN
 
\noindent\textbf{Team Leader:}\\
Ziluan Liu (liuziluan@huawei.com), Xueyi Zou (zouxueyi@huawei.com)
 
\noindent\textbf{Members:}\\
Xing Liu, Huawei Technologies Co., Ltd\\
Shuaijun Chen, Huawei Technologies Co., Ltd\\
Lei Zhao, Huawei Technologies Co., Ltd\\
Zhan Wang, Huawei Technologies Co., Ltd\\
Yuxuan Lin, Huawei Technologies Co., Ltd\\
Xu Jia, Huawei Technologies Co., Ltd\\
Ziluan Liu, Huawei Technologies Co., Ltd\\
Xueyi Zou, Huawei Technologies Co., Ltd\\

\subsection*{Webbzhou} \noindent\textbf{Title:} Guided Frequency Separation Network for Real World Super-Resolution

\noindent\textbf{Team Leader:}\\
Yuanbo Zhou (webbozhou@gmail.com)

\noindent\textbf{Members:}\\
Yuanbo Zhou,Fuzhou University 
Tong Tong, Fuzhou University, Imperial Vision Technology\\
Qinquan Gao, Fuzhou University, Imperial Vision Technology\\
Wei Deng, Imperial Vision Technology

\subsection*{Samsung-SLSI-MSL} \noindent\textbf{Title:} Real-World Super-Resolution using Generative Adversarial Networks

\noindent\textbf{Team Leader:}\\
Haoyu Ren, Amin Kheradmand (co-leader) (haoyu.ren@samsung.com)

\noindent\textbf{Members:}\\
Haoyu Ren     SOC R\&D, Samsung Semiconductor, Inc., USA\\
Amin Kheradmand    SOC R\&D, Samsung Semiconductor, Inc., USA\\
Mostafa El-Khamy   SOC R\&D, Samsung Semiconductor, Inc., USA\\
Shuangquan Wang   SOC R\&D, Samsung Semiconductor, Inc., USA \\
Dongwoon Bai   SOC R\&D, Samsung Semiconductor, Inc., USA \\
Jungwon Lee   SOC R\&D, Samsung Semiconductor, Inc., USA

\subsection*{ BMIPL-UNIST-YH-1} \noindent\textbf{Title:} Unpaired Domain-Adaptive Image Super-Resolution using Cycle-Consistent Adversarial Networks

\noindent\textbf{Team Leader:}\\
YongHyeok Seo (syh4661@unist.ac.kr)

\noindent\textbf{Members:}\\
SeYoung Chun, Ulsan national institute of science and technology

\subsection*{SVNIT1} \noindent\textbf{Title:} Unsupervised Real-World Single Image Super-Resolution (SR) using Generative Adversarial Networks (GAN) and Variational Auto-Encoders (VAE) 

\noindent\textbf{Team Leader:}\\
Kalpesh Prajapati (kalpesh.jp89@gmail.com)

\noindent\textbf{Members:}\\
Heena, Patel, Sardar Vallabhbhai National Institute Of Technology, Surat\\
Vishal, Chudasama, Sardar Vallabhbhai National Institute Of Technology, Surat\\
Kishor, Upla, Norwegian University of Science and Technology, Gj{\o}vik, Norway\\
Raghavendra, Ramachandra, Norwegian University of Science and Technology, Gj{\o}vik, Norway\\
Kiran, Raja, Norwegian University of Science and Technology, Gj{\o}vik, Norway\\
Christoph, Busch, Norwegian University of Science and Technology, Gj{\o}vik, Norway

\subsection*{SVNIT2} \noindent\textbf{Title:} Unsupervised Real-World Single Image Super-Resolution (SR)

\noindent\textbf{Team Leader:}\\
Kalpesh Prajapati (kalpesh.jp89@gmail.com)

\noindent\textbf{Members:}\\
Heena, Patel, Sardar Vallabhbhai National Institute Of Technology, Surat\\
Vishal, Chudasama, Sardar Vallabhbhai National Institute Of Technology, Surat\\
Kishor, Upla, Norwegian University of Science and Technology, Gj{\o}vik, Norway\\
Raghavendra, Ramachandra, Norwegian University of Science and Technology, Gj{\o}vik, Norway\\
Kiran, Raja, Norwegian University of Science and Technology, Gj{\o}vik, Norway\\
Christoph, Busch, Norwegian University of Science and Technology, Gj{\o}vik, Norway

\subsection*{InnoPeak-SR} \noindent\textbf{Title:} Deep Residual Channel Attention Generative Adversarial Networks for Image Super-Resolution and Noise Reduction

\noindent\textbf{Team Leader:}\\
Jie Cai (jie.cai@innopeaktech.com)

\noindent\textbf{Members:}\\
Jie Cai, InnoPeak Technology\\
Zibo Meng, InnoPeak Technology\\
Chiu Man Ho, InnoPeak Technology

{\small
\bibliographystyle{ieee_fullname}
\bibliography{references}

\begin{thebibliography}{10}\itemsep=-1pt

\bibitem{abdelhamed2020ntire}
Abdelrahman Abdelhamed, Mahmoud Afifi, Radu Timofte, Michael Brown, et~al.
\newblock {NTIRE} 2020 challenge on real image denoising: Dataset, methods and
  results.
\newblock In {\em The IEEE Conference on Computer Vision and Pattern
  Recognition (CVPR) Workshops}, June 2020.

\bibitem{ahn2018fast}
Namhyuk Ahn, Byungkon Kang, and Kyung-Ah Sohn.
\newblock Fast, accurate, and lightweight super-resolution with cascading
  residual network.
\newblock In {\em ECCV}, 2018.

\bibitem{ahn2018image}
Namhyuk Ahn, Byungkon Kang, and Kyung-Ah Sohn.
\newblock Image super-resolution via progressive cascading residual network.
\newblock In {\em CVPR}, 2018.

\bibitem{ancuti2020ntire}
Codruta~O. Ancuti, Cosmin Ancuti, Florin-Alexandru Vasluianu, Radu Timofte,
  et~al.
\newblock {NTIRE} 2020 challenge on nonhomogeneous dehazing.
\newblock In {\em The IEEE Conference on Computer Vision and Pattern
  Recognition (CVPR) Workshops}, June 2020.

\bibitem{arad2020ntire}
Boaz Arad, Radu Timofte, Yi-Tun Lin, Graham Finlayson, Ohad Ben-Shahar, et~al.
\newblock {NTIRE} 2020 challenge on spectral reconstruction from an rgb image.
\newblock In {\em The IEEE Conference on Computer Vision and Pattern
  Recognition (CVPR) Workshops}, June 2020.

\bibitem{begin2004blind}
Isabelle Begin and FR Ferrie.
\newblock Blind super-resolution using a learning-based approach.
\newblock In {\em ICPR}, 2004.

\bibitem{Bell19InternalGAN}
Sefi Bell{-}Kligler, Assaf Shocher, and Michal Irani.
\newblock Blind super-resolution kernel estimation using an internal-gan.
\newblock In {\em NeurIPS}, pages 284--293, 2019.

\bibitem{bulat2018learn}
Adrian Bulat, Jing Yang, and Georgios Tzimiropoulos.
\newblock To learn image super-resolution, use a gan to learn how to do image
  degradation first.
\newblock {\em arXiv preprint arXiv:1807.11458}, 2018.

\bibitem{cai2020residual}
Jie Cai, Zibo Meng, and Chiu~Man Ho.
\newblock Residual channel attention generative adversarial network for image
  super-resolution and noise reduction.
\newblock In {\em CVPR Workshops}, 2020.

\bibitem{AITA-Noah}
Shuaijun Chen, Enyan Dai, Zhen Han, Xu Jia, Ziluan Liu, Xing Liu, Xueyi Zou,
  Chunjing Xu, Jianzhuang Liu, and Qi Tian.
\newblock Unsupervised image super-resolution with an indirect supervised path.
\newblock In {\em CVPRW}, 2020.

\bibitem{Chen17}
Yunjin Chen and Thomas Pock.
\newblock Trainable nonlinear reaction diffusion: {A} flexible framework for
  fast and effective image restoration.
\newblock {\em {IEEE} Trans. Pattern Anal. Mach. Intell.}, 39(6):1256--1272,
  2017.

\bibitem{chu2017cyclegan}
Casey Chu, Andrey Zhmoginov, and Mark Sandler.
\newblock Cyclegan, a master of steganography.
\newblock {\em arXiv preprint arXiv:1712.02950}, 2017.

\bibitem{DaiTG15JointlyOptimizedRegressorsForSR}
Dengxin Dai, Radu Timofte, and Luc~Van Gool.
\newblock Jointly optimized regressors for image super-resolution.
\newblock {\em Comput. Graph. Forum}, 34(2):95--104, 2015.

\bibitem{dong2014learning}
Chao Dong, Chen~Change Loy, Kaiming He, and Xiaoou Tang.
\newblock Learning a deep convolutional network for image super-resolution.
\newblock In {\em ECCV}, 2014.

\bibitem{dong2016image}
Chao Dong, Chen~Change Loy, Kaiming He, and Xiaoou Tang.
\newblock Image super-resolution using deep convolutional networks.
\newblock {\em TPAMI}, 38(2):295--307, 2016.

\bibitem{fan2017balanced}
Yuchen Fan, Honghui Shi, Jiahui Yu, Ding Liu, Wei Han, Haichao Yu, Zhangyang
  Wang, Xinchao Wang, and Thomas~S Huang.
\newblock Balanced two-stage residual networks for image super-resolution.
\newblock In {\em CVPR}, 2017.

\bibitem{freeman2002example}
William~T Freeman, Thouis~R Jones, and Egon~C Pasztor.
\newblock Example-based super-resolution.
\newblock {\em IEEE Computer graphics and Applications}, 2002.

\bibitem{manuelFS}
Manuel Fritsche, Shuhang Gu, and Radu Timofte.
\newblock Frequency separation for real-world super-resolution.
\newblock In {\em 2019 {IEEE/CVF} International Conference on Computer Vision
  Workshops, {ICCV} Workshops 2019, Seoul, Korea (South), October 27-28, 2019},
  pages 3599--3608, 2019.

\bibitem{fuoli2020ntire}
Dario Fuoli, Zhiwu Huang, Martin Danelljan, Radu Timofte, et~al.
\newblock {NTIRE} 2020 challenge on video quality mapping: Methods and results.
\newblock In {\em The IEEE Conference on Computer Vision and Pattern
  Recognition (CVPR) Workshops}, June 2020.

\bibitem{gu2019blind}
Jinjin Gu, Hannan Lu, Wangmeng Zuo, and Chao Dong.
\newblock Blind super-resolution with iterative kernel correction.
\newblock In {\em CVPR}, 2019.

\bibitem{AIM2019ESR}
Shuhang Gu, Martin Danelljan, Radu Timofte, et~al.
\newblock Aim 2019 challenge on image extreme super-resolution: Methods and
  results.
\newblock In {\em ICCV Workshops}, 2019.

\bibitem{div8k}
Shuhang Gu, Andreas Lugmayr, Martin Danelljan, Manuel Fritsche, Julien Lamour,
  and Radu Timofte.
\newblock {DIV8K:} diverse 8k resolution image dataset.
\newblock In {\em 2019 {IEEE/CVF} International Conference on Computer Vision
  Workshops, {ICCV} Workshops 2019, Seoul, Korea (South), October 27-28, 2019},
  pages 3512--3516, 2019.

\bibitem{WGAN-GP}
Ishaan Gulrajani, Faruk Ahmed, Mart{\'{\i}}n Arjovsky, Vincent Dumoulin, and
  Aaron~C. Courville.
\newblock Improved training of wasserstein gans.
\newblock In {\em Advances in Neural Information Processing Systems 30: Annual
  Conference on Neural Information Processing Systems 2017, 4-9 December 2017,
  Long Beach, CA, {USA}}, pages 5767--5777, 2017.

\bibitem{UISRhan2019}
Zhen Han, Enyan Dai, Xu Jia, Xiaoying Ren, Shuaijun Chen, Chunjing Xu,
  Jianzhuang Liu, and Qi Tian.
\newblock Unsupervised image super-resolution with an indirect supervised path.
\newblock {\em CoRR}, abs/1910.02593, 2019.

\bibitem{samsung}
Mostafa El-Khamy Shuangquan Wang Dongwoon Bai Jungwon~Lee Haoyu~Ren,
  Amin~Kheradmand.
\newblock Real-world super-resolution using generative adversarial networks.
\newblock In {\em CVPRW}, 2020.

\bibitem{haris2018deep}
Muhammad Haris, Gregory Shakhnarovich, and Norimichi Ukita.
\newblock Deep back-projection networks for super-resolution.
\newblock In {\em CVPR}, 2018.

\bibitem{HouZYC11BM3D}
Yingkun Hou, Chunxia Zhao, Deyun Yang, and Yong Cheng.
\newblock Comments on "image denoising by sparse 3-d transform-domain
  collaborative filtering".
\newblock {\em {IEEE} Trans. Image Processing}, 20(1):268--270, 2011.

\bibitem{huang2015single}
Jia-Bin Huang, Abhishek Singh, and Narendra Ahuja.
\newblock Single image super-resolution from transformed self-exemplars.
\newblock In {\em CVPR}, 2015.

\bibitem{huang2018densely}
Yiwen Huang and Ming Qin.
\newblock Densely connected high order residual network for single frame image
  super resolution.
\newblock {\em arXiv preprint arXiv:1804.05902}, 2018.

\bibitem{IgnatovKTVG17DPED}
Andrey Ignatov, Nikolay Kobyshev, Radu Timofte, Kenneth Vanhoey, and Luc~Van
  Gool.
\newblock Dslr-quality photos on mobile devices with deep convolutional
  networks.
\newblock In {\em ICCV}, pages 3297--3305. {IEEE} Computer Society, 2017.

\bibitem{WESPE}
Andrey Ignatov, Nikolay Kobyshev, Radu Timofte, Kenneth Vanhoey, and Luc~Van
  Gool.
\newblock {WESPE:} weakly supervised photo enhancer for digital cameras.
\newblock In {\em 2018 {IEEE} Conference on Computer Vision and Pattern
  Recognition Workshops, {CVPR} Workshops 2018, Salt Lake City, UT, USA, June
  18-22, 2018}, pages 691--700, 2018.

\bibitem{ignatov2018pirm}
Andrey Ignatov, Radu Timofte, Thang Van~Vu, Tung~Minh Luu, Trung~X Pham, Cao
  Van~Nguyen, Yongwoo Kim, Jae-Seok Choi, Munchurl Kim, Jie Huang, et~al.
\newblock Pirm challenge on perceptual image enhancement on smartphones:
  Report.
\newblock {\em arXiv preprint arXiv:1810.01641}, 2018.

\bibitem{irani1991improving}
Michal Irani and Shmuel Peleg.
\newblock Improving resolution by image registration.
\newblock {\em CVGIP}, 1991.

\bibitem{isola2017image}
Phillip Isola, Jun-Yan Zhu, Tinghui Zhou, and Alexei~A Efros.
\newblock Image-to-image translation with conditional adversarial networks.
\newblock 2017.

\bibitem{Impressionism}
Xiaozhong Ji, Yun Cao, Ying Tai, Chengjie Wang, Jilin Li, and Feiyue Huang.
\newblock Real world super-resolution via kernel estimation and noise
  injection.
\newblock In {\em CVPRW}, 2020.

\bibitem{kim2020unsupervised}
Gwantae Kim, Kanghyu Lee, Junyeop Lee, Jeongki Min, Bokyeung Lee, Jaihyun Park,
  David~K. Han, and Hanseok Ko.
\newblock Unsupervised real-world super resolution with cycle generative
  adversarial network and domain discriminator.
\newblock In {\em CVPR Workshops}, 2020.

\bibitem{kim2018task}
Heewon Kim, Myungsub Choi, Bee Lim, and Kyoung~Mu Lee.
\newblock Task-aware image downscaling.
\newblock {\em ECCV}, 2018.

\bibitem{kim2016accurate}
Jiwon Kim, Jung Kwon~Lee, and Kyoung Mu~Lee.
\newblock Accurate image super-resolution using very deep convolutional
  networks.
\newblock In {\em CVPR}, 2016.

\bibitem{LaiHA017DeepLapPyr}
Wei{-}Sheng Lai, Jia{-}Bin Huang, Narendra Ahuja, and Ming{-}Hsuan Yang.
\newblock Deep laplacian pyramid networks for fast and accurate
  super-resolution.
\newblock In {\em {CVPR}}, pages 5835--5843. {IEEE} Computer Society, 2017.

\bibitem{lai2017deep}
Wei-Sheng Lai, Jia-Bin Huang, Narendra Ahuja, and Ming-Hsuan Yang.
\newblock Deep laplacian pyramid networks for fast and accurate
  super-resolution.
\newblock In {\em CVPR}, 2017.

\bibitem{ledig2017photo}
Christian Ledig, Lucas Theis, Ferenc Husz{\'a}r, Jose Caballero, Andrew
  Cunningham, Alejandro Acosta, Andrew~P Aitken, Alykhan Tejani, Johannes Totz,
  Zehan Wang, et~al.
\newblock Photo-realistic single image super-resolution using a generative
  adversarial network.
\newblock {\em CVPR}, 2017.

\bibitem{UDN}
Stamatios Lefkimmiatis.
\newblock Universal denoising networks : {A} novel {CNN} architecture for image
  denoising.
\newblock In {\em 2018 {IEEE} Conference on Computer Vision and Pattern
  Recognition, {CVPR} 2018, Salt Lake City, UT, USA, June 18-22, 2018}, pages
  3204--3213, 2018.

\bibitem{lim2017EDSR}
Bee Lim, Sanghyun Son, Heewon Kim, Seungjun Nah, and Kyoung~Mu Lee.
\newblock Enhanced deep residual networks for single image super-resolution.
\newblock {\em CVPR}, 2017.

\bibitem{liu2020unsupervised}
Zhisong Liu, Wan-Chi Siu, Marie-Paule Cani, Li-Wen Wang, and Chu-Tak Li.
\newblock Unsupervised real image super-resolution via generative variational
  autoencoder.
\newblock In {\em CVPR Workshops}, 2020.

\bibitem{lugmayrICCVW2019}
Andreas Lugmayr, Martin Danelljan, and Radu Timofte.
\newblock Unsupervised learning for real-world super-resolution.
\newblock In {\em ICCV Workshops}, 2019.

\bibitem{AIM2019RWSRchallenge}
Andreas Lugmayr, Martin Danelljan, Radu Timofte, et~al.
\newblock Aim 2019 challenge on real-world image super-resolution: Methods and
  results.
\newblock In {\em ICCV Workshops}, 2019.

\bibitem{lugmayr2020ntire}
Andreas Lugmayr, Martin Danelljan, Radu Timofte, et~al.
\newblock {NTIRE} 2020 challenge on real-world image super-resolution: Methods
  and results.
\newblock In {\em The IEEE Conference on Computer Vision and Pattern
  Recognition (CVPR) Workshops}, June 2020.

\bibitem{MaYY017NRQM}
Chao Ma, Chih{-}Yuan Yang, Xiaokang Yang, and Ming{-}Hsuan Yang.
\newblock Learning a no-reference quality metric for single-image
  super-resolution.
\newblock {\em Comput. Vis. Image Underst.}, 158:1--16, 2017.

\bibitem{michaeli2013nonparametric}
Tomer Michaeli and Michal Irani.
\newblock Nonparametric blind super-resolution.
\newblock In {\em ICCV}, 2013.

\bibitem{MGBPv2}
Pablo~Navarrete Michelini, Wenbin Chen, Hanwen Liu, and Dan Zhu.
\newblock Mgbpv2: Scaling up multi-grid back-projection networks.
\newblock In {\em 2019 {IEEE/CVF} International Conference on Computer Vision
  Workshops, {ICCV} Workshops 2019, Seoul, Korea (South), October 27-28, 2019},
  pages 3399--3407, 2019.

\bibitem{mittal2011brisque}
A Mittal, AK Moorthy, and AC Bovik.
\newblock Referenceless image spatial quality evaluation engine.
\newblock In {\em 45th Asilomar Conference on Signals, Systems and Computers},
  volume~38, pages 53--54, 2011.

\bibitem{MittalSB13NIQE}
Anish Mittal, Rajiv Soundararajan, and Alan~C. Bovik.
\newblock Making a "completely blind" image quality analyzer.
\newblock {\em {IEEE} Signal Process. Lett.}, 20(3):209--212, 2013.

\bibitem{NDBCM15piqe}
Venkatanath N., Praneeth D., Maruthi~Chandrasekhar Bh., Sumohana~S.
  Channappayya, and Swarup~S. Medasani.
\newblock Blind image quality evaluation using perception based features.
\newblock In {\em {NCC}}, pages 1--6. {IEEE}, 2015.

\bibitem{nah2020ntire}
Seungjun Nah, Sanghyun Son, Radu Timofte, Kyoung~Mu Lee, et~al.
\newblock {NTIRE} 2020 challenge on image and video deblurring.
\newblock In {\em The IEEE Conference on Computer Vision and Pattern
  Recognition (CVPR) Workshops}, June 2020.

\bibitem{TeamAY}
Ahn Namhyuk, Jaejun Yoo, and Kyung-Ah Sohn.
\newblock Simusr: A simple but strong baseline for unsupervised image
  super-resolution.
\newblock In {\em CVPR Workshops}, 2020.

\bibitem{park2003super}
Sung~Cheol Park, Min~Kyu Park, and Moon~Gi Kang.
\newblock Super-resolution image reconstruction: a technical overview.
\newblock {\em IEEE signal processing magazine}, 2003.

\bibitem{SVNIT1}
Kalpesh~J Prajapati, Vishal Chudasama, Heena Patel, Kishor Upla, Raghavendra
  Ramachandra, Kiran Raja, and Christoph Busch.
\newblock Unsupervised single image super-resolution network (usisresnet) for
  real-world data using generative adversarial network.
\newblock In {\em CVPRW}, 2020.

\bibitem{Radford16}
Alec Radford, Luke Metz, and Soumith Chintala.
\newblock Unsupervised representation learning with deep convolutional
  generative adversarial networks.
\newblock In {\em 4th International Conference on Learning Representations,
  {ICLR} 2016, San Juan, Puerto Rico, May 2-4, 2016, Conference Track
  Proceedings}, 2016.

\bibitem{RenEL18}
Haoyu Ren, Mostafa El{-}Khamy, and Jungwon Lee.
\newblock Dn-resnet: Efficient deep residual network for image denoising.
\newblock In {\em Computer Vision - {ACCV} 2018 - 14th Asian Conference on
  Computer Vision, Perth, Australia, December 2-6, 2018, Revised Selected
  Papers, Part {V}}, pages 215--230, 2018.

\bibitem{shocher2018zssr}
Assaf Shocher, Nadav Cohen, and Michal Irani.
\newblock Zero-shot” super-resolution using deep internal learning.
\newblock In {\em CVPR}, 2018.

\bibitem{SunH12SRFromInternetScaleSceneMatching}
Libin Sun and James Hays.
\newblock Super-resolution from internet-scale scene matching.
\newblock In {\em ICCP}, 2012.

\bibitem{timofte2017ntire}
Radu Timofte, Eirikur Agustsson, Luc Van~Gool, Ming-Hsuan Yang, Lei Zhang, Bee
  Lim, Sanghyun Son, Heewon Kim, Seungjun Nah, Kyoung~Mu Lee, et~al.
\newblock Ntire 2017 challenge on single image super-resolution: Methods and
  results.
\newblock {\em CVPR Workshops}, 2017.

\bibitem{Timofte2014a+}
Radu Timofte, Vincent De~Smet, and Luc Van~Gool.
\newblock A+: Adjusted anchored neighborhood regression for fast
  super-resolution.
\newblock In {\em ACCV}, pages 111--126. Springer, 2014.

\bibitem{Timofte16SevenWays}
Radu Timofte, Rasmus Rothe, and Luc~Van Gool.
\newblock Seven ways to improve example-based single image super resolution.
\newblock In {\em {CVPR}}, pages 1865--1873. {IEEE} Computer Society, 2016.

\bibitem{Timofte13AnchNeighReg}
Radu Timofte, Vincent~De Smet, and Luc~Van Gool.
\newblock Anchored neighborhood regression for fast example-based
  super-resolution.
\newblock In {\em ICCV}, pages 1920--1927, 2013.

\bibitem{MLP-SR}
Rao~Muhammad Umer, Gian~Luca Foresti, and Christian Micheloni.
\newblock Deep generative adversarial residual convolutional networks for
  real-world super-resolution.
\newblock In {\em CVPRW}, 2020.

\bibitem{wang2018esrgan}
Xintao Wang, Ke Yu, Shixiang Wu, Jinjin Gu, Yihao Liu, Chao Dong, Chen~Change
  Loy, Yu Qiao, and Xiaoou Tang.
\newblock Esrgan: Enhanced super-resolution generative adversarial networks.
\newblock {\em ECCV}, 2018.

\bibitem{WangBSS04SSIM}
Zhou Wang, Alan~C. Bovik, Hamid~R. Sheikh, and Eero~P. Simoncelli.
\newblock Image quality assessment: from error visibility to structural
  similarity.
\newblock {\em {IEEE} Trans. Image Processing}, 13(4):600--612, 2004.

\bibitem{YangY13SimpleFuncSR}
Chih{-}Yuan Yang and Ming{-}Hsuan Yang.
\newblock Fast direct super-resolution by simple functions.
\newblock In {\em ICCV}, pages 561--568, 2013.

\bibitem{YangWHM08SRAsSparseRepresentationOfRawPatches}
Jianchao Yang, John Wright, Thomas~S. Huang, and Yi Ma.
\newblock Image super-resolution as sparse representation of raw image patches.
\newblock In {\em CVPR}, 2008.

\bibitem{YangWHM10SRViaSparseRep}
Jianchao Yang, John Wright, Thomas~S. Huang, and Yi Ma.
\newblock Image super-resolution via sparse representation.
\newblock {\em {IEEE} Trans. Image Processing}, 19(11):2861--2873, 2010.

\bibitem{URDGN}
Xin Yu and Fatih Porikli.
\newblock Ultra-resolving face images by discriminative generative networks.
\newblock In {\em ECCV}, pages 318--333, 2016.

\bibitem{yuan2020demoireing}
Shanxin Yuan, Radu Timofte, Ales Leonardis, Gregory Slabaugh, et~al.
\newblock {NTIRE} 2020 challenge on image demoireing: Methods and results.
\newblock In {\em The IEEE Conference on Computer Vision and Pattern
  Recognition (CVPR) Workshops}, June 2020.

\bibitem{yuan2018unsupervised}
Yuan Yuan, Siyuan Liu, Jiawei Zhang, Yongbing Zhang, Chao Dong, and Liang Lin.
\newblock Unsupervised image super-resolution using cycle-in-cycle generative
  adversarial networks.
\newblock {\em CVPR Workshops}, 2018.

\bibitem{zhang2020ntire}
Kai Zhang, Shuhang Gu, Radu Timofte, et~al.
\newblock {NTIRE} 2020 challenge on perceptual extreme super-resolution:
  Methods and results.
\newblock In {\em The IEEE Conference on Computer Vision and Pattern
  Recognition (CVPR) Workshops}, June 2020.

\bibitem{zhang2018unreasonable}
Richard Zhang, Phillip Isola, Alexei~A Efros, Eli Shechtman, and Oliver Wang.
\newblock The unreasonable effectiveness of deep features as a perceptual
  metric.
\newblock {\em CVPR}, 2018.

\bibitem{rcan}
Yulun Zhang, Kunpeng Li, Kai Li, Lichen Wang, Bineng Zhong, and Yun Fu.
\newblock Image super-resolution using very deep residual channel attention
  networks.
\newblock In {\em Computer Vision - {ECCV} 2018 - 15th European Conference,
  Munich, Germany, September 8-14, 2018, Proceedings, Part {VII}}, pages
  294--310, 2018.

\bibitem{rnan}
Yulun Zhang, Kunpeng Li, Kai Li, Bineng Zhong, and Yun Fu.
\newblock Residual non-local attention networks for image restoration.
\newblock In {\em 7th International Conference on Learning Representations,
  {ICLR} 2019, New Orleans, LA, USA, May 6-9, 2019}, 2019.

\bibitem{RDN}
Yulun Zhang, Yapeng Tian, Yu Kong, Bineng Zhong, and Yun Fu.
\newblock Residual dense network for image super-resolution.
\newblock In {\em 2018 {IEEE} Conference on Computer Vision and Pattern
  Recognition, {CVPR} 2018, Salt Lake City, UT, USA, June 18-22, 2018}, pages
  2472--2481, 2018.

\bibitem{Zhou2020GuideFS}
Yuanbo Zhou, Wei Deng, Tong Tong, and Qinquan Gao.
\newblock Guided frequency separation network for real-world super-resolution.
\newblock In {\em CVPR Workshops}, 2020.

\bibitem{zhu2017unpaired}
Jun-Yan Zhu, Taesung Park, Phillip Isola, and Alexei~A Efros.
\newblock Unpaired image-to-image translation using cycle-consistent
  adversarial networks.
\newblock {\em ICCV}, 2017.

\end{thebibliography}
}

\end{document}